\documentclass[journal]{IEEEtran}
\usepackage{xkeyval, color, textcomp}

\usepackage{multirow, makecell, siunitx}

\usepackage[numbers,sort&compress]{natbib}
\usepackage{times}
\usepackage{multibib}
   \usepackage[pdftex]{graphicx}
   \graphicspath{figures}
   \DeclareGraphicsExtensions{.pdf,.jpeg,.png}
\RequirePackage{amsmath}
\RequirePackage{bm}
\RequirePackage{esvect}
\RequirePackage{amssymb}
\RequirePackage{relsize}
\RequirePackage{mathtools}

\DeclareRobustCommand{\buv}[1]{\widehat{\mathbf{#1}}}
\DeclareRobustCommand{\bv}[1]{\vv{\mathbf{#1}}}
\DeclareRobustCommand{\IP}{\textsc{\tiny IP}}
\DeclareRobustCommand{\SF}{\textsc{\tiny SF}}
\usepackage{soul}

\usepackage[switch]{lineno}

\usepackage{array}
\newcolumntype{L}{>{\centering\arraybackslash}m{2cm}}
\ifCLASSOPTIONcompsoc
 \usepackage[caption=false,font=normalsize,labelfont=sf,textfont=sf]{subfig}
\else
 \usepackage[caption=false,font=footnotesize]{subfig}
\fi
\usepackage{fixltx2e}
\makeatletter
\newlength{\sfp@hseplen}\newlength{\sfp@vseplen}
\define@cmdkey{subfigpos}[sfp@]{pos}[ul]{}% \sfp@pos
\define@cmdkey{subfigpos}[sfp@]{font}[\small]{}% \sfp@font
\define@cmdkey{subfigpos}[sfp@]{vsep}[0\baselineskip]{\setlength{\sfp@vseplen}{\sfp@vsep}}% \sfp@vsep
\define@cmdkey{subfigpos}[sfp@]{hsep}[10pt]{\setlength{\sfp@hseplen}{\sfp@hsep}}% \sfp@hsep
\newcommand{\subfigimg}[3][,]{%
  \setkeys{Gin,subfigpos}{pos,font,vsep,hsep,#1}% Set (default) keys
  \setbox1=\hbox{\includegraphics{#3}}% Store image in box
  \ifnum\pdfstrcmp{\sfp@pos}{ul}=0% UPPER LEFT placement of subfig label
    \leavevmode\rlap{\usebox1}% Print image
    \rlap{\hspace*{\sfp@hsep}\raisebox{\dimexpr\ht1-\sfp@vsep}{\sfp@font{#2}}}% Print label
    \phantom{\usebox1}% Insert appropriate spacing
  \else\ifnum\pdfstrcmp{\sfp@pos}{ur}=0% UPPER RIGHT placement of subfig label
    \leavevmode\usebox1% Print image
    \llap{\raisebox{\dimexpr\ht1-\sfp@vsep}{\sfp@font{#2}}\hspace*{\sfp@hsep}}% Print label
  \else\ifnum\pdfstrcmp{\sfp@pos}{lr}=0% LOWER RIGHT placement of subfig label
    \leavevmode\usebox1% Print image
    \llap{\raisebox{\sfp@vsep}{\sfp@font{#2}}\hspace*{\sfp@hsep}}% Print label
  \else% Assume LOWER LEFT placement of subfig label
    \leavevmode\rlap{\usebox1}% Print image
    \rlap{\hspace*{\sfp@hseplen}\raisebox{\sfp@vsep}{\sfp@font{#2}}}% Print label
    \phantom{\usebox1}% Insert appropriate spacing
  \fi\fi\fi
}
\makeatother

\begin{document}
\title{Improving blood vessel tortuosity measurements via highly sampled numerical integration of the Frenet-Serret equations}

\author{Alexander~Brummer,
	David~Hunt,
        Van~Savage% <-this % stops a space
%\thanks{Manuscript received Month XX, XXXX; revised Month XX, XXXX.  This work was supported in part by the National Science Foundation under Grant 1254159}
\thanks{This work was supported in part by the National Science Foundation grant 1254159}
\thanks{A. Brummer and V. Savage are with the Institute for Quantitative and Computational Biosciences, and the Departments of Computational Medicine, and Ecology and Evolutionary Biology, University of California, Los Angeles, Los Angeles, CA, 90095 USA e-mail: abrummer@ucla.edu.}% <-this % stops a space
\thanks{D. Hunt is with the Institute for Learning and Brain Sciences, University of Washington, Seattle, WA, 98195 USA.}%
\thanks{V. Savage is also with the Santa Fe Institute, Santa Fe, NM, 87501.}% <-this % stops a space
}

\maketitle

%\linenumbers


% Abstract has a 250 word limit, no citations, single paragrah.  Must contain three or four key words or phrases
\begin{abstract}
Measures of vascular tortuosity\textemdash how curved and twisted a vessel is\textemdash are associated with a variety of vascular diseases.  Consequently, measurements of vessel tortuosity that are accurate and comparable across modality, resolution, and size are greatly needed.  Yet in practice, precise and consistent measurements are problematic\textemdash mismeasurements, inability to calculate, or contradictory and inconsistent measurements occur within and across studies.  Here, we present a new method of measuring vessel tortuosity that ensures improved accuracy.  Our method relies on numerical integration of the Frenet-Serret equations. By reconstructing the three-dimensional vessel coordinates from tortuosity measurements, we explain how to identify and use a minimally-sufficient sampling rate based on vessel radius while avoiding errors associated with oversampling and overfitting. Our work identifies a key failing in current practices of filtering asymptotic measurements and highlights inconsistencies and redundancies between existing tortuosity metrics. We demonstrate our method by applying it to manually constructed vessel phantoms with known measures of tortuousity, and 9,000 vessels from medical image data spanning human cerebral, coronary, and pulmonary vascular trees, and the carotid, abdominal, renal, and iliac arteries.
\end{abstract}


% \fi
\IEEEpeerreviewmaketitle

%%%%%%%%%------------INTRODUCTION-------------%%%%%%%%%%%

\section{Introduction}
\IEEEPARstart{R}{ecent} work in both clinical and mathematical modeling studies has shown that measures of vessel tortuosity\textemdash the extent of `curliness', `squiggliness', or `wiggliness'\textemdash serve as biomarkers of diseases such as atherosclerosis, hypertension, arteriovenous malformations, recovery from stroke or stent implantation, and classification of tumors and their response to intervention \cite{han_jvascres_2012, folarin_etal_microvascres_2010, shelton_etal_ultrasoundmedbio_2015, huang_etal_ieeetransmedim_2008, rao_etal_ieeebiomedeng_2016, bullitt_etal_ieeetransmedim_2003, bullitt_etal_medimcomcad_2005, gessner_etal_radiology_2012, lindsey_etal_moleimgbio_2017, piccinelli_etal_ieeetransmedim_2009, oloumi_etal_jmedim_2016, alilou_etal_scireports_2018, kobayashi_etal_japanesecirculation_2015}.  In the research literature there exists many different definitions of tortuosity, with researchers constructing measures designed specifically to target particular biomarkers and for a given cohort \cite{bullitt_etal_ieeetransmedim_2003, bullitt_etal_medimcomcad_2005, folarin_etal_microvascres_2010, gessner_etal_radiology_2012, rao_etal_ieeebiomedeng_2016, huang_etal_ieeetransmedim_2008, lindsey_etal_moleimgbio_2017, piccinelli_etal_ieeetransmedim_2009, shelton_etal_ultrasoundmedbio_2015, oloumi_etal_jmedim_2016, alilou_etal_scireports_2018, kobayashi_etal_japanesecirculation_2015, hart_etal_intjmedinfo_1999, grisan_etal_ieeeengmedbio_2003, bogunovic_etal_medimg_2012}.  Using two components of tortuosity\textemdash curvature and torsion\textemdash it is possible to completely reconstruct the measured vessel, thereby providing a systematic check for measurement error.  Yet, we have not found a single study that exploits this capability.  Furthermore, these two mathematical measures of vessels can be linked to local pressures and stresses that act on vessels, connecting physical mechanisms to morphological features \cite{han_jbiomech_2009a, han_jbiomech_2009b, gammack_etal_jfluidmech_2001}.  

As part of the measurement process, researchers have used sampling rates\textemdash the number of points constituting a vessel\textemdash nearly equal to the voxel dimensions, or resolution limits, for their images.  This practice has proven sufficient for binary differentiation between diseased and healthy vessels as long as there is no significant variation in modality, resolution, and vessel size \cite{bullitt_etal_ieeetransmedim_2003, bullitt_etal_acadrad_2005, gessner_etal_radiology_2012}.  However, we show that measurements made at voxel-level sampling rates generally underestimate common tortuosity metrics and obscure existing equivalencies between different proposed metrics, thereby reducing the diagnostic and prognostic potential of these valuable biomarkers.  This is of interest since growing experimental evidence demonstrates a causal link between vascular endothelial growth factor signaling and tumor angiogenesis \cite{pandey_etal_hypertension_2018, huang_etal_amjcancres_2018, apte_etal_cell_2019}.  As fluid shear stress and pressure can be mathematically expressed in terms of curvature and torsion \cite{berger_etal_annrevfluidmech_1983, gammack_etal_jfluidmech_2001}, improved accuracy in these two measures of tortuosity may better inform underlying disease pathology.

We present a method based on numerical integration that uses curvature and torsion measured at sub-voxel sampling rates.  From this we reconstruct vessel centerline coordinates to an accuracy related to the vessel radius, thereby allowing for comparisons across modality, resolution, and size.  To test this method we first examine manually constructed vessel phantoms of semi-circles (constant, non-zero curvature yet zero torsion), helices (constant, non-zero curvature and torsion), and salkowski curves (constant curvature and non-constant torsion) \cite{monterde_compaidgeomdes_2009}, all of which have tortuosity metrics that can be expressed analytically.  Following this, we examine previously published data of the common, external, and internal carotid arteries \cite{kamenskiy_etal_jbiomecheng_2012}; the abdominal aorta, associated right renal artery, and both left and right common iliac arteries \cite{oflynn_etal_annalsbiomedeng_2007}; complete coronary arterial trees \cite{vorobstova_etal_annalsbiomedeng_2016}; the anterior, posterior, left, and right middle cerebral vascular trees \cite{bullitt_etal_neurobiologyaging_2010}; and pulmonary arterial and venous trees from clinical imaging of patients with and without pulmonary hypertension \cite{helmberger_etal_plosone_2014} (see Fig.~\ref{fig:data_viz}).  Excluding the patients with pulmonary hypertension, all data are from healthy human patients.

%%%%%%%%%------------BACKGROUND-------------%%%%%%%%%%%

\section{Background}

\subsection{Curvature, Torsion, and the Frenet-Serret Coordinate Frame} 

Differential geometry was developed to deal with details of curved surfaces \cite{kuhnel_differential_geometry_2006, patrikalakis_etal_shape_interrogation_2009}. Here we provide the standard set of techniques borrowed from differential geometry to estimate the tortuosity of vessels.

Spatial curves are described in terms of position and vectors, $\bv{r}(s_j)$ that assume Cartesian coordinates\textemdash$\bv{r}(s_j) = x(s_j)\buv{x} + y(s_j)\buv{y} + z(s_j)\buv{z}$ with $(x(s_j), y(s_j), z(s_j))$ defined relative to the origin of the medical images. The subscript $j$ represents either the indexed voxel-space of the original images or a subsampled space.  We choose $s_j$ to be the arc length, defined numerically as the sum of the euclidean distances between all neighboring points from the first to the $j^{th}$ point, so $s_j = \sum_{i=1}^{i=j-1}||\bv{r}_{i+1} - \bv{r}_i ||$.

The first and second derivatives of points along a spatial curve define the osculating plane such that changes in the curves shape can be described in two ways\textemdash in-plane changes and out-of-plane changes.  Curvature, $\kappa(s_j)$, measures the rate of in-plane changes with respect to the curve's arc length $s_j$.  Torsion, $\tau(s_j)$, measures the rate of out-of-plane changes.  Together these quantities are fundamental descriptors of any continuously differentiable curve that can be used to reconstruct the spatial $x_j, y_j,$ and $z_j$-coordinates.  To calculate curvature and torsion, it is essential to introduce the Frenet-Serret coordinate frame.

\begin{figure}[!t]
    \centering  
    \includegraphics[width=3.45in]{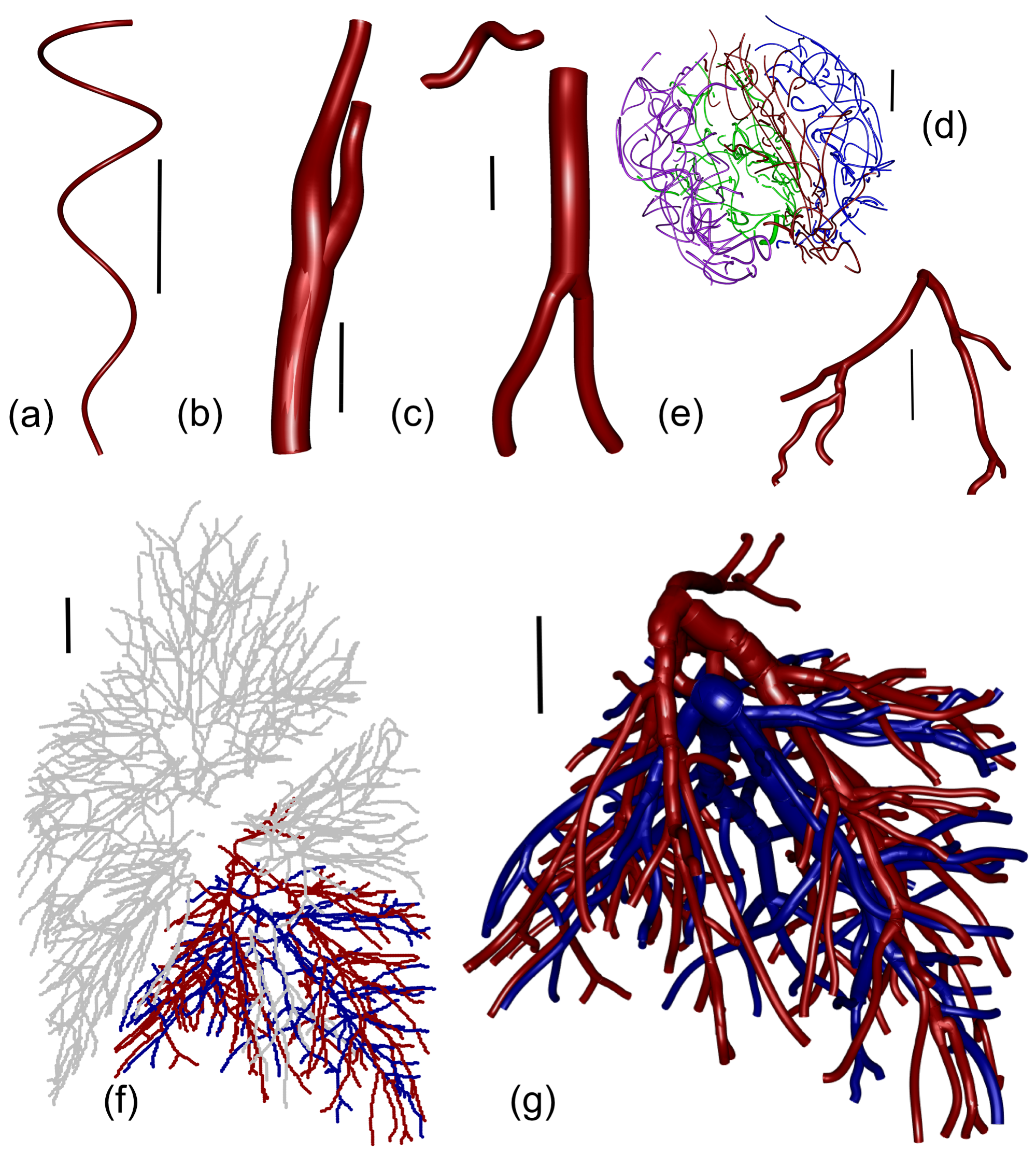}
    \footnotesize
    \caption{Visualizations of vessel phantom, vessels, and vessel trees studied.  ({\bf a}) A segment of a Salkowski curve, or spiral, noted for having constant curvature and non-constant torsion (see Supplementary Material \ref{sec:si_salk} and \cite{monterde_compaidgeomdes_2009}).  ({\bf b}) Common, external, and internal carotid arteries from Kamenskiy et al.~\cite{kamenskiy_etal_jbiomecheng_2012}. ({\bf c}) Abdominal, iliac, and renal arteries from O'Flynn et al.~\cite{oflynn_etal_annalsbiomedeng_2007}.  ({\bf d}) Posterior (green), anterior (brown), left middle (purple), and right middle (blue) cerebral arterial trees from Bullitt et al.~\cite{bullitt_etal_neurobiologyaging_2010}.  ({\bf e}) Coronary arterial trees from Vorobstova et al.~\cite{vorobstova_etal_annalsbiomedeng_2016}.  ({\bf f-g}) Pulmonary arterial (red) and venous (blue) trees from patients with pulmonary hypertension from Helmberger et al.~\cite{helmberger_etal_plosone_2014}. Solid black bars represent 20 $mm$ scales.}
    \label{fig:data_viz}
\end{figure}

\begin{figure}[!t]
    \centering  
    \includegraphics[width=3.45in]{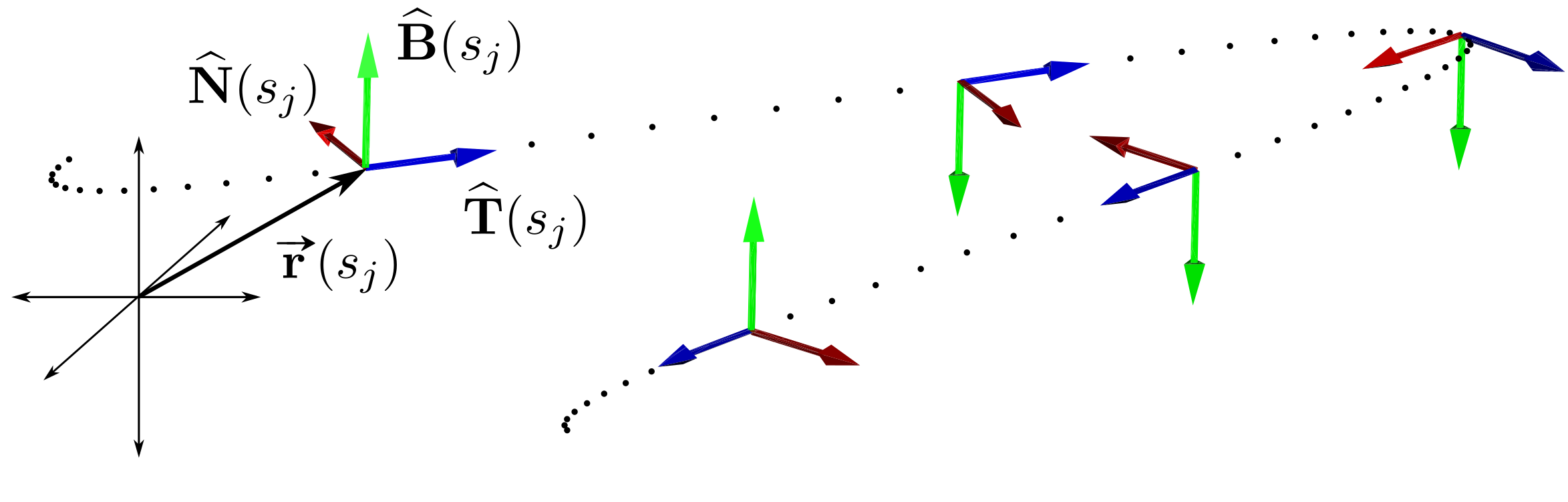}
    \footnotesize
    \caption{Origin of coordinates, position vector, $\bv{r}(s_j)$, and unit tangent, $\buv{T}(s_j)$, normal, $\buv{N}(s_j)$, and binormal, $\buv{B}(s_j)$, vectors (in blue, red, and green respectively) drawn on simulated 2D-data (one period of sinusoidal curve).}
    \label{fig:frame_coords}
\end{figure}

The Frenet-Serret (FS) frame is a moving coordinate system defined at all points along a curve.  This frame exists wherever the second derivative is continuous.  The FS-frame can be calculated from the position vector of the curve, $\bv{r}(s_j)$.  The unit-vectors that constitute the FS-frame are the tangent, $\buv{T}(s_j)$, the normal, $\buv{N}(s_j)$, and the binormal, $\buv{B}(s_j)$, vectors  (Figure \ref{fig:frame_coords}).  The tangent vector is the normalized, first-derivative of the position vector with respect to arc length, $\buv{T}(s_j) = \bv{r}'(s_j)/||\bv{r}'(s_j)||$, and the normal vector is the normalized, second-derivative of the position vector with respect to arc length, $\buv{N}(s_j) = \bv{r}''(s_j)/||\bv{r}''(s_j)||$.  These two unit-vectors point respectively in the instantaneous directions of velocity and acceleration of an object in motion.  Finally, the binormal vector is the cross-product of the tangent and normal vectors, $\buv{B}(s_j) = \buv{T}(s_j)\times\buv{N}(s_j)$ that points in the direction of angular velocity for rigid body rotation.

\begin{figure*}[!t]
    \centering
    \subfigimg[width=2.6in, pos = ll]{\textbf{(a)}}{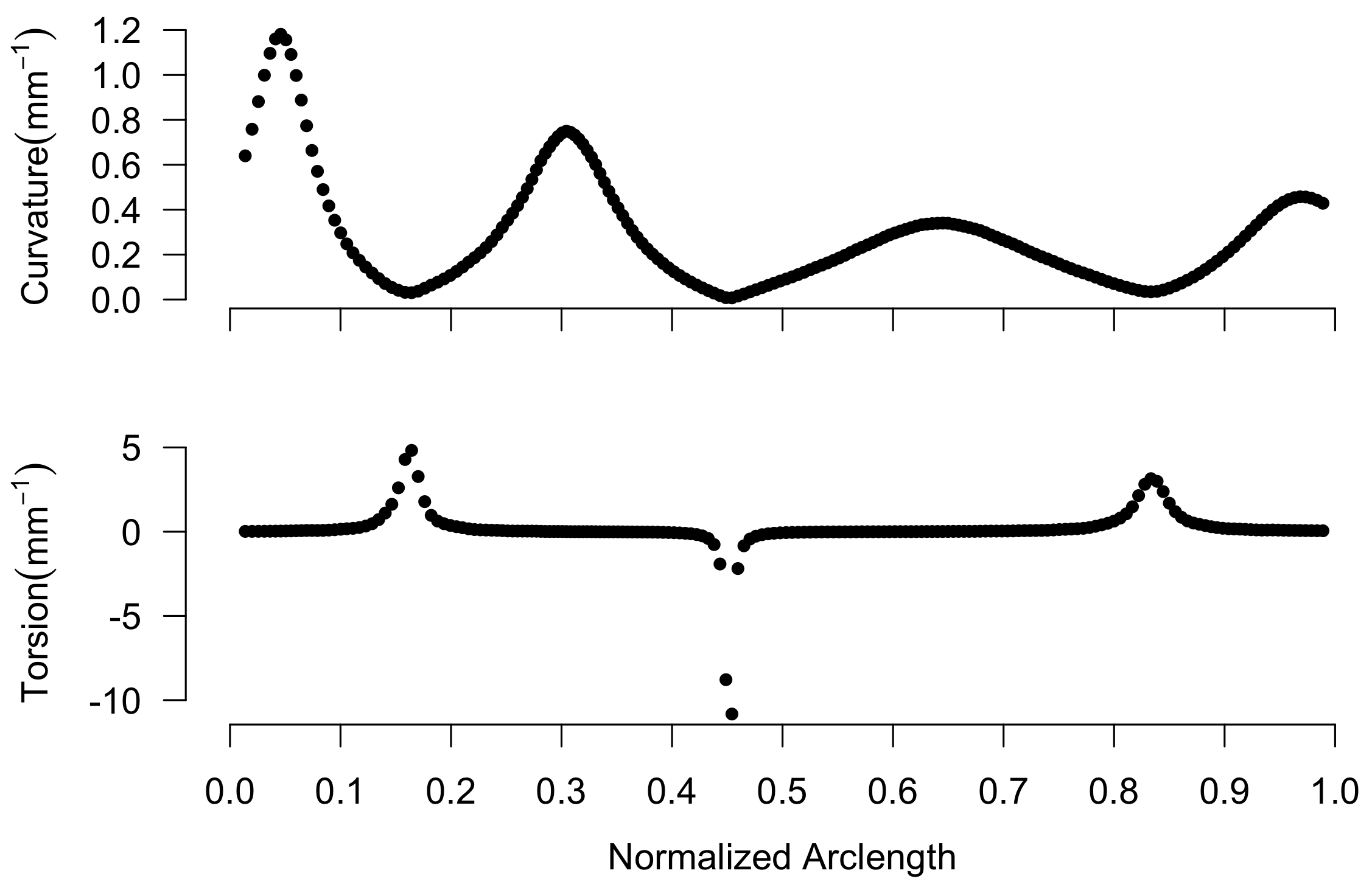}
    \hfill
    \subfigimg[width=2.0in, pos = ll]{\textbf{(b)}}{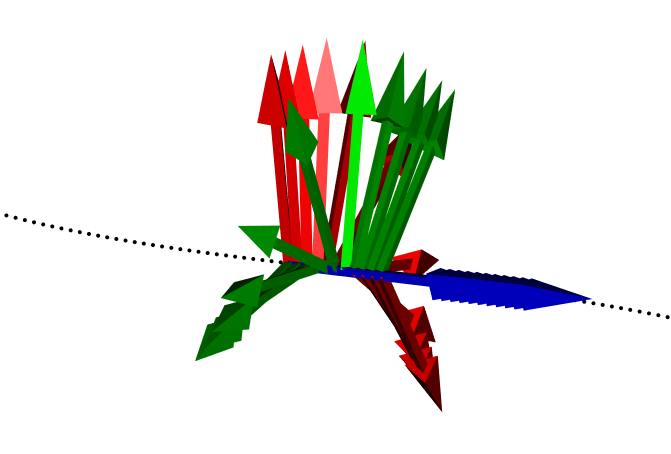}
    \hfill
    \subfigimg[width=2.3in, height = 1.75in, pos = ll]{\textbf{(c)}}{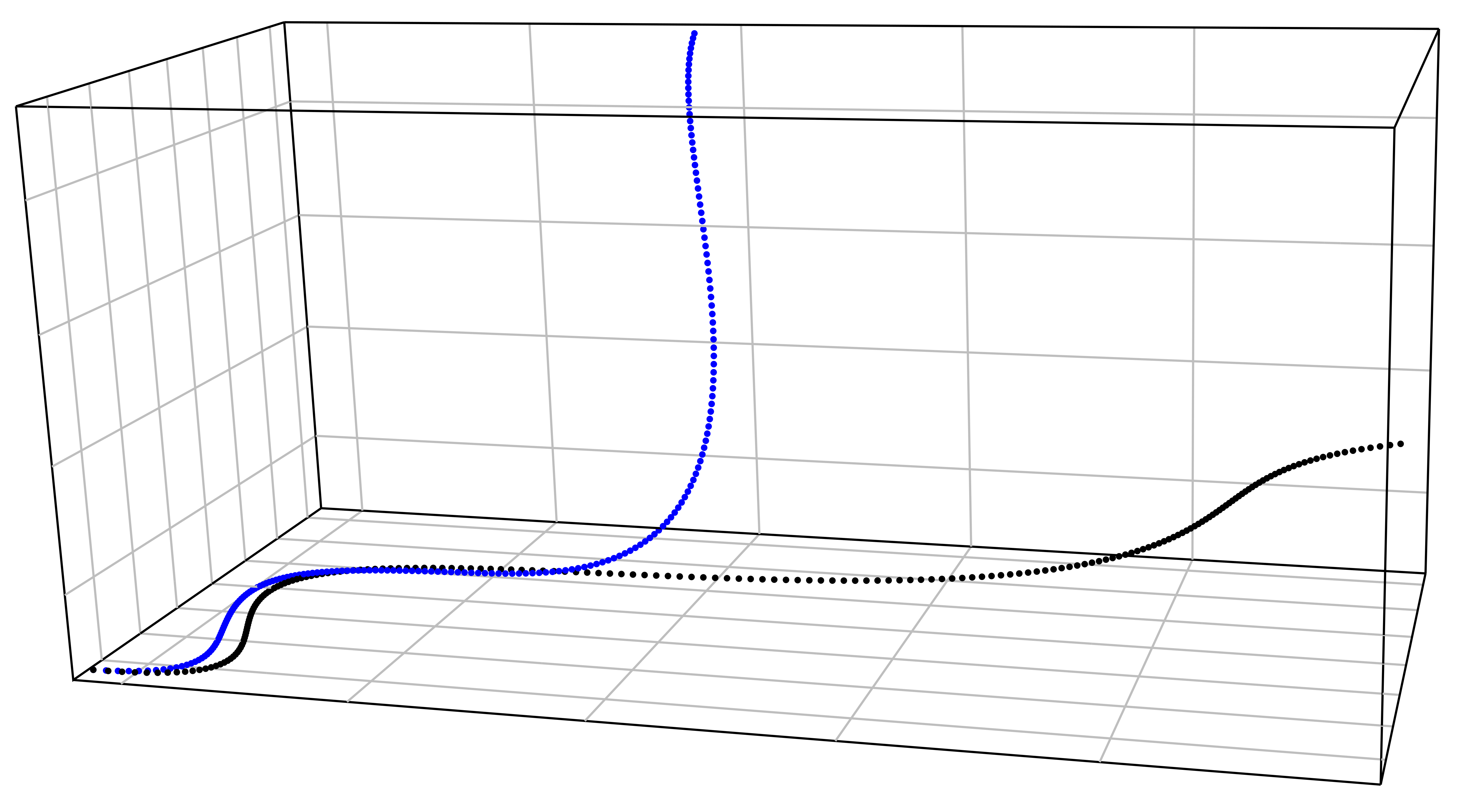}
    \\
    \subfigimg[width=2.6in, pos = ll]{\textbf{(d)}}{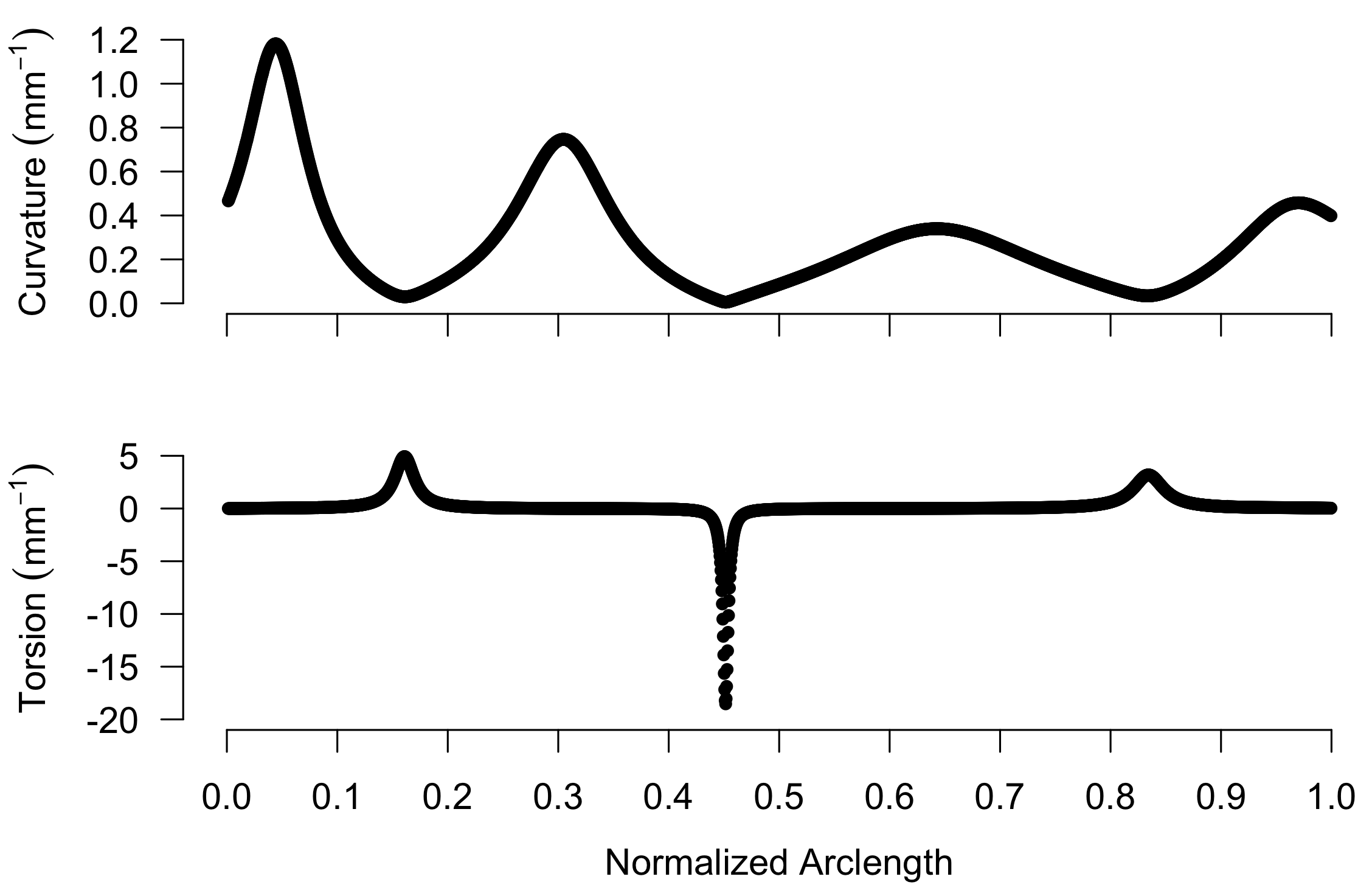}
    \hfill
    \subfigimg[width=2.0in, pos = ll]{\textbf{(e)}}{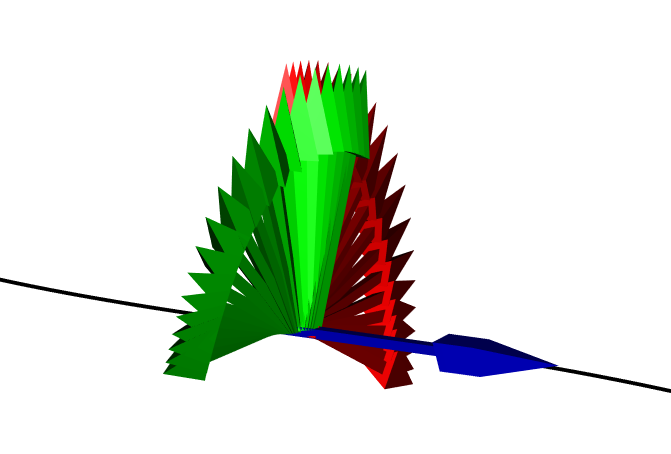}
    \hfill
    \subfigimg[width=2.3in, pos = ll]{\textbf{(f)}}{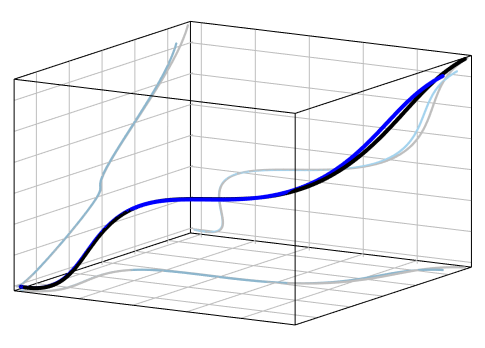}
    \footnotesize
    \caption{Demonstration of reconstruction at original \textbf{(a)-(c)} and 10X \textbf{(d)-(f)} sampling rates.  Graphs of curvature and torsion versus normalized arc length at RAW sampling \textbf{(a)} and 10X sampling \textbf{(d)}, note the doubling of maximal torsion value.  In \textbf{(b)} and \textbf{(e)} we show the discrete rotations of the Frenet-Serret frames through the nearly straight vessel region corresponding to the negative torsion spike from \textbf{(a)} and \textbf{(d)} (unit-vectors not drawn to scale).  In \textbf{(c)} and \textbf{(f)} we show spatial reconstructions under the same conditions.  Black lines represent the original (interpolated) vessels, blue lines represent the reconstructed vessel.  Grey and light-blue lines represent 2D-planar projections of black and blue for additional perspective.  The effects of low sampling rates on reconstruction error are shown to be a result of poorly resolved regions where spikes in torsion occur.  Data is for the MARG2 coronary artery from patient B in the Vorobstova et al.~dataset \cite{vorobstova_etal_annalsbiomedeng_2016}}
    \label{fig:tortuosity_prelim}
    \end{figure*}

Curvature $\kappa(s_j)$ and torsion $\tau(s_j)$ are the rates of change of the tangent and binormal vectors, which point in the direction of the normal vector. This is expressed formulaically as,

\begin{align}
    \frac{d}{ds_j}\buv{T}(s_j) & = \nu_j \kappa(s_j)\buv{N}(s_j)
    \label{eq:curv_FS}\\ 
    \nonumber\\
    \frac{d}{ds_j}\buv{B}(s_j) & = - \nu_j \tau(s_j)\buv{N}(s_j)
    \label{eq:tors_FS}
\end{align}

\noindent where $\nu_j$ is the ``speed'' of the curve at $s_j$, defined as $\nu_j = ||d\bv{r}_j/ds_j||$.  Of note, curvature and torsion can be interpreted as the rates of rotation of the Frenet-Serret frame about the binormal and tangent unit-vector axes, respectively.  This interpretation will be beneficial when we investigate the relationship between increased sampling rates and smooth rotations of the Frenet-Serret frame (see Fig.~\ref{fig:tortuosity_prelim}).  Finally, the rate of change of the normal vector in terms of curvature and torsion is,

\begin{equation}
    \frac{d}{ds_j}\buv{N}(s_j) = -\nu_j\kappa(s_j)\buv{T}(s_j) + \nu_j\tau(s_j)\buv{B}(s_j)
    \label{eq:curvplustors_FS}
\end{equation}

Using Eqs.~(\ref{eq:curv_FS})-(\ref{eq:curvplustors_FS}) and the definitions of the FS-frame, curvature and torsion can be expressed solely in terms of derivatives of the position vector.

\begin{align}
    \kappa(s_j) & =  \frac{\left \| \bv{r}'(s_j) \times \bv{r}''(s_j) \right \|}{\left \|  \bv{r}'(s_j) \right \|^3} 
    \label{eq:curv_pos}\\
    \nonumber\\
    \tau(s_j) & =  \frac{\bv{r}'(s_j) \cdot \left ( \bv{r}''(s_j) \times \bv{r}'''(s_j)\right )}{\left \| \bv{r}'(s_j) \times \bv{r}''(s_j) \right \|^2}
    \label{eq:tors_pos}
\end{align}

\noindent Note that curvature is always a positive quantity, whereas torsion is a signed quantity.  Positive and negative values of torsion denote right- and left-handed rotations about the direction of the tangent vector.  Furthermore, the definitions of the FS-frame vectors and the forms of Eqs.~(\ref{eq:curv_FS})--(\ref{eq:tors_pos}) are commonly used when working in a non-arc length parameterization.  Analytically speaking, curves expressed in the arc length parameterization have unit-speed, or $\nu = ||d\bv{r}(s)/ds|| = 1$ (see Bogunovic et al.~in \cite{bogunovic_etal_medimg_2012} or Lorthois et al.~in \cite{lorthois_etal_microvascres_2014}).  However, when working with discretized curves originating from data\textemdash as we are\textemdash this unit-speed condition is not guaranteed.

%\subsection{Analytic examples of generating curves from curvature and torsion}  
%
%While the theme of this paper is \textit{numerically} reconstructing curves from measures of curvature and torsion, we would be remiss to not point out several instances of \textit{analytic} reconstructions for simple scenarios.  
%\textcolor{red}{Present here example of reconstruction of coordinates for circle and helix using constant curvature/torsion values, and stepping through the FS-frame.  Will be nice way to easy into abstract/complex reconstruction section that appears later.  Could even briefly mention how FS theorem gives exact solutions for these scenarios (remember differentiating a second time for the helix solutions), but for ``s'' dependent functions for curvature and torsion, need numerical methods to find solutions.  Also can reference Salkowski curves here too \cite{monterde_compaidgeomdes_2009}.}

\subsection{Tortuosity metrics}
\label{sec:tortuosity_metrics}

A multitude of tortuosity metrics are used in the literature \cite{han_jvascres_2012, bullitt_etal_ieeetransmedim_2003, bullitt_etal_medimcomcad_2005, folarin_etal_microvascres_2010, gessner_etal_radiology_2012, rao_etal_ieeebiomedeng_2016, huang_etal_ieeetransmedim_2008, lindsey_etal_moleimgbio_2017, piccinelli_etal_ieeetransmedim_2009, shelton_etal_ultrasoundmedbio_2015, oloumi_etal_jmedim_2016, alilou_etal_scireports_2018, kobayashi_etal_japanesecirculation_2015, hart_etal_intjmedinfo_1999, grisan_etal_ieeeengmedbio_2003, bogunovic_etal_medimg_2012, kamenskiy_etal_jbiomecheng_2012, oflynn_etal_annalsbiomedeng_2007, vorobstova_etal_annalsbiomedeng_2016, bullitt_etal_neurobiologyaging_2010, lorthois_etal_microvascres_2014, oflynn_etal_annalsbiomedeng_2010, bullitt_etal_medimageanalysis_2005, gelman_etal_iovs_2005, meng_etal_surgradanat_2008, helmberger_etal_plosone_2014}. We now briefly review the most commonly used measures.

\subsubsection{Distance Metric ($\mathcal{D}$)}

The distance metric is a simple ratio of the total vessel arc length to the distance between endpoints.
\begin{equation}
    \mathcal{D} = \frac{\mathlarger{\sum}_{j=0}^{N-1}||\bv{r}_j - \bv{r}_{j+1}||}{||\bv{r}_0 - \bv{r}_N||}
\end{equation}
This metric measures any deviation from perfect straightness but fails to differentiate between 2D and 3D curves.  This measure is bounded below by a value of 1 corresponding to perfectly straight vessels and approaches infinity when the endpoints are co-located such as in a complete loop or cycle.  An alternative definition is $\mathcal{D}_1$, and defined as $\mathcal{D}_1 = \mathcal{D} - 1$ (see \cite{oflynn_etal_annalsbiomedeng_2007, piccinelli_etal_ieeetransmedim_2009, oflynn_etal_annalsbiomedeng_2010, bogunovic_etal_medimg_2012} and Supplementary Table \ref{tab:oflynn_data}).

\subsubsection{Inflection-Count Metric ($\mathcal{IC}$)}

The inflection-count metric simply counts the number of inflection points along a vessel.  This metric is argued to be useful for distinguishing between curves with single arcs versus those with multiple arcs \cite{bullitt_etal_ieeetransmedim_2003}.  Two methods of identifying inflection points have been proposed by Bullitt et al.~in \cite{bullitt_etal_ieeetransmedim_2003}.  One is to identify and count all points where the total curvature is a local minimum.  Total curvature, $\kappa_{tot}$, is defined by Hart et al.~in \cite{hart_etal_intjmedinfo_1999} as $\kappa_{tot} = \int\kappa(s)ds$ (also see Folarin et al.~in \cite{folarin_etal_microvascres_2010}).  This is not a well-defined equation for detecting inflection points however as $\kappa_{tot}$ monotonically increases with arc length and can have no local minima.  Thus, we instead count the number of local minima in curvature, $\kappa(s_j)$.  This metric will be denoted as $\mathcal{IC}_\kappa$.

The other method is to identify and count the local maxima of $\Delta\bv{N}\cdot\Delta\bv{N}$ when $\Delta\bv{N}\cdot\Delta\bv{N}$ is greater than 1.0 \cite{bullitt_etal_ieeetransmedim_2003, huang_etal_ieeetransmedim_2008}.  This latter method will be denoted as $\mathcal{IC}_N$.  For both metrics the value of $1$ is added to the count and multiplied by the distance metric.  Thus, for scenarios where the curve makes a broad arc, the inflection-count metrics will still return values no less than the distance metric.  This practice is standard convention.

\subsubsection{Sum-of-Angles Metric ($\mathcal{SOA}$)} 

This metric integrates angular changes in orientation between (at least) four neighboring points along a vessel.  It is widely used in lieu of curvature and/or torsion \cite{bullitt_etal_ieeetransmedim_2003, bullitt_etal_medimageanalysis_2005, bullitt_etal_medimcomcad_2005, oloumi_etal_jmedim_2016, gelman_etal_iovs_2005, rao_etal_ieeebiomedeng_2016, folarin_etal_microvascres_2010, gessner_etal_radiology_2012, huang_etal_ieeetransmedim_2008, lorthois_etal_microvascres_2014, shelton_etal_ultrasoundmedbio_2015}.  It was originally proposed by Bullitt et al.~in \cite{bullitt_etal_ieeetransmedim_2003} as a geometric variation to the total curvature, $\kappa_{tot}$, proposed by Hart et al.~in \cite{hart_etal_intjmedinfo_1999}, by incorporating torsion into the integration.  The motivation for this metric is to identify vessels with high frequency, low amplitude oscillations that are not well quantified by either the distance or inflection-count metrics.  The $\mathcal{SOA}$ metric is, 
\begin{equation}
\begin{multlined}
\mathcal{SOA}(s_j) = \Biggl \lbrace \biggl[ \cos^{-1}\biggl( \frac{\Delta\bv{r}(s_{j})}{|\Delta\bv{r}(s_{j})|} \cdot \frac{\Delta\bv{r}(s_{j+1})}{|\Delta\bv{r}(s_{j+1})|}\biggr) \biggr]^2 \\
+ \biggl[ \cos^{-1}\biggl( \frac{\Delta\bv{r}(s_{j-1})\times\Delta\bv{r}(s_j)}{|\Delta\bv{r}(s_{j-1})\times\Delta\bv{r}(s_j)|} \cdot \\
\frac{\Delta\bv{r}(s_j)\times\Delta\bv{r}(s_{j+1})}{|\Delta\bv{r}(s_j)\times\Delta\bv{r}(s_{j+1})|} \biggr) \biggr]^2  \Biggr \rbrace^{1/2}
\label{eq:soam}
\end{multlined}
\end{equation}
where the $\Delta$ symbol represents forward differences \textemdash $\Delta \bv{r}(s_j) = \bv{r}(s_{j+1}) - \bv{r}(s_j)$.  In Eq.~(\ref{eq:soam}), the first term within the square-root captures in-plane changes in orientation, and the second term captures out-of-plane changes.  Only three unique indices appear in Eq.~(\ref{eq:soam}) ($j-1$, $j$, and $j+1$), yet at least four points are needed ($j-1$, $j$, $j+1$, and $j+2$) because the simplest discretized derivatives each require a neighboring point.  Eq.~(\ref{eq:soam}) is a local definition of the $\mathcal{SOA}$, and in typical applications the reported value is the summed total divided by the total arc length \textemdash $\mathcal{SOA} = \sum_{j=2}^{j=N-2}\mathcal{SOA}(s_j)/\sum_{j=1}^{j=N}s_j$.  Importantly, all resulting values from the inverse cosine expressions must be modulated by the value $\pi$ \cite{bullitt_etal_ieeetransmedim_2003}.

\subsubsection{Statistical Features of Curvature and Torsion}

Some studies treat the metrics of curvature and torsion, Eqs.~(\ref{eq:curv_pos}) and (\ref{eq:tors_pos}), as statistical distributions and use various statistical features\textemdash mean, root-mean-square, or maximum\textemdash as singular measures of tortuosity \cite{oflynn_etal_annalsbiomedeng_2007, oflynn_etal_annalsbiomedeng_2010, lorthois_etal_microvascres_2014, gelman_etal_iovs_2005}, while other studies use integrated measures or weighted-averages \cite{bogunovic_etal_medimg_2012, hart_etal_intjmedinfo_1999, vorobstova_etal_annalsbiomedeng_2016, grisan_etal_ieeeengmedbio_2003, meng_etal_surgradanat_2008}. Moreover, measures with physical dimensions of inverse length are often defined to be dimensionless by integrating curvature and/or torsion along arc length.  A third metric that is commonly measured and studied as a distribution is the combined curvature torsion ($\mathcal{CCT}_j = \mathsmaller{\sqrt{\kappa_j^2 + \tau_j^2}}$).  This quantity is the magnitude of the Darboux vector\textemdash $\Omega_j = \kappa_j \buv{B}_j + \tau_j \buv{T}_j$\textemdash a so-called rotation vector of the Frenet-Serret frame for a rigid-body moving along a curve.  Thus, the combined curvature torsion quantity represents the angular speed of the Frenet-Serret frame\cite{patrikalakis_etal_shape_interrogation_2009}.

\section{Methods}
\label{methods}
\subsection{Data Acquisition}
We analyze manually constructed vessel phantoms and reanalyze data from five independent studies.  Imaging and segmentation routines for each dataset are summarized below.  Original sampling rates employed for each study are reported.  When necessary, all data were smoothed and interpolated as instructed in the respective publications to reproduce the originally published results.  Further information on the steps taken to reproduce originally published results, as well as our own measured values, can be found in the Supplemental Material Section \ref{app:data_reproduction}, and all vessel centerline coordinates and radii studied are available online at IEEE{\it DataPort} \cite{brummer_etal_ieeedataport_2019}.

\subsubsection{Vessel Phantoms} These consisted of three classes: semi-circles, helices, and Salkowski curves.  Semi-circles and helices were chosen as they are commonly used phantoms in the literature \cite{oflynn_etal_annalsbiomedeng_2007, bullitt_etal_ieeetransmedim_2003, lorthois_etal_microvascres_2014}.  Salkowski curves were chosen as they have non-trivial shapes (Fig.~\ref{fig:data_viz}({\bf a})) with constant curvature and non-constant torsion\cite{monterde_compaidgeomdes_2009}.  Importantly, tortuosity metrics for all three classes of vessel phantom shapes can be expressed analytically, providing standards against which our proposed methods can be validated.  Gaussian noise was added to each phantom to simulate variation normally associated with imprecision in segmentation algorithms.  The noise width was also used as a proxy for vessel diameter.  Following this, each noisy phantom was then splined and resampled as described in Section \ref{interpolate_sample}.  Thus, 1000 individual phantoms were generated for each vessel shape.  For further details on phantom generation and expected tortuousity values, see Supplementary Material Section \ref{app:phantoms}.

\subsubsection{Carotid arteries}
This data from Kamenskiy et al.~represents the averaged coordinates of the common, external, and internal carotid arteries for 16 patients \cite{kamenskiy_etal_jbiomecheng_2012} (Fig.~\ref{fig:data_viz}({\bf b})).  This is the smallest dataset studied, with only two vessels total, as the common and internal carotid arteries were merged into one continuous vessel.  Images were acquired with X-ray computed tomography angiography (CTA) on a Philips Brilliance 64, resulting in an in-plane resolution of 0.488 mm/pixel and a slice thickness of 1 mm.  Axial segmentation was performed first on a per-slice basis using a semi-automatic edge tracking algorithm implemented on custom Matlab software \cite{selzer_etal_atherosclerosis_1994}. Following this, segmented slices were merged into continuous digitized vessels using the Computer Aided Design (CAD) software Solid Works.  The analysis employed a sampling rate of 1 point/mm.

\subsubsection{Abdominal arteries}
This data from O'Flynn et al.~ consists of the abdominal aorta, the right renal artery, and the left and right common iliac arteries from one adult male individual (Fig.~\ref{fig:data_viz}({\bf c})) \cite{oflynn_etal_annalsbiomedeng_2007}.  Images were acquired with contrast-enhanced magnetic resonance angiography (MRA) on a Siemens Symphony at 1.5 Tesla, resulting in an in-plane resolution of 0.703 mm/pixel and a slice thickness of 1.5 mm.  Images were re-sliced to create iso-cubic voxels with dimensions of 0.703 mm$\times$0.703 mm$\times$0.703 mm.  Image segmentation was performed using a region growing algorithm implemented with the Mimics and Image Pro software.  The analysis employed a sampling rate of 5 points/mm.

\subsubsection{Cerebral Arteries}
This data from Bullitt et al.~consists of the primary cerebral arterial trees (Fig.~\ref{fig:data_viz}({\bf d})) of 42 adults ranging in age from 18 to 49, of which 18 were male and 24 were female. \cite{bullitt_etal_neurobiologyaging_2010}.  This is the largest dataset analyzed, with approximately 6,000 individual vessels.  The original study had 100 individuals, but segmented data existed for only 42.  Images were acquired with overlapping 3D-time-of-flight MRA on a Siemens Allegra at 3 Tesla, resulting in voxel sizes of 0.5 mm$\times$0.5 mm$\times$0.8 mm.  Image segmentation was performed using a custom software written in C++ that implements an image intensity ridge traversal with dynamic scaling algorithm \cite{aylward_etal_ieeetransmedim_2002}.  The analysis employed a sampling rate of 2 points/mm.

\subsubsection{Coronary Arteries}
This data from Vorobstova et al.~consists of coronary arterial trees (Fig.~\ref{fig:data_viz}({\bf e})) for three individual adult patients, resulting in 23 vessels \cite{vorobstova_etal_annalsbiomedeng_2016}.  Images were acquired with CTA, with no further details provided in the original manuscript.  Segmentation was performed using the publicly available software ITK-SNAP \cite{yushkevich_etal_neuroimage_2006}, with manual improvements, smoothing, and analyses performed with the CAD software MeshLab, Geomagic Studio, and ANSYS Fluent 14.5.  The sampling rate used in the original study was not stated, thus we determined that a sampling rate of 10 points/mm would best reproduce originally published results.

\subsubsection{Pulmonary Arterial and Venous Trees}
This data from Helmberger et al.~consists of pulmonary arterial and venous vessel trees (Fig.~\ref{fig:data_viz}({\bf f-g})) from 25 patients with and without pulmonary hypertension \cite{helmberger_etal_plosone_2014}.  This dataset consists of approximately 3,000 vessels.  Images were acquired via thoracic X-ray computed tomography (CT) on a Siemens, 128-slice, dual-energy CT scanner, resulting in voxel dimensions of 0.6 mm$\times$0.6 mm$\times$0.6 mm.  Artery/vein detection was performed using the algorithm of Payer et al.~that simultaneously leverages the anatomical properties of (1) uniform distributions of arterial and venous vessels and (2) proximity of pulmonary arteries to bronchi \cite{payer_etal_medimganal_2016}.  Data provided consisted of .mhd formatted files encoded for arteries or veins.  Thus, a final in-house segmentation and skeletonization routine using Angicart had to be performed to extract vessel coordinates and radii.  Following Angicart, original vessel sampling rates were 2 points/mm.

\subsubsection{Angicart Software for Segmenting Pulmonary Vessels}
Segmentation of pulmonary vessels was accomplished using the open-source software Angicart, developed by authors of this paper, David Hunt at the University of California, Los Angeles \cite{newberry_etal_ploscompbio_2015}.  Angicart segments images of vasculature to identify a network skeletonization for analysis of vessels and the junctions joining them.  It is a fully automated software (as defined by Myatt et al.~in \cite{myatt_etal_frontneuroinfo_2012}) that only requires vessels of interest to be brighter on a greyscale than surrounding tissues.  Output consists of vessel radii, lengths, connectivity, and centerline coordinates.  Results from Angicart output have been published previously for human thoracic MRA data \cite{newberry_etal_ploscompbio_2015} and micro-CT mouse lung data \cite{tekin_etal_ploscompbio_2016}.  Updates to the original Angicart methods, including a preliminary coarse segmentation to accurately address the vascular topology, are described in Supplementary Material Section \ref{angicart_description}.

\subsection{Interpolating and Increased Sampling}
\label{interpolate_sample}
To increase sampling rates, we interpolate based on fits of a linear combination of polynomial curves (B-splines) that are piecewise-continuous to their fourth derivative.  Continuity of higher-order derivatives is essential to ensure that asymptotic features of torsion estimates are not due to discontinuity in any of the required derivatives of the position vector.  We imposed Neumann boundary conditions on the vessels (matching first derivatives) instead of Dirichlet boundary conditions to yield a better fit to the original vessel as determined by visual inspection.  We chose penalized-splines (P-splines) to perform the interpolation as these incorporate a penalization to avoid overfitting \cite{eilers_etal_statsci_1996, eilers_etal_wirescompstat_2010}.  The penalization is such that overfitting is mitigated even as a greater number of P-splines per vessel are incorporated \cite{aguilera_etal_mathcompmod_2013}.  This feature is important as some vessels may require a large number of P-splines in order to yield accurate interpolation due to increased length and tortuosity.  Yet, we found that no more than 20 P-splines were needed for the longest vessels studied.  As the coordinates of both the raw and sub-sampled data are nearly equally spaced point-to-point, the knots used to join the P-splines are spaced uniformly along the arc length of the vessels \cite{lee_cad_1989}.  Following interpolation, the fitted data are then sampled to the desired level.

A side effect of using many P-splines and Neumann boundary conditions is the tendency of the ends of the vessels to exhibit high frequency spatial oscillations.  When subsampled, these produce highly discontinuous derivatives, thus affecting curvature and torsion measurements.  As these oscillations are transient, we filtered the first and last 10\% of each vessel length following interpolation and sub-sampling.
 
\subsection{Calculation of Curvature and Torsion}

Curvature $\kappa(s_j)$ and torsion $\tau(s_j)$ are calculated using Eqs.~(\ref{eq:curv_pos}) and (\ref{eq:tors_pos}) with fifth-order, centered difference approximations to the first three derivatives of the position vector ($\bv{r}'(s_j), \bv{r}''(s_j)$, and $\bv{r}'''(s_j)$).
{\it Difference methods} are chosen to ensure accuracy up to the 4\textsuperscript{th} derivative of the position vector, and to present an approach that is independent of the choice of interpolation method (splining, polynomial fitting, or Fourier series approximations).  The {\it difference methods} are presented in full in Supplementary Material Section \ref{app:differences}.

\subsection{Numerical Integration of Frenet-Serret Equations}
\label{numerical}

%\begin{figure}[!t]
%    \centering  
%    \subfigimg[width=3.45in, pos = ll]{\textbf{(a)}}{figures/analysis/max_error_vs_resolution_unfiltered.png}
%    \vspace{1pt}
%    \\
%     \subfigimg[width=3.45in, pos = ll]{\textbf{(b)}}{figures/analysis/max_error_vs_resolution_filtered.png}
%    \footnotesize
%    \caption{Graphs of reconstruction error versus sampling resolution for \textbf{(a)} unfiltered and \textbf{(b)} filtered data.  Points are averages of dataset (geometric average for Bullitt et al.) and error bars are standard deviations (back-transformed for Bullitt et al.).  Horizontal strips denote one standard deviation in values of vessel minima used to set the threshold for acceptable reconstruction error.  In \textbf{(b)}, the number of vessels completely filtered at 100X resolution for the datasets are (in order of Kamenskiy, O'Flynn, Vorobstova, then Bullitt) 1/2, 4/4, 16/24, 4799/5838.}
%    \label{fig:recon_error_unfil}
%\end{figure}

Reconstructions of vessel centerline coordinates from measures of curvature and torsion can be done by integrating the Frenet-Serret equations. The value of this approach is that it demonstrates the accuracy of the measures of curvature and torsion, particularly in relation to asymptotic regions where curvature approaches zero.  The procedure for performing numerical integration of the Frenet-Serret equations begins with re-writing Eqs.~(\ref{eq:curv_FS})-(\ref{eq:curvplustors_FS}) as a linear system of first-order, ordinary differential equations as follows,
\begin{equation}
\frac{d}{ds_j}\begin{pmatrix} \buv{T}(s_j) \\ \buv{N}(s_j) \\ \buv{B}(s_j) \end{pmatrix} = 
\begin{pmatrix} \textbf{0} & \nu_j\kappa_j\textbf{I} & \textbf{0} \\  -\nu_j\kappa_j\textbf{I} & \textbf{0} & \nu_j\tau_j\textbf{I} \\ \textbf{0} & -\nu_j\tau_j\textbf{I} & \textbf{0} \end{pmatrix} 
\begin{pmatrix} \buv{T}(s_j) \\ \buv{N}(s_j) \\ \buv{B}(s_j) \end{pmatrix}
\label{eq:fs_system}
\end{equation}
where $\textbf{I}$ is the $3\times3$ identity matrix, and $\textbf{0}$ is the $3 \times 3$ matrix of zeros.  Eq.~\ref{eq:fs_system} is expressed with a vector of unit vectors and a matrix of identity and zero matrices to illustrate the anti-symmetric structure of the Frenet-Serrat equations.  For notational convenience, this equation can be expressed in a compact manner as, $d\bm{\Phi}(s_j)/ds_j = \textbf{M}(s_j) \bm{\Phi}(s_j)$, where $\bm{\Phi}(s_j)$ is the $9\times1$ vector of FS-frame unit-vector components, and $\textbf{M}(s_j)$ is the $9\times9$ matrix of curvature and torsion values as written in Eq.~(\ref{eq:fs_system}).

The conditions for solving Eq.~(\ref{eq:fs_system}) are unique in that the solution is known entirely.  This includes the boundary conditions of both Dirichlet $(\bm{\Phi}(0),\bm{\Phi}(N))$ and Neumann $(\bm{\Phi}'(0),\bm{\Phi'}(N))$ type.  Thus, one can choose whether to use boundary value problem (BVP) or initial value problem (IVP) methods.  We have chosen to use IVP methods given that the BVP methods require solving a sparse, $9N\times9N$ non-linear, inverse-matrix problem (where $N$ represents the number of centerline vessel coordinates) and would require the combination of implicit and iterative methods to yield stable convergence given the eigenvalue structure of $\bm{M}(s_j)$.  On the other hand, as an IVP we need only solve $N$, $9\times9$ problems (one for each of the $N$ coordinates along the vessel centerline) using the same methods described.  Thus, the IVP method scales with the number of centerline coordinates, while the BVP method scales as the square of the number of centerline coordinates.

We solve Eq.~(\ref{eq:fs_system}) as an IVP using a combination of backward Euler and Newton's method.  The only non-zero eigenvalues of $\textbf{M}(s_j)$ are the imaginary values $\lambda_\pm = \pm i \nu_j\sqrt{\kappa_j^2 + \tau_j^2}$.  Thus, to ensure stability we adopt the backward Euler method to discretize the derivative on the left-hand-side of Eq.~(\ref{eq:fs_system}) as $d\bm{\Phi}(s_j)/ds_j = \left(\bm{\Phi}(s_j) - \bm{\Phi}(s_{j-1})\right)/\left(s_j - s_{j-1}\right)$.  However, the backward Euler method applied to Eq.~(\ref{eq:fs_system}) requires solving $\bm{\Phi}(s_j) = \bm{\Phi}(s_{j-1}) + (s_j - s_{j-1})\bm{M}(s_j)\bm{\Phi}(s_j)$, which is a non-linear equation.  Thus, we use Newton's method to iteratively seek a convergent solution for $\bm{\Phi}(s_j)$ \cite{leveque_finite_difference_2007}.

Upon integration of Eq.~(\ref{eq:fs_system}), the position vector $\bv{r}(s_j)$ can be calculated as
\begin{equation}
    \bv{r}'(s_j) = ||\bv{r}'(s_j)||\buv{T}(s_j)
    \label{eq:nonlinear_position}
\end{equation}
Eq.~(\ref{eq:nonlinear_position}) is non-linear and attempts to integrate it using the same methods as for Eq.~(\ref{eq:fs_system}) are unstable.  However, upon discretization with a forward Euler method, Eq.~(\ref{eq:nonlinear_position}) can be linearized\textemdash $\bv{r}'(s_j) = \buv{T}(s_j)$\textemdash and easily integrated.  The methods of integration for Eqs.~(\ref{eq:fs_system}) and (\ref{eq:nonlinear_position}) are presented in detail in Supplementary Material Section \ref{app:numerical}.

Once Eq.~(\ref{eq:nonlinear_position}) is integrated, a point-wise error, $\epsilon_j$, can be calculated for comparison against the original (measured, splined, and subsampled) position vector, $\bv{r}(s_j)$, and the reconstructed position vector, $\bv{R}(s_j)$.
\begin{equation}
    \epsilon_j = ||\bv{r}(s_j)-\bv{R}(s_j)||
    \label{eq:error}
\end{equation}
To establish a threshold for satisfactory reconstruction, the maximum of the point-wise error $\epsilon_j$ was compared against the minimum vessel radius (see Fig.~\ref{fig:recon_error_unfil}).  In addition to calculating the point-wise error, it is visually instructive to graph original and reconstructed vessels to identify where breakdowns in reconstruction occur (see Fig.~\ref{fig:tortuosity_prelim}).

In the study of vascular retinopathy, the dimension representing image depth is often eliminated (thus, helical curves become strictly sinusoidal).  While this approach complicates interpretation of curvature and torsion as indicators of fluid flow, it is a diagnostic practice with a long history of success (see \cite{hart_etal_intjmedinfo_1999, oloumi_etal_jmedim_2016, grisan_etal_ieeeengmedbio_2003, gelman_etal_iovs_2005, lisowska_etal_ieeeengmedbio_2014}).  We briefly summarize the needed changes to apply these methods to strictly two-dimensional data. First, set torsion to be zero and reduce the dimensionality of the the Frenet-Serret equations by eliminating the binormal vector.  Curvature must now be allowed to change sign to account for inflection points.  As zeroing torsion does not change the fact that all non-zero eigenvalues will be imaginary, then the same methods of numerical integration described herein should apply.

\subsection{Measured tortuosity metrics and filtering}
\label{metrics_filtering}
For a vessel composed of $N$ points, we will examine the \textit{dimensionless} metrics of total curvature, $\mathcal{TC}$, total torsion, $\mathcal{TT}$, and total combined curvature and torsion, $\mathcal{TCCT}$, expressed as,
\begin{align}
 \mathcal{TC} & = \sum_{j=1}^{N-1}\kappa_j(s_{j+1}-s_j) \\
 \mathcal{TT} & = \sum_{j=1}^{N-1}|\tau_j|(s_{j+1}-s_j) \\
 \mathcal{TCCT} & = \sum_{j=1}^{N-1}\mathsmaller{\sqrt{\kappa_j^2 + \tau_j^2}}(s_{j+1}-s_j)
\end{align}
 
We also examine the arc length normalized values of these metrics\textemdash $\mathcal{TC}/\mathcal{L}$, $\mathcal{TT}/\mathcal{L}$, $\mathcal{TCCT}/\mathcal{L}$\textemdash where $\mathcal{L} = s_N$.  When examining the manually defined vessel phantoms, these six metrics allow for comparison against analytically derived expected values (see Supplementary Material \ref{sec:si_salk}).  We also examine the \textit{dimensionful} metrics of average curvature, $\mathcal{AC}$, average torsion, $\mathcal{AT}$, and average combined curvature and torsion, $\mathcal{ACCT}$, expressed as,

\begin{align}
\mathcal{AC} & = \frac{1}{N}\sum_{j=1}^{N}\kappa_j \\
\mathcal{AT} & = \frac{1}{N}\sum_{j=1}^{N}|\tau_j| \\
\mathcal{ACCT} & = \frac{1}{N}\sum_{j=1}^{N}\mathsmaller{\sqrt{\kappa_j^2 + \tau_j^2}}
\end{align}
We also measure both definitions of inflection point counts, $\mathcal{IC}_N$ and $\mathcal{IC}_\kappa$, as well as the sum-of-angles metric $\mathcal{SOA}$.

A disadvantage of the FS-frame is that it is not defined for straight sections of vessels or planar sections of infinitesimally small length corresponding to points of inflection.  When straight or singular sections of vessels are problematic, an empirical approach to handling them is to filter these regions \cite{bullitt_etal_ieeetransmedim_2003} or remove the vessel entirely from the study \cite{gessner_etal_radiology_2012}.  Such a method for regional filtering was proposed by Bullitt et al.~in \cite{bullitt_etal_ieeetransmedim_2003}.  In that case the filter was applied to what was incorrectly called the acceleration vector, but in fact was a centered, second-order difference of the position vector without division by the square of the step size.  Specifically, for any vessel point $s_j$, if $\bv{r}(s_{j+1}) - 2\bv{r}(s_j) + \bv{r}(s_{j-1}) < 10^{-6}$cm, then neither curvature nor torsion were calculated.  We show that such filtering methods, when applied to the metrics of curvature and torsion, can in fact remove essential information that is necessary for an accurate characterization of the vessel for the purposes of reconstruction.

\section{Results}

\begin{table*}[!t]
\renewcommand{\arraystretch}{1.3}
% if using array.sty, it might be a good idea to tweak the value of
% \extrarowheight as needed to properly center the text within the cells
\caption{Phantom vessel reconstruction error and average radius (in {\it mm}) for semi-circle, helix, and salkowski curves at sampling rates of $1X$, $2X$, $10X$, and $100X$.  Original ($1X$) sampling rates were: 1.06 points$/mm$ (circle), 0.978 points$/mm$ (helix), and 1.138 points$/mm$ (salkowski curve).  At sampling rates of $10X$ the reconstruction error becomes less than the average vessel radius, indicating sufficient overlap between the original and reconstructed vessel.  At higher sampling rates overfitting can occur (see Table \ref{tab:phan_tortuosity}).}
\label{tab:phan_recon}
\centering
\sisetup{
table-space-text-post = $\substack{+0.0000\\-0.0000}$
}%see table-space-text-post in siunitx documentation
\begin{tabular}{L|S[table-format=4.4]S[table-format = 3.4]S[table-format=2.4]S[table-format=2.4]|S[table-format=2.4]}
  \hline
\multicolumn{1}{L|}{\multirow{2}{*}{Vessel Shape}} & \multicolumn{4}{c|}{Reconstruction Error (in $mm$) at Sampling Rates of:} & \multicolumn{1}{L}{\multirow{2}{*}{Average Radius}} \\
 \multicolumn{1}{L|}{} & \multicolumn{1}{c}{$1X$} & \multicolumn{1}{c}{$2X$} & \multicolumn{1}{c}{$10X$} & \multicolumn{1}{c|}{$100X$} & \multicolumn{1}{L}{} \\  \hline
 Semi-Circle & 2.0170$\substack{+0.0085\\-0.0082}$ & 1.1786$\substack{+0.0098\\-0.0084}$ & 0.2725$\substack{+0.0033\\-0.0026}$ & 0.0615$\substack{+0.0005\\-0.0004}$ & 0.4245$\substack{+0.4617\\-0.0799}$\\ \hline
Helix & 2.9494$\substack{+0.1687\\-0.1583}$ & 2.1028$\substack{+0.0808\\-0.0918}$ & 0.4777$\substack{+0.0182\\-0.0177}$ & 0.1002$\substack{+0.0041\\-0.0029}$ & 1.4220$\substack{+0.2886\\-0.2344}$  \\ \hline
 Salkowski & 445.3347$\substack{+14.4961\\-18.0033}$ & 24.2025$\substack{+8.4533\\-6.1328}$ & 0.8342$\substack{+0.1201\\-0.0434}$ & 0.1101$\substack{+0.020\\-0.006}$ & 2.5334$\substack{+0.4161\\-0.1069}$  \\ 
   \hline 
\end{tabular}
\end{table*}

\begin{table*}[!t]
\renewcommand{\arraystretch}{1.3}
% if using array.sty, it might be a good idea to tweak the value of
% \extrarowheight as needed to properly center the text within the cells
\caption{Tortuosity values for Salkowski curve at various sampling rates compared against analytic values.  See Supplementary Material \ref{sec:si_salk} for analytic expressions.  Agreement between measured and analytic values occurs at the $10X$ sampling rate.  Evidence of over-fitting vessel centerline coordinates is evident by tortuosity measurements overestimating the analytic values at $100X$ sampling rate.}
\label{tab:phan_tortuosity}
\centering
\sisetup{
table-space-text-post = $\substack{+0.0000\\-0.0000}$
}%see table-space-text-post in siunitx documentation
\begin{tabular}{p{2.5cm}|S[table-format=4.4]S[table-format = 3.4]S[table-format=2.4]S[table-format=2.4]|S[table-format=2.2]}
  \hline
\multicolumn{1}{c|}{\multirow{2}{*}{Tortuosity Metric}} & \multicolumn{4}{c|}{Tortuosity Measurement at Sampling Rates of:} & \multicolumn{1}{L}{\multirow{2}{*}{Analytical Value}} \\
 \multicolumn{1}{c|}{} & \multicolumn{1}{c}{$1X$} & \multicolumn{1}{c}{$2X$} & \multicolumn{1}{c}{$10X$} & \multicolumn{1}{c|}{$100X$} & \multicolumn{1}{L}{} \\  \hline
$\mathcal{TC}$ & 9.1853$\substack{+0.0722\\-0.0815}$ & 11.4441$\substack{+0.3616\\-0.0868}$ & 12.5053$
\substack{+0.1241\\-0.1463}$ & 12.8485$
\substack{+0.1129\\-0.1291}$ & 12.8 \\ \hline
$\mathcal{TT}$ & 4.0178$\substack{+0.0424\\-0.0480}$ & 7.1485$\substack{+0.1261\\-0.1247}$ & 8.8061$\substack{+0.1892\\-0.1951}$ & 9.2257$\substack{+0.1921\\-0.2142}$ & 8.95  \\ \hline
$\mathcal{TCCT}$ & 10.1199$\substack{+0.0672\\-0.0750}$ & 13.9311$\substack{+0.4292\\-0.0100}$ & 16.0751$\substack{+0.3384\\-0.1972}$ & 16.6818$\substack{+0.3867\\-0.1850}$ & 16.31\\ 
   \hline 
$\mathcal{TC}/\mathcal{L}$ $(mm^{-1})$ & 0.7976$\substack{+0.0062\\-0.0076}$ & 0.8967$\substack{+0.0070\\-0.0060}$ & 0.9781$\substack{+0.0072\\-0.0068}$ & 0.9976$\substack{+0.0072\\-0.0089}$ & 1.0 \\ \hline
$\mathcal{TT}/\mathcal{L}$ $(mm^{-1})$ & 0.3489$\substack{+0.0037\\-0.0043}$ & 0.5586$\substack{+0.0087\\-0.0099}$ & 0.6883$\substack{+0.0137\\-0.0159}$ & 0.7162$\substack{+0.0159\\-0.0176}$ & 0.699 \\ \hline
$\mathcal{TCCT}/\mathcal{L}$ $(mm^{-1})$ & 0.8788$\substack{+0.0061\\-0.0067}$ & 1.0911$\substack{+0.0120\\-0.0076}$ & 1.2550$\substack{+0.0245\\-0.0129}$ & 1.2956$\substack{+0.0279\\-0.0138}$ & 1.274 \\ \hline
\end{tabular}
\end{table*}

\subsection{Reconstruction}  We find that increasing the sampling rate\textemdash decreasing the size of $|\Delta s_j|$\textemdash universally improves reconstruction accuracy of vessels when compared to the minimum vessel radius.  This is demonstrated in both the phantom and clinical vessels.  Furthermore, the effect of filtering is shown to result in poorer reconstruction accuracy, a feature that we demonstrate on the clinical data below.

\begin{figure}[!t]
    \centering  
    \subfigimg[width=3.45in, pos = ll]{\textbf{(a)}}{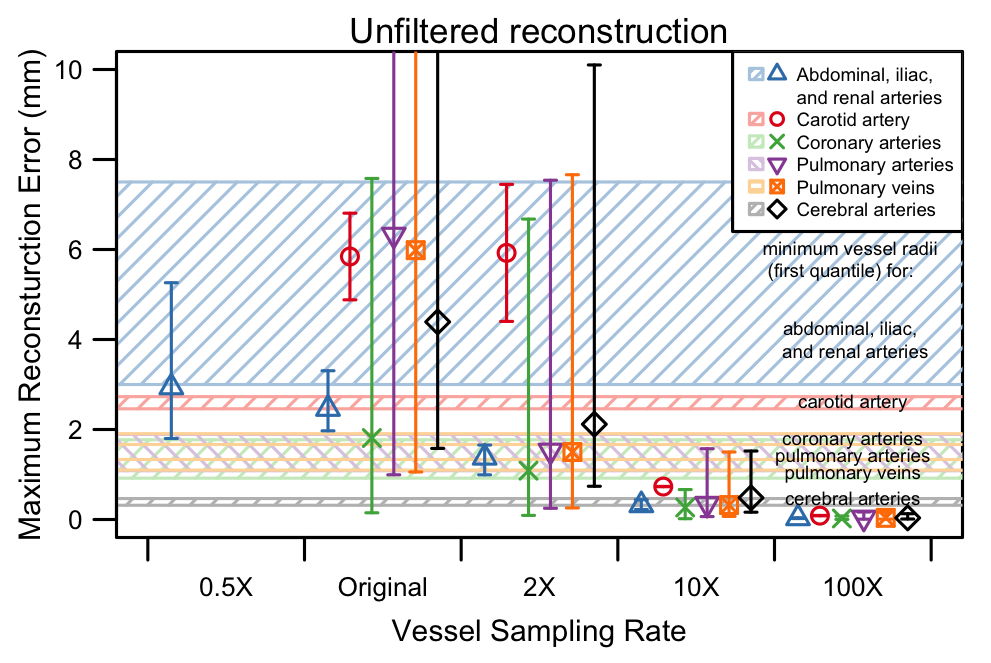}
    \vspace{1pt}
    \\
     \subfigimg[width=3.45in, pos = ll]{\textbf{(b)}}{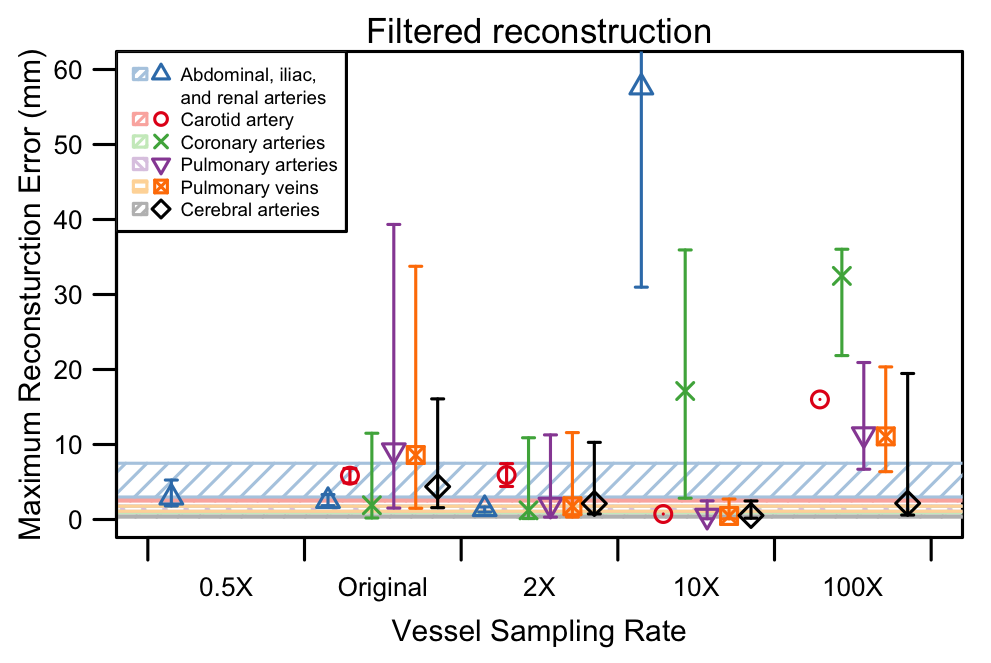}
    \footnotesize
    \caption{Graphs of reconstruction error versus sampling rate for \textbf{(a)} unfiltered vessel coordinates and \textbf{(b)} filtered vessel coordinates, where filtering is performed when vessel segments are approximately straight (see Section \ref{metrics_filtering}).  Points and error bars represent medians and 18.5\%-81.5\% quantiles, respectively.  Horizontal strips denote 18.5\%-81.5\% quantiles in minimum vessel radius for each separate dataset.  In {\bf (a)} reconstruction error decreases with increasing sampling rate, eventually becoming less than the vessel radius.  The cross-over between reconstruction error (points) and vessel radius (strips) determines the sufficient sampling rate.  Overall, we find that sampling rates of 10-100 points/mm are sufficient for vessels with an average minimum radius larger than 1 mm, and 100-1000 points/mm are sufficient for less than 1mm.  In {\bf (b)} filtering is shown to increase reconstruction error far beyond the vessel radius.}
    \label{fig:recon_error_unfil}
\end{figure}

\subsubsection{Phantom Vessels}  Table \ref{tab:phan_recon} demonstrates that reconstruction errors became less than the average vessel radius once sampling rates of approximately 10 points/mm were used for all phantom vessel shapes considered\textemdash semi-circle, helix, and Salkowski curve.  Of particular importance are the reconstruction errors for the Salkowski curve that are factors of 100 and 10 times larger than the average vessels radii for sampling rates of $1X$ and $2X$, respectively.  This is due to the fact that Salkowski curves exhibit non-constant torsion expressed as $\tau(s) = \tan(s)$.  Care was taken when defining the Salkowski curves to avoid the near-asymptotic regime of the tangent function.

\begin{figure}[!t]
    \centering  
    \subfigimg[width=2.6in, pos = ll]{\textbf{(a)}}{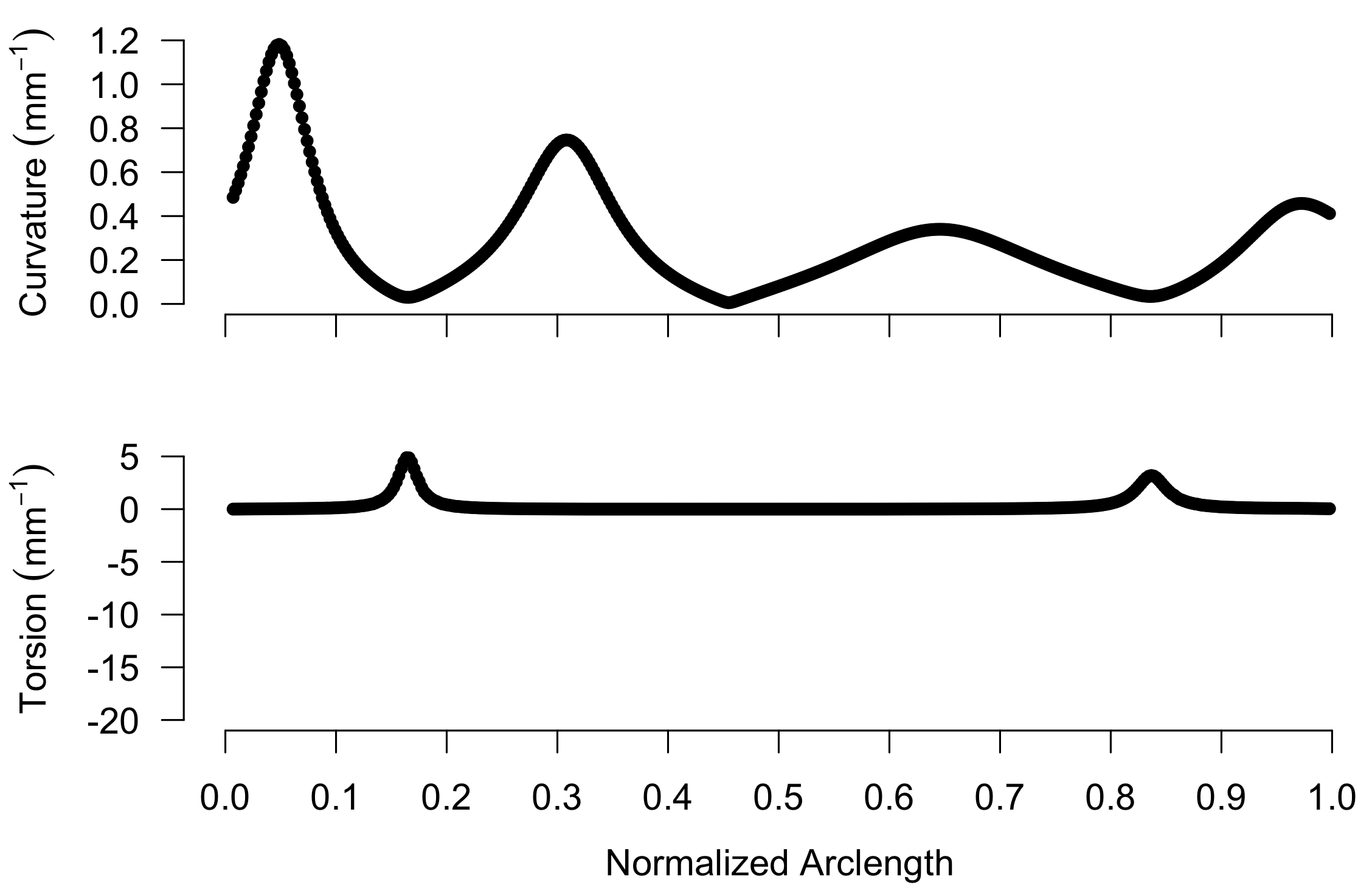}
    \vspace{1pt}
    \\
     \subfigimg[width=2.6in, pos = ll]{\textbf{(b)}}{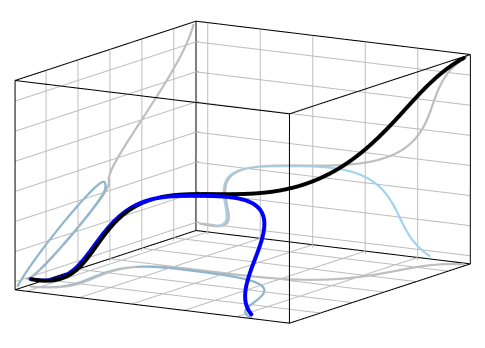}
    \footnotesize
    \caption{Graphs of curvature and torsion \textbf{(a)} and results of vessel reconstruction \textbf{(b)} when filtering is applied at 10X sampling rate.  In black is the original vessel and in blue is the reconstructed vessel.  Note that removal of the negative spike in torsion has resulted in a significantly erroneous rotation of the vessel along the axis of the tangent vector at approximately the midpoint of this vessel.  Data is for the MARG2 coronoary artery from patient B in the Vorobstova et al.~dataset \cite{vorobstova_etal_annalsbiomedeng_2016}}
    \label{fig:filter_recon}
\end{figure}

\subsubsection{Clinical Vessels}
Qualitative results of the integration reconstruction procedure are presented in Figure \ref{fig:tortuosity_prelim} for one of the coronary arteries.  There we can see an example where increasing the sampling rate improves reconstruction accuracy by providing a greater resolution of the middle torsion spike (Fig.~\ref{fig:tortuosity_prelim}\textbf{(a)}, and \textbf{(d)}).  Increased sampling at this location in the vessel smoothes the rotation of the Frenet-Serret frame (Fig.~\ref{fig:tortuosity_prelim}\textbf{(b)}, and \textbf{(e)}) at a location where the vessel is approximately straight (Fig.~\ref{fig:tortuosity_prelim}\textbf{(c)}, and \textbf{(f)}).  Across all vessels studied (Fig.~\ref{fig:recon_error_unfil}), sampling rates on the order of $1-10$ points/mm are incapable of providing sufficient curvature and torsion measurements to accurately reconstruct the vessels.  We find that increasing sampling rates to $10-100$ points/mm is sufficient for vessels with an average minimum radius larger than 1 mm, and that $100-1000$ points/mm is sufficient for vessels with an average minimum radius less than $1$ mm.

\subsubsection{Filtering}
The effect of filtering is shown to universally result in mischaracterization of the original vessel in terms of erroneous reconstructions, sometimes by as much as 10 times the radius of the vessel itself (see Figs.~\ref{fig:filter_recon} and \ref{fig:recon_error_unfil}\textbf{(b)}).  This is demonstrated with and without the use of increased sampling.  In fact, since the filtering threshold proposed by Bullitt et al.~does not account for the step-size between points in calculating the acceleration vector (the factor of $1/4\Delta s^2$), the effects of filtering are exacerbated by increasing sampling rates because entire vessels are filtered at that threshold.

\subsection{Tortuosity metrics, subsampling, and correlation}
The absence of expected or known values of tortuosity represents an inherent challenge in examining how tortuosity metrics of real vessels depend on the sampling rate.  As the Salkowski curve represents a non-trivial phantom vessel with analytically known values of tortuosity metrics, we can infer some general trends in this relationship.

\subsubsection{Phantom Vessels}
In Table \ref{tab:phan_tortuosity} are the measured tortuosity metrics of total curvature, $\mathcal{TC}$, total torsion, $\mathcal{TT}$, and total combined curvature and torsion $\mathcal{TCCT}$ for Salkowski curves measured with increasing sampling rate, and compared against the analytical values.  Also presented are arc length normalized versions of each of the three metrics.  For all six metrics we find that the measured values are lower than the expected analytical value for the initial sampling rate and increase with increasing sampling rate. Only once sampling rates of 10X to 100X are used do the measured values agree with the analytic values.  For the non-arc length normalized metrics ($\mathcal{TC}$, $\mathcal{TT}$, $\mathcal{TCCT}$) the sampling rate of 100X results in values greater than the expected analytic values.  However, the arc length normalized metrics ($\mathcal{TC}/\mathcal{L}$, $\mathcal{TT}/\mathcal{L}$, $\mathcal{TCCT}/\mathcal{L}$) do not exhibit the same degree of overestimation.  This suggests that the 100X sampling rate results in an oversampled vessel, and that in general arc length normalized tortuosity metrics may be more robust to errors associated with oversampling.

\subsubsection{Clinical Vessels}
\begin{figure*}[!t]
    \centering
    \subfigimg[width=2.3in, pos = ll]{\textbf{(a)}}{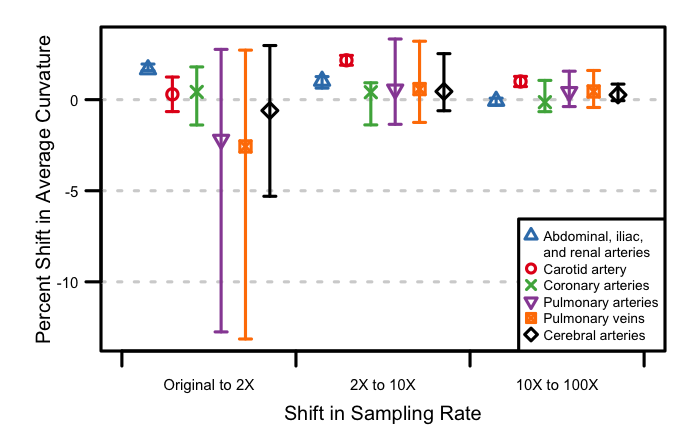}
    \hfill
    \subfigimg[width=2.3in, pos = ll]{\textbf{(b)}}{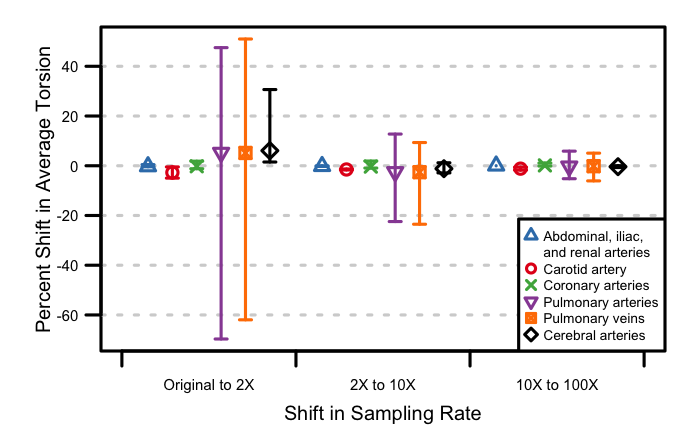}
    \hfill
    \subfigimg[width=2.3in, pos = ll]{\textbf{(c)}}{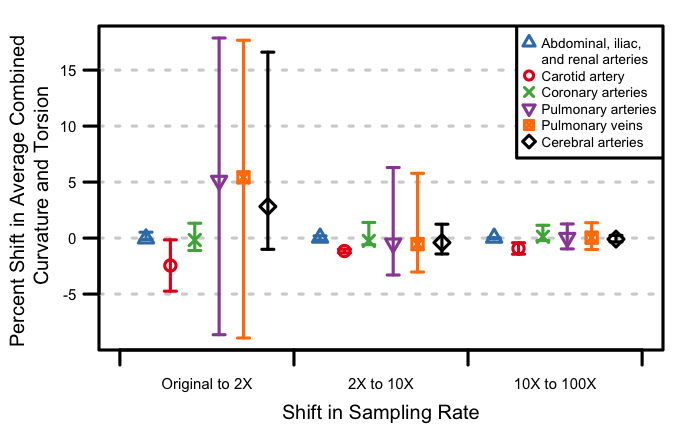}
    \vspace{10pt}
    \\
    \subfigimg[width=2.3in, pos = ll]{\textbf{(d)}}{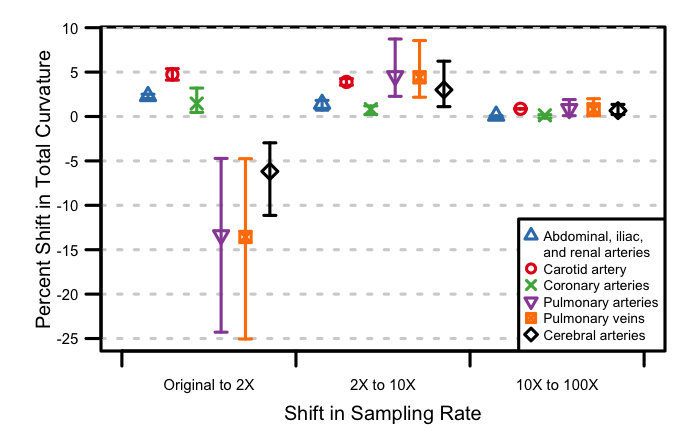}
    \hfill
    \subfigimg[width=2.3in, pos = ll]{\textbf{(e)}}{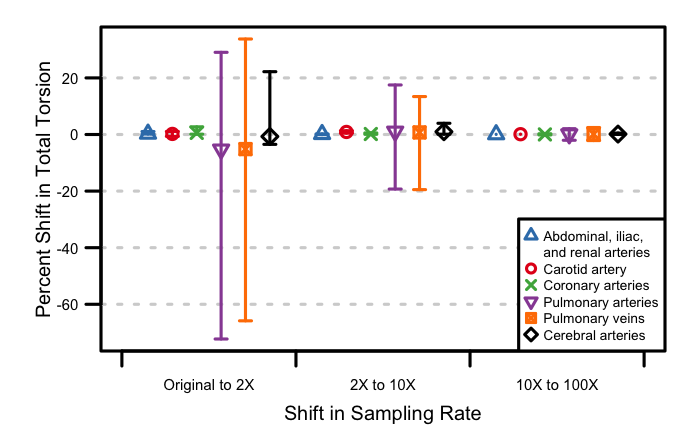}
    \hfill
    \subfigimg[width=2.3in, pos = ll]{\textbf{(f)}}{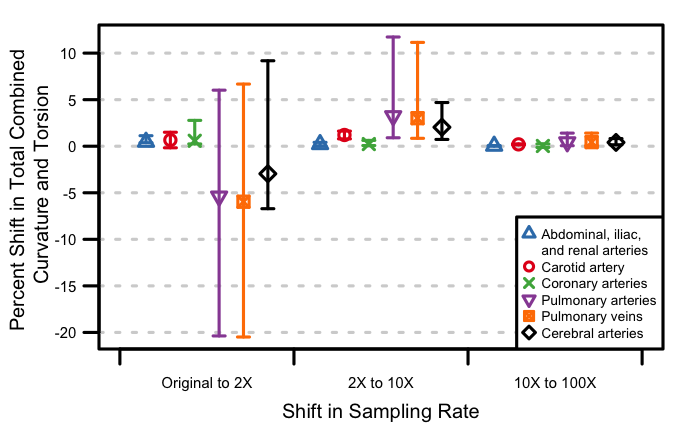}
    \vspace{10pt}
    \\
        \subfigimg[width=2.3in, pos = ll]{\textbf{(g)}}{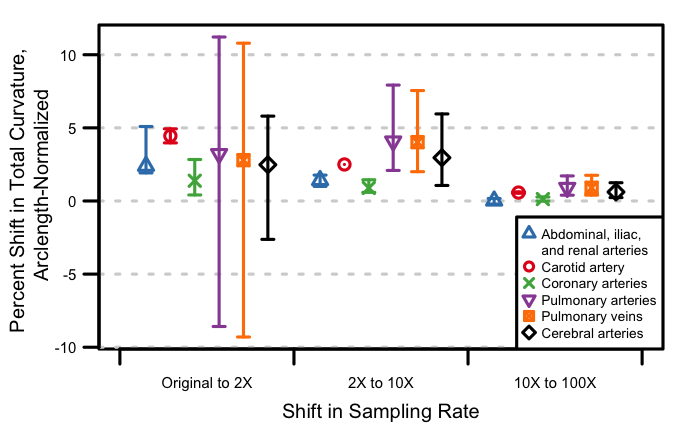}
    \hfill
    \subfigimg[width=2.3in, pos = ll]{\textbf{(h)}}{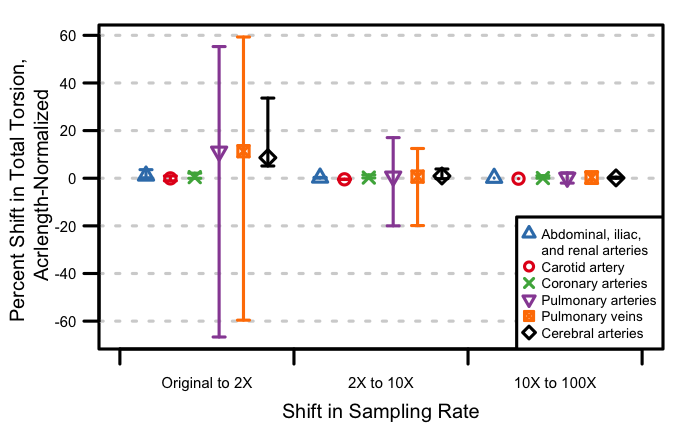}
    \hfill
    \subfigimg[width=2.3in, pos = ll]{\textbf{(i)}}{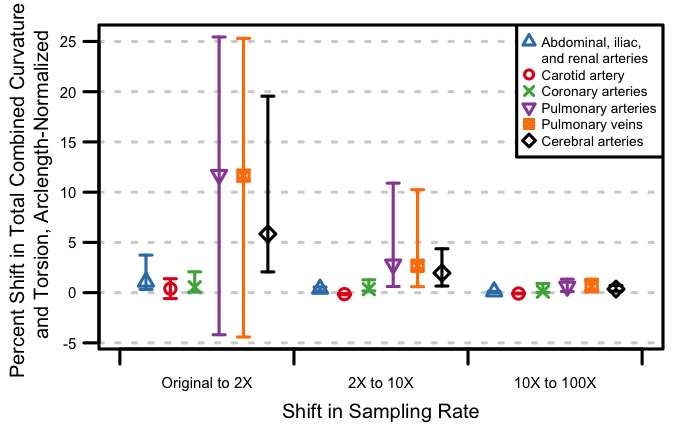}
    \vspace{10pt}
    \\
    \subfigimg[width=2.3in, pos = ll]{\textbf{(j)}}{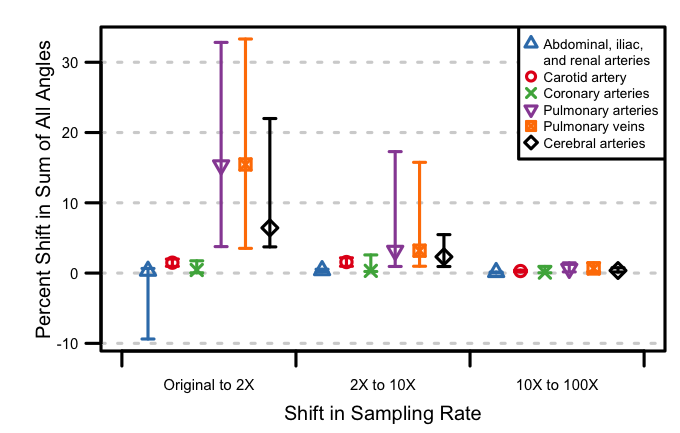}
    \hfill
    \subfigimg[width=2.3in, pos = ll]{\textbf{(k)}}{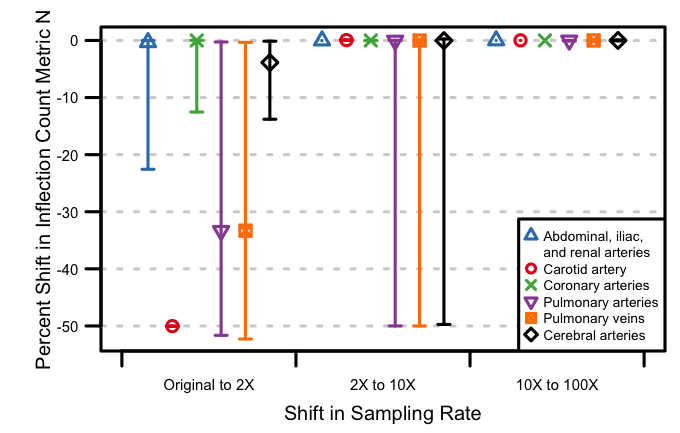}
    \hfill
    \subfigimg[width=2.3in, pos = ll]{\textbf{(l)}}{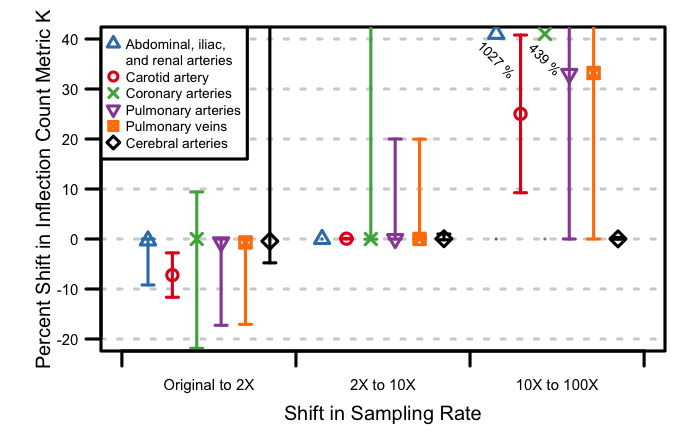}
    \footnotesize
    \caption{Graphs of percent shift in tortuosity measures for each step-wise change in sampling rate.  For each tortuosity measure graphed in {\bf (a)-(l)}, percent shifts are calculated for each vessel and summarized for each data set using quantiles of 18.5\%, 50\%, and 82.5\%.  Tortuosity metrics were calculated as defined in Section \ref{sec:tortuosity_metrics}.  Positive percent shifts indicate increases in sampling rates increased the magnitude of the tortuosity metric, while negative percent shifts indicate increases in sampling rates decreased the magnitude.  Note that the median percent shifts for both the average ($\mathcal{AC}$, $\mathcal{AT}$, $\mathcal{ACCT}$) and total metrics ($\mathcal{TC}$, $\mathcal{TT}$, $\mathcal{TCCT}$) vary in sign, while the median percent shifts for the arc length normalized metrics ($\mathcal{TC}/\mathcal{L}$, $\mathcal{TT}/\mathcal{L}$, $\mathcal{TCCT}/\mathcal{L}$, $\mathcal{SOA}$) are consistently positive, indicating that the tortuosity metrics increased in magnitude with each increase in sampling rate.  This trend of increasing tortuosity magnitude with increasing sampling rate is consistent with the analysis of the Salkowski curve for the same metrics (see Table \ref{tab:phan_tortuosity}).  Also of note is the significant decrease in uncertainty in all tortuosity metrics (excluding the inflection counts) as sampling rates are increased.}
    \label{fig:tortuosity_sampling}
    \end{figure*}

In the clinical vessels we also observe sampling rate-dependent changes in tortuosity metrics (see Figure \ref{fig:tortuosity_sampling}).  For all metrics examined we find that variation in tortuosity measurements due to increases in sampling rates is significantly reduced between the 10X and 100X the original sampling rates.  Furthermore, for the specific tortuosity metrics normalized by arc length ($\mathcal{TC}/\mathcal{L}$, $\mathcal{TT}/\mathcal{L}$, $\mathcal{TCCT}/\mathcal{L}$, and $\mathcal{SOA}$), we find that increases in sampling rates resulted in increased magnitudes of tortuosity, suggesting the potential for underestimations in previous studies. Finally, increased sampling identifies strong correlations between many metrics, indicating a high level of redundancy in commonly used metrics.

Across all tortuosity metrics other than the inflection counts, increases in sampling rates result in significant decreases in the observed variation in measured tortuosity.  This indicates a substantial amount of uncertainty exists in tortuosity measurements made at lower sampling rates.  Whether calculating the average, total, or arc length normalized totals of tortuosity metrics, the torsion only metrics exhibited considerably more variation than the curvature only metrics as a result of increasing sampling rates (see Fig.~\ref{fig:tortuosity_sampling} {\bf (a-b)}, {\bf (d-e)}, and {\bf (g-h)}).  This difference is expected since calculating torsion requires 2\textsuperscript{nd}- and 3\textsuperscript{rd}-order derivatives of the vessel centerline coordinates, thereby making torsion-only tortuosity metrics more sensitive to discontinuities due to under-sampling.

Within the cerebral arteries and pulmonary vessels, the measurements of average and total curvature, total torsion, and total combined curvature and torsion ($\mathcal{AC}$, $\mathcal{TC}$, $\mathcal{TT}$, $\mathcal{TCCT}$) first decreased in magnitude with increased sampling (Original to 2X), then increased in magnitude (2X to 10X, see Figure \ref{fig:tortuosity_sampling} {\bf (a)} and {\bf (d-f)}).  This is unlike nearly all other datasets considered for these metrics where increased sampling rates result in increases in tortuosity (see Figure \ref{fig:tortuosity_sampling} {\bf (a-j)}.  This difference may be due to the large abundance of shorter and initially under-sampled vessels in the cerebral and pulmonary vessel data compared to the coronary arteries of comparable size (see reconstruction error at Original sampling in Figure \ref{fig:recon_error_unfil}).  Specifically, limited sampling in shorter may lead to exceedingly high values of curvature that decrease in magnitude with increased sampling (Original to 2X).  After which, further increases in sampling lead to the common trend of increases in magnitude (2X to 10X).  Given the correlation between vessel length and radius, this scenario reinforces the value of a sampling rate determined by vessel radius.

By examining the arc length normalized metrics of total curvature, torsion, total combined curvature and torsion, and sum-of-angles ($\mathcal{TC}/\mathcal{L}$, $\mathcal{TT}/\mathcal{L}$, $\mathcal{TCCT}/\mathcal{L}$, $\mathcal{SOA}$) we see that increases in sampling rates regularly result in increases in tortuosity metrics (see Figure \ref{fig:tortuosity_sampling}), a trend also observed in the Salkowski phantom vessel (see Table \ref{tab:phan_tortuosity}).  This result seems to indicate that (1) systematic underestimation may have occurred in previous studies utilizing these metrics and (2) accounting for vessel length in tortuosity metrics may be a viable way of controlling for covariation between vessel length, sampling rates, and tortuosity metrics.  This latter point is in agreement with the results for the phantom Salkowski vessel.

For all metrics but the inflection-count metrics, insignificant changes ($<5\%$) occur as the sampling rate is increased from $10X$ to $100X$.  Since the vessels considered in this study had original sampling rates of 1-10 points/mm, this result suggests that for vessels whose radius is between 0.1-10 mm, sampling rates of 10-100 points/mm are necessary for the purposes of providing accurate and precise summary statistics of tortuosity measures.  This range of sampling rates is in agreement with that needed for the measurements of Salkowski curve tortuosity to converge to the analytic values.

The two different methods of calculating and counting inflection points, $\mathcal{IC}_\kappa$ and $\mathcal{IC}_N$ have contrasting changes as sampling rates are increased (Figs.~\ref{fig:tortuosity_sampling}\textbf{(k)} and \ref{fig:tortuosity_sampling}\textbf{(l)}).  For the inflection-count metric based on largest nearest-neighbor changes in the direction of the normal vector, $\mathcal{IC}_N$, all percent shifts are negative, indicating that the number of inflection points detected is decreasing with increased sampling.  This effect is due to the fact that the threshold used to detect a ``large enough" change in the direction of the normal vector is not defined in terms of the sampling rate.  Thus, increased sampling rates result in naturally limiting the size of $|\Delta \buv{N}_j \cdot \Delta \buv{N}_j|$ until eventually no pair of points will cross the threshold.

For the inflection-count metric based on detecting local minima of the curvature $\mathcal{IC}_\kappa$, increased sampling leads to gross increases in the measured values.  In particular, as high as $439\%$ and $1027\%$ for the coronary arteries and abdominal, iliac, and renal arteries, respectively.  This occurs because uniformly distributed, under-sampling heavily influences the number of points that comprise the regions where maxima and minima occur (or where $|d^2\kappa(s_j)/ds_j^2|$ is greatest).

The divergent behavior of these two different ways of detecting and counting inflection points suggests that either the definitions for both of these methods need be refined to account for sampling rates or that alternative approaches be proposed that are not so heavily dependent on sampling rates.

Given the extent to which many tortuosity metrics are defined in terms of curvature and torsion\textemdash either directly or indirectly\textemdash we examined the correlation coefficients and significance of correlation between all measurements at the $100X$ sampling rate, with the addition of all of the dimensionless metrics being normalized by total arc length $\mathcal{L}$.  These results are presented in Table \ref{tab:cor_coeff_norm} and are ordered by correlation clustering.

%\begin{table*}[!t]
%\caption{\textcolor{red}{Update this table for the new data included from helmberger et al.}Coefficients of covariance for tortuosity metrics considered.  ***$p<0.001$.  Only measurements made at the sampling rate of $100X$ were considered.  All correlation coefficients are rounded to two significant digits.}
%\label{tab:cor_coeff}
%\centering
%\begin{tabular}{llllllllll}
%  \hline
%Variables & 1 & 2 & 3 & 4 & 5 & 6 & 7 & 8 & 9 \\ 
%  \hline
%1. $\mathcal{AC}$ & -- &  &  &  &  &  &  &  &  \\ 
%2. $\mathcal{AT}$ & 0.48*** & -- &  &  &  &  &  &  &  \\ 
%3. $\mathcal{ACCT}$ & 0.73*** & 0.93*** & -- &  &  &  &  &  &  \\ 
%4. $\mathcal{SOA}$ & 0.73*** & 0.93*** & 1*** & -- &  &  &  &  &  \\ 
%5. $\mathcal{TT}$ & -0.13*** & 0.42*** & 0.28*** & 0.28*** & -- &  &  &  &  \\
%6. $\mathcal{TCCT}$ & 0.1*** & 0.18*** & 0.22*** & 0.22*** & 0.84*** & -- &  &  &  \\ 
%7. $\mathcal{IC}_\kappa$ & 0.29*** & -0.086*** & 0.079*** & 0.076*** & 0.26*** & 0.57*** & -- &  &  \\ 
%8. $\mathcal{TC}$ & 0.38*** & -0.21*** & 0.017 & 0.014 & 0.25*** & 0.69*** & 0.75*** & -- &  \\ 
%9. $\mathcal{IC}_N$ & 0.44*** & -0.24*** & 0.01 & 0.0085 & -0.075*** & 0.34*** & 0.76*** & 0.78*** & -- \\ 
%   \hline
%\end{tabular}
%\end{table*}

\begin{figure}[!t]
    \centering  
    \includegraphics[width = 3.45in]{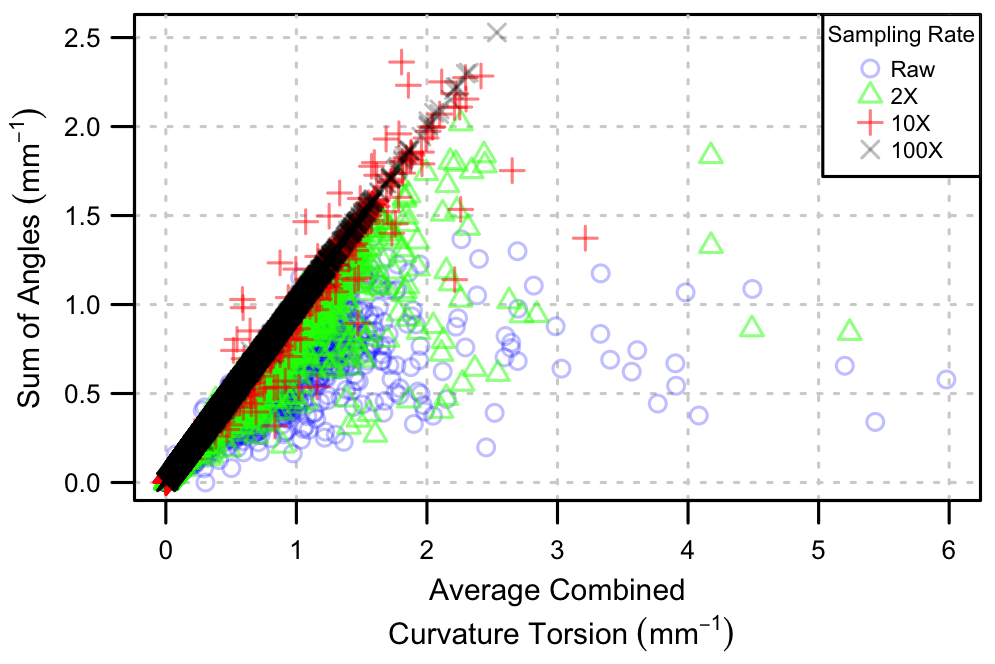}
    \footnotesize
    \caption{Graph of sum-of-angles (SOA) metric versus average combined curvature and torsion for varying sampling rates.  Convergence along the one-to-one line demonstrates that these two metrics are identical measures at sufficiently high sampling rates.  All vessels examined in this study are included.}
    \label{fig:soa_acct_samp}
\end{figure}

Of particular note in Table \ref{tab:cor_coeff_norm} is the complete correlation between the metrics for average combined curvature and torsion, $\mathcal{ACCT}$, and the sum-of-angles, $\mathcal{SOA}$.  To examine further, we graphed the measured values for $\mathcal{SOA}$ versus $\mathcal{ACCT}$ for each of the sampling rates in Fig.~\ref{fig:soa_acct_samp}.  We find that the correlation between $\mathcal{SOA}$ and $\mathcal{ACCT}$ strengthens as the sampling rate is increased and attains a nearly one-to-one relationship.  Specifically, we find that in the limit of high sampling rates, the SOA metric and the average combined curvature and torsion metrics produce the same values.  The relationship is so striking that it motivates an analytical exploration between the two metrics.  We present such an exploration in Supplementary Material Section \ref{app:metric_equivalence}, and report here the main result.  Namely, we find that in the high sampling limit the definition of $\mathcal{SOA}_j$ can be expressed as,

\begin{align}
\mathcal{SOA}_j = \sqrt{\cos^{-1}\left( 1 - \frac{\Delta s^2\kappa_j^2}{2} \right)^2 + \cos^{-1}\left( 1 - \frac{\Delta s^2\tau_j^2}{2} \right)^2}
\end{align}

\noindent Using the approximation that $\cos^{-1}(1-\psi)^2 \approx 2\psi$ for $|\psi| < 1$, and that $\mathcal{SOA} = \sum_{j=1}^N\mathcal{SOA}_j/\sum_{j=1}^N \Delta s$, then,

\begin{align}
\mathcal{SOA} = \frac{1}{N}\sum_{j=1}^N\sqrt{\kappa_j^2 + \tau_j^2}
\end{align}

\noindent which is the exact definition of the average combined curvature-torsion metric, $\mathcal{ACCT}$.

In addition to the $\mathcal{SOA}$ and $\mathcal{ACCT}$ metrics, other pairs of metrics are found to have correlations of unity in Table \ref{tab:cor_coeff_norm}.  In particular, all of the dimensionless metrics normalized by total arc length are highly correlated with their average, dimensionful analogues.  As with the $\mathcal{SOA}$ and $\mathcal{ACCT}$ metrics, this motivates an analytical comparison (Supplemental Material \ref{app:all_metric_equivalence}) in which {\it we demonstrate an exact equivalency between total metrics normalized by arc length and their average analogues occurs at high sampling rates}.  Specifically, $\mathcal{AC} = \mathcal{TC}/\mathcal{L}$, $\mathcal{AT} = \mathcal{TT}/\mathcal{L}$, and $\mathcal{ACCT} = \mathcal{TCCT}/\mathcal{L}$.

\section{Discussion}

We present a numerical approach to reconstructing vessel coordinates from measures of curvature and torsion.  Using this approach we show the importance that the full spectrum of measured values of curvature and torsion play in accurately reconstructing the vessel.  We have shed light on how common tortuosity metrics depend on sampling rates in both phantom and clinical vessels, identifying tortuosity metrics normalized by vessel arc length as having a high likelihood of increased accuracy and resilience to over sampling.  Based on these findings, we provide a range of recommended sampling rates ($10-100$ points/mm) in relation to the diameter of the vessels studied ($1-10$ mm).  For vessels outside of this range, our numerical approach provides a methodology for determining ideal sampling rates in future studies.  Finally, we identify equivalencies between commonly used metrics, suggesting the need for future work to identify unique descriptors of curves that can directly inform properties of fluid flow. 

Our analysis raises several questions regarding the motivation and efficacy of established measures of vessel tortuosity, as well as suggesting new avenues of research for the continued improvement of image-based biomarkers. The numerical integration of the Frenet-Serret equations relies solely on finite-difference methods for evaluating derivatives of the vessel position vector, and therefore the vessel curvature and torsion.  While we chose to use the combinations of backward Euler, Newton's method, and forward Euler to perform the numerical integration of Eqs.~(\ref{eq:fs_system}) and (\ref{eq:nonlinear_position}), future work could focus on examining the performance tradeoffs of alternative numerical methods (e.g. boundary value problem approaches, adaptive time-stepping, or higher-order backward differentiation methods \cite{leveque_finite_difference_2007}).  Similarly, an alternative approach could be working in the parallel-transport frame instead of the Frenet-Serret frame.  Although this frame is less physically intuitive, the basis-vectors remain well-defined through straight sections of the vessel and inflection points \cite{bogunovic_etal_medimg_2012}.  Because these two coordinate systems are related by orthogonal transformations, the eigenvalues of the frame matrices are identical, suggesting similar performances during numerical integration.  A different coordinate system that is both well-defined and with different eigenvalues is the quaternion frame, more commonly used in computer graphics and aerospace engineering because it requires less computer memory and computation time \cite{hanson_etal_ieeetransviscompgraph_1995}. Future studies built within the quaternion frame may enable applications to even larger or higher-resolution images, opening up further research questions and directions.

Our investigation into the dependence of established tortuosity metrics on vessel interpolation sampling rates leads to some surprising results.  One unexpected finding of our analysis is the redundancy of existing metrics.  Specifically, that integrated measures\textemdash once normalized by arc length\textemdash equate to their average analogues. This result is most notably demonstrated by the sum-of-angles (SOA) becoming equal to the average combined curvature and torsion metric.  This is demonstrated both analytically as well as empirically at high sampling rates in Table \ref{tab:cor_coeff_norm}.  Because of these exact equalities among tortuosity metrics, it is worth considering which sets of metrics give independent and complementary information.

Interestingly, the two operational definitions of the inflection-count metric lead to diverging results as sampling rate is increased.  This is despite the intent to capture the same feature of a curve\textemdash a point of inflection.  Motivation for the $\mathcal{IC}_\kappa$ approach (searching for local minima in curvature) is that, analytically speaking, inflection points for spatial curves are defined as instantaneous locations where there is zero curvature \cite{kuhnel_differential_geometry_2006, patrikalakis_etal_shape_interrogation_2009}.  Yet, in real data, it is very challenging to numerically define what constitutes a true zero\textemdash within the resolution limit of the image, or numerical precision of the computer, software, and/or algorithm. This is why, in practice, any calculation must rely on local minima and not exact zeroes.  Motivation for the $\mathcal{IC}_N$ approach (searching for large rotations of the FS-frame) comes from the fact that as the FS-frame passes through an inflection point in a 2D curve generated from data, it will undergo a sudden rotation of at most $\pi$ depending on the sampling rate being used.  These two definitions of inflection points are intended to be consistent, as was demonstrated by Bullitt et al.~ for simulated 2D circular arcs and sinusoidal curves \cite{bullitt_etal_ieeetransmedim_2003}.  However, consistency for non-planar (3D) vessels is far from obvious given that the local minima of curvature may correspond to either true inflection points or to regions where the vessel simply changes planar orientation.  Indeed our results show these two approaches are inconsistent with one another as sampling rates are increased, suggesting that these two metrics should be revisited.

Based on this work, we question the practice of using statistical features of curvature and torsion\textemdash moments of distributions or integrated
summed totals and averages\textemdash as descriptors of vessel shape.  While diagnostically validated, summarizing a vessel with singular measures of curvature and torsion can be misinterpreted as suggesting constancy in those values\textemdash associated with circles and helices\textemdash thereby obscuring the actual vessel shape and any connection to underlying biology.  Furthermore, the growing evidence of biological processes rooted in mechanotransduction\textemdash electrochemical signaling events that induce, or are induced by, changes in local tissue deformations\textemdash motivates a greater need to understand local variation in vessel shape.  

Potential links between measures of shape and fluid shear stress are the Dean and Germano numbers.  These numbers relate local radius, curvature, torsion, viscosity, and blood flow for circular (Dean number) and helical (Germano number) pipes, and they are analogous to Reynolds numbers for quickly classifying flow behavior \cite{dean_lonedindubjsci_1927, dean_lonedindubjsci_1928, germano_jfluidmech_1989, formaggia_etal_cardiovascular_mathematics_2009}.  Recent theoretical and computational work has examined scenarios of non-constant curvature and/or torsion, thereby expanding the known morphospace of curves and their connection to fluid shear stress, pressure and bending and twisting moments \cite{gammack_etal_jfluidmech_2001, monterde_compaidgeomdes_2009, benecke_etal_appmathmod_2005}.  For example, the ratio of torsion to curvature\textemdash when compared with vessel radius, viscosity, and flow rate\textemdash can serve as an indicator of changes in the formation of vortices associated with secondary flow patterns and subsequently changes in the local stresses on vessel walls \cite{gammack_etal_jfluidmech_2001, vorobstova_etal_annalsbiomedeng_2016}.

A promising alternative to singular statistical measures as classifiers are modern machine learning approaches, such as statistical shape analysis, that can take advantage of the full feature space of curvature and torsion \cite{srivastava_etal_ieeetpami_2011}.  Here, populations of sample shapes are generalized by definition of a metric that can smoothly match, deform, and compare different shapes.  These methods have shown promise in identifying structural differences in the corpus callosum associated with Schiozphrenia \cite{shatanu_etal_neuroimage_2013}.  Separately, machine learning performance can be improved by transforming vessel data from raw spatial coordinates to variables indicative of biomechanics and physiology (e.g. the Dean and Germano numbers)\cite{brummer_etal_comparison_insub}.

In this paper we demonstrate the value of maintaining highly-resolved measures of curvature and torsion as unique descriptors of individual vessels.  We have done this by highlighting the utility of these measures in the spatial reconstruction of vessels and by bringing attention to redundancies and inconsistencies between existing metrics.

\section{Conclusion}

	We show that under-sampled measurements of curvature and torsion lead to inaccurate reconstructions of vessels.  In particular, we use numerical integration to demonstrate that higher rates of sampling of the Frenet-Serret equations are necessary to accurately reconstruct a curve from its curvature and torsion values.  For vessels greater than 1 mm in radius, we find that sampling rates between $10-100$ points/mm are sufficient. For vessels smaller than 1 mm in radius, we find $100-1000$ points/mm are sufficient. (Sufficient here means within an error bound equal to the minimum of the vessel radius.) 
	 
	We show in both non-trivial phantom vessels and clinical vessels that a consequence of higher sampling is that many tortuosity measures undergo significant increases in magnitude, suggesting possible underestimates in previous work.  Furthermore, we find that the currently-defined methods for identifying inflection points lead to diverging results as sampling rates are increased. We also show redundancy between the sum-of-angles (SOA) metric and the average combined curvature and torsion metric.  Taken together, these results suggest that currently-used methods need revision, and that previous efforts to classify vessel types could benefit from our approach.
	
	Our results should help inform and motivate future work at the interface of theory and measurement to characterize curves, especially for vessels from medical images.  Potential applications for increased accuracy in vessel characterization range from tracking patient response after stent implantation to diagnosing vascular diseases afflicting tissues that span multiple scales.

\section*{Acknowledgments}

The authors would like to thank Abhay Pandit and Oliver Carroll of the C\'URAM-SFI Research Center for Medical Devices, National University of Ireland, Galway, and Padraig O'Flynn for providing the abdominal, iliac, and renal arteries data; Pavlos Vlachos for providing the coronary arterial tree data; Alexey Kamenskiy for providing the carotid arteries data; Stephen Aylward for providing the cerebral vascular data; and Savinien Bonheur and Michael Pienn of the Ludwig Boltzmann Institute for Lung Vascular Research, Graz, Austria for providing the pulmonary data.  The authors also thank the same individuals for comments on earlier drafts.  The cerebral vascular data from the MR brain images from healthy volunteers used in this paper were collected and made available by the CASILab at The University of North Carolina at Chapel Hill and were distributed by the MIDAS Data Server at Kitware, Inc.

\section*{Data Availability}

The datasets that support the findings of this study are available at IEEE{\it DataPort} \cite{brummer_etal_ieeedataport_2019}.

\ifCLASSOPTIONcaptionsoff
  \newpage
\fi

% \IEEEtriggeratref{8}



\bibliographystyle{IEEEtran}
%\bibliography{IEEEabrv,f32_nih}

\begin{thebibliography}{10}
\providecommand{\url}[1]{#1}
\csname url@samestyle\endcsname
\providecommand{\newblock}{\relax}
\providecommand{\bibinfo}[2]{#2}
\providecommand{\BIBentrySTDinterwordspacing}{\spaceskip=0pt\relax}
\providecommand{\BIBentryALTinterwordstretchfactor}{4}
\providecommand{\BIBentryALTinterwordspacing}{\spaceskip=\fontdimen2\font plus
\BIBentryALTinterwordstretchfactor\fontdimen3\font minus
  \fontdimen4\font\relax}
\providecommand{\BIBforeignlanguage}[2]{{%
\expandafter\ifx\csname l@#1\endcsname\relax
\typeout{** WARNING: IEEEtran.bst: No hyphenation pattern has been}%
\typeout{** loaded for the language `#1'. Using the pattern for}%
\typeout{** the default language instead.}%
\else
\language=\csname l@#1\endcsname
\fi
#2}}
\providecommand{\BIBdecl}{\relax}
\BIBdecl

\bibitem{han_jvascres_2012}
H.-C. Han, ``Twisted blood vessels: symptoms, etiology and biomechanical
  mechanisms,'' \emph{Journal of Vascular Research}, vol.~49, no.~3, pp.
  185--197, 2012.

\bibitem{folarin_etal_microvascres_2010}
A.~Folarin, M.~Konerding, J.~Timonen, S.~Nagl, and R.~Pedley,
  ``Three-dimensional analysis of tumour vascular corrosion casts using
  stereoimaging and micro-computed tomography,'' \emph{Microvascular Research},
  vol.~80, no.~1, pp. 89 -- 98, 2010.

\bibitem{shelton_etal_ultrasoundmedbio_2015}
S.~E. Shelton, Y.~Z. Lee, M.~Lee, E.~Cherin, F.~S. Foster, S.~R. Aylward, and
  P.~A. Dayton, ``Quantification of microvascular tortuosity during tumor
  evolution using acoustic angiography,'' \emph{Ultrasound in Medicine and
  Biology}, vol.~41, no.~7, pp. 1896 -- 1904, 2015.

\bibitem{huang_etal_ieeetransmedim_2008}
S.-F. Huang, R.-F. Chang, W.~K. Moon, Y.-H. Lee, D.-R. Chen, and J.~S. Suri,
  ``Analysis of tumor vascularity using three-dimensional power doppler
  ultrasound images,'' \emph{IEEE Transactions on Medical Imaging}, vol.~27,
  no.~3, pp. 320--330, 2008.

\bibitem{rao_etal_ieeebiomedeng_2016}
S.~R. Rao, S.~E. Shelton, and P.~A. Dayton, ``The ``fingerprint'' of cancer
  extends beyond solid tumor boundaries: Assessment with a novel ultrasound
  imaging approach,'' \emph{IEEE Transactions on Biomedical Engineering},
  vol.~63, no.~5, pp. 1082--1086, May 2016.

\bibitem{bullitt_etal_ieeetransmedim_2003}
E.~Bullitt, G.~Gerig, S.~M. Pizer, W.~Lin, and S.~R. Aylward, ``Measuring
  tortuosity of the intracerebral vasculature from mra images,'' \emph{IEEE
  Transactions on Medical Imaging}, vol.~22, no.~9, pp. 1163--1171, Sept 2003.

\bibitem{bullitt_etal_medimcomcad_2005}
E.~Bullitt, D.~Zeng, G.~Gerig, S.~Aylward, S.~Joshi, J.~K. Smith, W.~Lin, and
  M.~G. Ewend, ``Vessel tortuosity and brain tumor malignancy: A blinded
  study,'' \emph{Academic Radiology}, vol.~12, no.~10, pp. 1232 -- 1240, 2005.

\bibitem{gessner_etal_radiology_2012}
R.~C. Gessner, S.~R. Aylward, and P.~A. Dayton, ``Mapping microvasculature with
  acoustic angiography yields quantifiable differences between healthy and
  tumor-bearing tissue volumes in a rodent model,'' \emph{Radiology}, vol. 264,
  no.~3, pp. 733--740, 2012.

\bibitem{lindsey_etal_moleimgbio_2017}
B.~D. Lindsey, S.~E. Shelton, F.~S. Foster, and P.~A. Dayton, ``Assessment of
  molecular acoustic angiography for combined microvascular and molecular
  imaging in preclinical tumor models,'' \emph{Molecular Imaging and Biology},
  vol.~19, no.~2, pp. 194--202, 2017.

\bibitem{piccinelli_etal_ieeetransmedim_2009}
M.~Piccinelli, A.~Veneziani, D.~A. Steinman, A.~Remuzzi, and L.~Antiga, ``A
  framework for geometric analysis of vascular structures: application to
  cerebral aneurysms,'' \emph{IEEE Transactions on Medical Imaging}, vol.~28,
  no.~8, pp. 1141--1155, 2009.

\bibitem{oloumi_etal_jmedim_2016}
F.~Oloumi, R.~M. Rangayyan, and A.~L. Ells, ``Computer-aided diagnosis of
  retinopathy in retinal fundus images of preterm infants via quantification of
  vascular tortuosity,'' \emph{Journal of Medical Imaging}, vol.~3, no.~4, p.
  044505, 2016.

\bibitem{alilou_etal_scireports_2018}
M.~Alilou, M.~Orooji, N.~Beig, P.~Prasanna, P.~Rajiah, C.~Donatelli,
  V.~Velcheti, S.~Rakshit, M.~Yang, F.~Jacono \emph{et~al.}, ``Quantitative
  vessel tortuosity: A potential ct imaging biomarker for distinguishing lung
  granulomas from adenocarcinomas,'' \emph{Scientific Reports}, vol.~8, no.~1,
  p. 15290, 2018.

\bibitem{kobayashi_etal_japanesecirculation_2015}
M.~Kobayashi, K.~Hoshina, S.~Yamamoto, Y.~Nemoto, T.~Akai, K.~Shigematsu,
  T.~Watanabe, and M.~Ohshima, ``Development of an image-based modeling system
  to investigate evolutional geometric changes of a stent graft in an abdominal
  aortic aneurysm,'' \emph{Circulation Journal}, vol.~79, no.~7, pp.
  1534--1541, 2015.

\bibitem{hart_etal_intjmedinfo_1999}
W.~E. Hart, M.~Goldbaum, B.~C{\^o}t{\'e}, P.~Kube, and M.~R. Nelson,
  ``Measurement and classification of retinal vascular tortuosity,''
  \emph{International Journal of Medical Informatics}, vol.~53, no. 2-3, pp.
  239--252, 1999.

\bibitem{grisan_etal_ieeeengmedbio_2003}
E.~Grisan, M.~Foracchia, and A.~Ruggeri, ``A novel method for the automatic
  evaluation of retinal vessel tortuosity,'' in \emph{Proceedings of the 25th
  Annual International Conference of the IEEE Engineering in Medicine and
  Biology Society}, vol.~1, September 2003, pp. 866--869.

\bibitem{bogunovic_etal_medimg_2012}
H.~Bogunovi{\'c}, J.~M. Pozo, R.~C{\'a}rdenes, M.~C. Villa-Uriol, R.~Blanc,
  M.~Piotin, and A.~F. Frangi, ``Automated landmarking and geometric
  characterization of the carotid siphon,'' \emph{Medical Image Analysis},
  vol.~16, no.~4, pp. 889--903, 2012.

\bibitem{han_jbiomech_2009a}
H.-C. Han, ``Blood vessel buckling within soft surrounding tissue generates
  tortuosity,'' \emph{Journal of Biomechanics}, vol.~42, no.~16, pp.
  2797--2801, 2009.

\bibitem{han_jbiomech_2009b}
------, ``The theoretical foundation for artery buckling under internal
  pressure,'' \emph{Journal of Biomechanical Engineering}, vol. 131, no.~12, p.
  124501, 2009.

\bibitem{gammack_etal_jfluidmech_2001}
D.~Gammack and P.~E. Hydon, ``Flow in pipes with non-uniform curvature and
  torsion,'' \emph{Journal of Fluid Mechanics}, vol. 433, pp. 357--382, 2001.

\bibitem{bullitt_etal_acadrad_2005}
E.~Bullitt, D.~Zeng, G.~Gerig, S.~Aylward, S.~Joshi, J.~K. Smith, W.~Lin, and
  M.~G. Ewend, ``Vessel tortuosity and brain tumor malignancy: a blinded
  study1,'' \emph{Academic Radiology}, vol.~12, no.~10, pp. 1232--1240, 2005.

\bibitem{pandey_etal_hypertension_2018}
A.~K. Pandey, E.~K. Singhi, J.~P. Arroyo, T.~A. Ikizler, E.~R. Gould, J.~Brown,
  J.~A. Beckman, D.~G. Harrison, and J.~Moslehi, ``Mechanisms of vegf (vascular
  endothelial growth factor) inhibitor--associated hypertension and vascular
  disease,'' \emph{Hypertension}, vol.~71, no.~2, pp. e1--e8, 2018.

\bibitem{huang_etal_amjcancres_2018}
Q.~Huang, X.~Hu, W.~He, Y.~Zhao, S.~Hao, Q.~Wu, S.~Li, S.~Zhang, and M.~Shi,
  ``Fluid shear stress and tumor metastasis,'' \emph{American Journal of Cancer
  Research}, vol.~8, no.~5, p. 763, 2018.

\bibitem{apte_etal_cell_2019}
R.~S. Apte, D.~S. Chen, and N.~Ferrara, ``Vegf in signaling and disease: beyond
  discovery and development,'' \emph{Cell}, vol. 176, no.~6, pp. 1248--1264,
  2019.

\bibitem{berger_etal_annrevfluidmech_1983}
S.~Berger, L.~Talbot, and L.~Yao, ``Flow in curved pipes,'' \emph{Annual review
  of fluid mechanics}, vol.~15, no.~1, pp. 461--512, 1983.

\bibitem{monterde_compaidgeomdes_2009}
J.~Monterde, ``Salkowski curves revisited: A family of curves with constant
  curvature and non-constant torsion,'' \emph{Computer Aided Geometric Design},
  vol.~26, no.~3, pp. 271 -- 278, 2009.

\bibitem{kamenskiy_etal_jbiomecheng_2012}
A.~V. Kamenskiy, J.~N. MacTaggart, I.~I. Pipinos, J.~Bikhchandani, and Y.~A.
  Dzenis, ``Three-dimensional geometry of the human carotid artery,''
  \emph{Journal of Biomechanical Engineering}, vol. 134, no.~6, p. 064502,
  2012.

\bibitem{oflynn_etal_annalsbiomedeng_2007}
P.~M. O'Flynn, G.~O'Sullivan, and A.~S. Pandit, ``Methods for three-dimensional
  geometric characterization of the arterial vasculature,'' \emph{Annals of
  Biomedical Engineering}, vol.~35, no.~8, pp. 1368--1381, 2007.

\bibitem{vorobstova_etal_annalsbiomedeng_2016}
N.~Vorobtsova, C.~Chiastra, M.~A. Stremler, D.~C. Sane, F.~Migliavacca, and
  P.~Vlachos, ``Effects of vessel tortuosity on coronary hemodynamics: an
  idealized and patient-specific computational study,'' \emph{Annals of
  Biomedical Engineering}, vol.~44, no.~7, pp. 2228--2239, 2016.

\bibitem{bullitt_etal_neurobiologyaging_2010}
E.~Bullitt, D.~Zeng, B.~Mortamet, A.~Ghosh, S.~R. Aylward, W.~Lin, B.~L. Marks,
  and K.~Smith, ``The effects of healthy aging on intracerebral blood vessels
  visualized by magnetic resonance angiography,'' \emph{Neurobiology of Aging},
  vol.~31, no.~2, pp. 290--300, 2010.

\bibitem{helmberger_etal_plosone_2014}
M.~Helmberger, M.~Pienn, M.~Urschler, P.~Kullnig, R.~Stollberger, G.~Kovacs,
  A.~Olschewski, H.~Olschewski, and Z.~Balint, ``Quantification of tortuosity
  and fractal dimension of the lung vessels in pulmonary hypertension
  patients,'' \emph{PloS one}, vol.~9, no.~1, 2014.

\bibitem{kuhnel_differential_geometry_2006}
W.~K{\"u}hnel, \emph{Differential geometry: Curves - Surfaces - Manifolds},
  2nd~ed.\hskip 1em plus 0.5em minus 0.4em\relax American Mathematical Society,
  2006.

\bibitem{patrikalakis_etal_shape_interrogation_2009}
N.~M. Patrikalakis and T.~Maekawa, \emph{Shape interrogation for computer aided
  design and manufacturing}.\hskip 1em plus 0.5em minus 0.4em\relax Springer
  Science \& Business Media, 2009.

\bibitem{lorthois_etal_microvascres_2014}
S.~Lorthois, F.~Lauwers, and F.~Cassot, ``Tortuosity and other vessel
  attributes for arterioles and venules of the human cerebral cortex,''
  \emph{Microvascular Research}, vol.~91, pp. 99 -- 109, 2014.

\bibitem{oflynn_etal_annalsbiomedeng_2010}
P.~M. O'Flynn, G.~O'Sullivan, and A.~S. Pandit, ``Geometric variability of the
  abdominal aorta and its major peripheral branches,'' \emph{Annals of
  Biomedical Engineering}, vol.~38, no.~3, pp. 824--840, 2010.

\bibitem{bullitt_etal_medimageanalysis_2005}
E.~Bullitt, K.~E. Muller, I.~Jung, W.~Lin, and S.~Aylward, ``Analyzing
  attributes of vessel populations,'' \emph{Medical Image Analysis}, vol.~9,
  no.~1, pp. 39 -- 49, 2005.

\bibitem{gelman_etal_iovs_2005}
R.~Gelman, M.~E. Martinez-Perez, D.~K. Vanderveen, A.~Moskowitz, and A.~B.
  Fulton, ``Diagnosis of plus disease in retinopathy of prematurity using
  retinal image multiscale analysis,'' \emph{Investigative Ophthalmology \&
  Visual Science}, vol.~46, no.~12, pp. 4734--4738, 2005.

\bibitem{meng_etal_surgradanat_2008}
S.~Meng, S.~H. Geyer, M.~P. Viana, W.~J. Weninger \emph{et~al.}, ``Objective
  characterization of the course of the parasellar internal carotid artery
  using mathematical tools,'' \emph{Surgical and Radiologic Anatomy}, vol.~30,
  no.~6, p. 519, 2008.

\bibitem{brummer_etal_ieeedataport_2019}
\BIBentryALTinterwordspacing
A.~B.~V. Savage, ``Healthy vessel centerline coordinates,'' 2019. [Online].
  Available: \url{http://dx.doi.org/10.21227/1ckk-b175}
\BIBentrySTDinterwordspacing

\bibitem{selzer_etal_atherosclerosis_1994}
\BIBentryALTinterwordspacing
R.~H. Selzer, H.~N. Hodis, H.~Kwong-Fu, W.~J. Mack, P.~L. Lee, C.~ran Liu, and
  C.~hua Liu, ``Evaluation of computerized edge tracking for quantifying
  intima-media thickness of the common carotid artery from b-mode ultrasound
  images,'' \emph{Atherosclerosis}, vol. 111, no.~1, pp. 1 -- 11, 1994.
\BIBentrySTDinterwordspacing

\bibitem{aylward_etal_ieeetransmedim_2002}
S.~R. Aylward and E.~Bullitt, ``Intialization, noise, singularities, and scale
  in ridge traversal for tubular object centerline extraction,'' \emph{IEEE
  Transactions on Medical Imaging}, vol.~21, no.~2, pp. 61--75, Feb 2002.

\bibitem{yushkevich_etal_neuroimage_2006}
P.~A. Yushkevich, J.~Piven, H.~C. Hazlett, R.~G. Smith, S.~Ho, J.~C. Gee, and
  G.~Gerig, ``User-guided 3d active contour segmentation of anatomical
  structures: significantly improved efficiency and reliability,''
  \emph{Neuroimage}, vol.~31, no.~3, pp. 1116--1128, 2006.

\bibitem{payer_etal_medimganal_2016}
C.~Payer, M.~Pienn, Z.~B{\'a}lint, A.~Shekhovtsov, E.~Talakic, E.~Nagy,
  A.~Olschewski, H.~Olschewski, and M.~Urschler, ``Automated integer
  programming based separation of arteries and veins from thoracic ct images,''
  \emph{Medical Image Analysis}, vol.~34, pp. 109 -- 122, 2016, special Issue
  on the 2015 Conference on Medical Image Computing and Computer Assisted
  Intervention.

\bibitem{newberry_etal_ploscompbio_2015}
M.~G. Newberry, D.~B. Ennis, and V.~M. Savage, ``Testing foundations of
  biological scaling theory using automated measurements of vascular
  networks,'' \emph{PLoS Computational Biology}, vol.~11, no.~8, p. e1004455,
  2015.

\bibitem{myatt_etal_frontneuroinfo_2012}
\BIBentryALTinterwordspacing
D.~Myatt, T.~Hadlington, G.~Ascoli, and S.~Nasuto, ``Neuromantic -- from
  semi-manual to semi-automatic reconstruction of neuron morphology,''
  \emph{Frontiers in Neuroinformatics}, vol.~6, p.~4, 2012.
\BIBentrySTDinterwordspacing

\bibitem{tekin_etal_ploscompbio_2016}
E.~Tekin, D.~Hunt, M.~G. Newberry, and V.~M. Savage, ``Do vascular networks
  branch optimally or randomly across spatial scales?'' \emph{PLoS
  Computational Biology}, vol.~12, no.~11, pp. 1--28, 11 2016.

\bibitem{eilers_etal_statsci_1996}
P.~H. Eilers and B.~D. Marx, ``Flexible smoothing with b-splines and
  penalties,'' \emph{Statistical Science}, pp. 89--102, 1996.

\bibitem{eilers_etal_wirescompstat_2010}
------, ``Splines, knots, and penalties,'' \emph{Wiley Interdisciplinary
  Reviews: Computational Statistics}, vol.~2, no.~6, pp. 637--653, 2010.

\bibitem{aguilera_etal_mathcompmod_2013}
A.~M. Aguilera and M.~Aguilera-Morillo, ``Comparative study of different
  b-spline approaches for functional data,'' \emph{Mathematical and Computer
  Modelling}, vol.~58, no. 7-8, pp. 1568--1579, 2013.

\bibitem{lee_cad_1989}
E.~T. Lee, ``Choosing nodes in parametric curve interpolation,''
  \emph{Computer-Aided Design}, vol.~21, no.~6, pp. 363--370, 1989.

\bibitem{leveque_finite_difference_2007}
R.~J. LeVeque, \emph{Finite difference methods for ordinary and partial
  differential equations: steady-state and time-dependent problems}.\hskip 1em
  plus 0.5em minus 0.4em\relax Siam, 2007, vol.~98.

\bibitem{lisowska_etal_ieeeengmedbio_2014}
A.~Lisowska, R.~Annunziata, G.~K. Loh, D.~Karl, and E.~Trucco, ``An
  experimental assessment of five indices of retinal vessel tortuosity with the
  ret-tort public dataset,'' in \emph{Engineering in Medicine and Biology
  Society (EMBC), 2014 36th Annual International Conference of the IEEE}.\hskip
  1em plus 0.5em minus 0.4em\relax IEEE, 2014, pp. 5414--5417.

\bibitem{hanson_etal_ieeetransviscompgraph_1995}
A.~J. Hanson and H.~Ma, ``Quaternion frame approach to streamline
  visualization,'' \emph{IEEE Transactions on Visualization and Computer
  Graphics}, vol.~1, no.~2, pp. 164--174, Jun 1995.

\bibitem{dean_lonedindubjsci_1927}
W.~R. Dean, ``Note on the motion of fluid in a curved pipe,'' \emph{The London,
  Edinburgh, and Dublin Philosophical Magazine and Journal of Science}, vol.~4,
  no.~20, pp. 208--223, 1927.

\bibitem{dean_lonedindubjsci_1928}
------, ``The stream-line motion of fluid in a curved pipe,'' \emph{The London,
  Edinburgh, and Dublin Philosophical Magazine and Journal of Science}, vol.~5,
  no.~30, pp. 673--695, 1928.

\bibitem{germano_jfluidmech_1989}
M.~Germano, ``The dean equations extended to a helical pipe flow,''
  \emph{Journal of Fluid Mechanics}, vol. 203, pp. 289--305, 1989.

\bibitem{formaggia_etal_cardiovascular_mathematics_2009}
L.~Formaggia, A.~Quarteroni, and A.~Veneziani, \emph{Cardiovascular
  Mathematics: Modeling and simulation of the circulatory system}.\hskip 1em
  plus 0.5em minus 0.4em\relax Springer Science \& Business Media, 2010,
  vol.~1.

\bibitem{benecke_etal_appmathmod_2005}
S.~Benecke and J.~H. Van~Vuuren, ``Modelling torsion in an elastic cable in
  space,'' \emph{Applied Mathematical Modelling}, vol.~29, no.~2, pp. 117--136,
  2005.

\bibitem{srivastava_etal_ieeetpami_2011}
A.~{Srivastava}, E.~{Klassen}, S.~H. {Joshi}, and I.~H. {Jermyn}, ``Shape
  analysis of elastic curves in euclidean spaces,'' \emph{IEEE Transactions on
  Pattern Analysis and Machine Intelligence}, vol.~33, no.~7, pp. 1415--1428,
  2011.

\bibitem{shatanu_etal_neuroimage_2013}
S.~H. Joshi, K.~L. Narr, O.~R. Philips, K.~H. Nuechterlein, R.~F. Asarnow,
  A.~W. Toga, and R.~P. Woods, ``Statistical shape analysis of the corpus
  callosum in schizophrenia,'' \emph{NeuroImage}, vol.~64, pp. 547 -- 559,
  2013.

\bibitem{brummer_etal_comparison_insub}
A.~B. Brummer, P.~Lymperopoulos, J.~Shen, E.~Tekin, L.~P. Bentley, A.~Gray,
  I.~Oliveras, B.~J. Enquist, and S.~V. M., ``Branching principles of animal
  and plant networks identified by combining extensive data, machine learning,
  and modeling,'' in Review. Available on arXiv:1903.04642.

\end{thebibliography}

\begin{thebibliography}{10}
\providecommand{\url}[1]{#1}
\csname url@samestyle\endcsname
\providecommand{\newblock}{\relax}
\providecommand{\bibinfo}[2]{#2}
\providecommand{\BIBentrySTDinterwordspacing}{\spaceskip=0pt\relax}
\providecommand{\BIBentryALTinterwordstretchfactor}{4}
\providecommand{\BIBentryALTinterwordspacing}{\spaceskip=\fontdimen2\font plus
\BIBentryALTinterwordstretchfactor\fontdimen3\font minus
  \fontdimen4\font\relax}
\providecommand{\BIBforeignlanguage}[2]{{%
\expandafter\ifx\csname l@#1\endcsname\relax
\typeout{** WARNING: IEEEtran.bst: No hyphenation pattern has been}%
\typeout{** loaded for the language `#1'. Using the pattern for}%
\typeout{** the default language instead.}%
\else
\language=\csname l@#1\endcsname
\fi
#2}}
\providecommand{\BIBdecl}{\relax}
\BIBdecl

\bibitem{s_monterde_compaidgeomdes_2009}
J.~Monterde, ``Salkowski curves revisited: A family of curves with constant
  curvature and non-constant torsion,'' \emph{Computer Aided Geometric Design},
  vol.~26, no.~3, pp. 271 -- 278, 2009.

\bibitem{s_newberry_etal_ploscompbio_2015}
M.~G. Newberry, D.~B. Ennis, and V.~M. Savage, ``Testing foundations of
  biological scaling theory using automated measurements of vascular
  networks,'' \emph{PLoS Computational Biology}, vol.~11, no.~8, p. e1004455,
  2015.

\bibitem{s_payer_etal_medimganal_2016}
C.~Payer, M.~Pienn, Z.~B{\'a}lint, A.~Shekhovtsov, E.~Talakic, E.~Nagy,
  A.~Olschewski, H.~Olschewski, and M.~Urschler, ``Automated integer
  programming based separation of arteries and veins from thoracic ct images,''
  \emph{Medical Image Analysis}, vol.~34, pp. 109 -- 122, 2016, special Issue
  on the 2015 Conference on Medical Image Computing and Computer Assisted
  Intervention.

\bibitem{s_leveque_finite_difference_2007}
R.~J. LeVeque, \emph{Finite difference methods for ordinary and partial
  differential equations: steady-state and time-dependent problems}.\hskip 1em
  plus 0.5em minus 0.4em\relax Siam, 2007, vol.~98.

\bibitem{s_aylward_etal_ieeetransmedim_2002}
S.~R. Aylward and E.~Bullitt, ``Intialization, noise, singularities, and scale
  in ridge traversal for tubular object centerline extraction,'' \emph{IEEE
  Transactions on Medical Imaging}, vol.~21, no.~2, pp. 61--75, Feb 2002.

\bibitem{s_bullitt_etal_medimganalysis_2001}
E.~Bullitt, S.~Aylward, K.~Smith, S.~Mukherji, M.~Jiroutek, and K.~Muller,
  ``Symbolic description of intracerebral vessels segmented from magnetic
  resonance angiograms and evaluation by comparison with x-ray angiograms,''
  \emph{Medical Image Analysis}, vol.~5, no.~2, pp. 157--169, 2001.

\bibitem{s_aguilera_etal_mathcompmod_2013}
A.~M. Aguilera and M.~Aguilera-Morillo, ``Comparative study of different
  b-spline approaches for functional data,'' \emph{Mathematical and Computer
  Modelling}, vol.~58, no. 7-8, pp. 1568--1579, 2013.

\bibitem{s_bullitt_etal_neurobiologyaging_2010}
E.~Bullitt, D.~Zeng, B.~Mortamet, A.~Ghosh, S.~R. Aylward, W.~Lin, B.~L. Marks,
  and K.~Smith, ``The effects of healthy aging on intracerebral blood vessels
  visualized by magnetic resonance angiography,'' \emph{Neurobiology of Aging},
  vol.~31, no.~2, pp. 290--300, 2010.

\bibitem{s_kamenskiy_etal_jbiomecheng_2012}
A.~V. Kamenskiy, J.~N. MacTaggart, I.~I. Pipinos, J.~Bikhchandani, and Y.~A.
  Dzenis, ``Three-dimensional geometry of the human carotid artery,''
  \emph{Journal of Biomechanical Engineering}, vol. 134, no.~6, p. 064502,
  2012.

\bibitem{s_oflynn_etal_annalsbiomedeng_2007}
P.~M. O'Flynn, G.~O'Sullivan, and A.~S. Pandit, ``Methods for three-dimensional
  geometric characterization of the arterial vasculature,'' \emph{Annals of
  Biomedical Engineering}, vol.~35, no.~8, pp. 1368--1381, 2007.

\bibitem{s_vorobstova_etal_annalsbiomedeng_2016}
N.~Vorobtsova, C.~Chiastra, M.~A. Stremler, D.~C. Sane, F.~Migliavacca, and
  P.~Vlachos, ``Effects of vessel tortuosity on coronary hemodynamics: an
  idealized and patient-specific computational study,'' \emph{Annals of
  Biomedical Engineering}, vol.~44, no.~7, pp. 2228--2239, 2016.

\bibitem{s_lorthois_etal_microvascres_2014}
S.~Lorthois, F.~Lauwers, and F.~Cassot, ``Tortuosity and other vessel
  attributes for arterioles and venules of the human cerebral cortex,''
  \emph{Microvascular Research}, vol.~91, pp. 99 -- 109, 2014.

\end{thebibliography}



% *** SPECIAL COMMANDS ***
%
\DeclareRobustCommand{\buv}[1]{\widehat{\mathbf{#1}}}
\DeclareRobustCommand{\bv}[1]{\vv{\mathbf{#1}}}
\DeclareRobustCommand{\IP}{\textsc{\tiny IP}}
\DeclareRobustCommand{\SF}{\textsc{\tiny SF}}
\setcounter{table}{0}
\setcounter{figure}{0}
\setcounter{equation}{0}
\setcounter{section}{0}

\renewcommand{\thetable}{S\arabic{table}}   
\renewcommand{\thefigure}{S\arabic{figure}}
\renewcommand{\theequation}{S\arabic{equation}}
\renewcommand*{\citenumfont}[1]{S#1}
\renewcommand*{\bibnumfmt}[1]{[S#1]}

%\appendices

\vspace{15pt}
\begin{center}
{\Large--Supplemental Material--}
\end{center}

\section{Phantom Vessels and Analytic Tortuosity}
\label{app:phantoms}
Here we describe the parameters for each class of phantom vessel.  In each class, gaussian noise was incorporated by drawing new vessel centerline coordinates, $\bv{R}(s_j)$, from the distribution defined by

\begin{align}
G(\bv{R}(s_j)) = \frac{e^{\left( -\frac{1}{2}(\bv{R}(s_j) - \bv{r}(s_j))^T {\bf\Sigma}^{-1}(\bv{R}(s_j) - \bv{r}(s_j))\right)}}{\sqrt{(2\pi)^3 |{\bf\Sigma}|}}
\end{align}

\noindent where $G(\bv{R}(s_j))$ is centered about the original centerline coordinate, $\bv{r}(s_j)$, and ${\bf\Sigma}$ is the coordinate covariance matrix, defined for our purposes as diagonal and with isotropic variance, ${\bf\Sigma} = \sigma\bf{I}$.  The value of $\sigma$ will be specified for each class below.

After incorporating gaussian noise, vessel phantoms were interpolated with P-splines and sub-sampled as described in the main text.  Effective radii for the vessel phantoms were defined as the maximum euclidean distance between the interpolated coordinates and coordinates with gaussian noise.  Thus, by specifying the value of $\sigma$, the effective radii of the phantom vessels can be controlled.

%In all three classes, 50 centerline coordinates were used to define the original vessel.

The benefit of the three classes of phantom vessels studied is that tortuosity metrics can be expressed analytically.  Below are generic definitions for the metrics of total curvature $(\mathcal{TC})$, total torsion $(\mathcal{TT})$, total combined curvature and torsion $(\mathcal{TCCT})$, and arclength $(\mathcal{L})$:

\begin{align}
	\mathcal{TC} & = \int_{s_0}^{s_f} \kappa(\tilde{s})d\tilde{s} \label{eq:tc}\\
	\mathcal{TT} & = \int_{s_0}^{s_f} \tau(\tilde{s})d\tilde{s} \label{eq:tt}\\
	\mathcal{TCCT} & = \int_{s_0}^{s_f} \sqrt{\kappa(\tilde{s})^2 + \tau(\tilde{s})^2}d\tilde{s} \label{eq:tcct}\\
	\mathcal{L} & = \int_{s_0}^{s_f} d\tilde{s} \label{eq:arclength}
\end{align}

\noindent where $d\tilde{s}$ represents an infinitesimal arc length, and $s_0$ and $s_f$ are the beginning and end measurement points along the length of the vessels.  Normalized versions of  Eqs.~(\ref{eq:tc}) - (\ref{eq:tcct}) simply require division by Eq.~(\ref{eq:arclength}).

\subsection{Circles}
The circular phantom vessels generated were defined to each have a radius of $1 mm$ and an arc length of $\pi mm$.  That is, we only studied semi-circular phantom vessels for this class.  Expressed in parameterized spatial coordinates, 

\begin{align}
\bv{r}(t)_c & = 
	\begin{cases}
		\cos(t)\buv{x} \\
		\sin(t)\buv{y}
	\end{cases}
\end{align}

\noindent with $t \in [0, \pi]$.  For each semi-circular phantom, noise was added such that $\sigma = 0.0005$, resulting in an average effective vessel radius of $\overline{r}_{c, eff} = 0.0969 mm\pm 0.0005 mm$ and an average sampling rate of 16$points/mm$.  Analytic curvature and torsion for the circles are $\kappa_c = 1 mm^{-1}$ and $\tau_c = 0$, with numerical values for tortuosity metrics presented in Table \ref{tab:phan_tort_vals}.

%Since curvature and torsion for a circle are constant with respect to arc length, then the expected values for the tortuosity metrics for the circle phantoms are:
%
%\begin{align}
%\mathcal{TC}_c & = (0.8\pi mm)\kappa_c = 2.513 \nonumber \\
%\mathcal{TT}_c & = (0.8\pi mm)\tau_c = 0 \nonumber \\
%\mathcal{TCCT}_c & = (0.8\pi mm)\sqrt{\kappa_c^2 + \tau_c^2} = 2.513 \nonumber \\
%\mathcal{TC/L}_c & =\kappa_c = 1 mm^{-1} \nonumber \\
%\mathcal{TT/L}_c & = \tau_c = 0 \nonumber \\
%\mathcal{TCCT/L}_c & = \sqrt{\kappa_c^2 + \tau_c^2} = 1 mm^{-1} \nonumber
%\end{align}
%
%\noindent where the factor of $0.8$ is due to removing the first and last 10\% of a vessel following interpolation.

\subsection{Helices}
The helical phantom vessels generated were defined to each have a radius of $a = 4 mm$ and a pitch of $ 2\pi b $, where $b = 2mm$  Expressed in parameterized spatial coordinates,

\begin{align}
\bv{r}(t)_h & = 
	\begin{cases}
		4 \cos(t)\buv{x} \\
		4 \sin(t)\buv{y} \\
		2 t\buv{z}
	\end{cases}
\end{align}

\noindent with $t \in [0, 10]$.  For each helical phantom, $\sigma = 0.1$, resulting in an average effective vessel radius of $\overline{r}_{h, eff} = 1.435 mm\pm 0.009 mm$ and an average sampling rate of 1$point/mm$.  Analytic tortuosity measures for the helical phantoms are $\kappa_h = a/(a^2 + b^2) = 0.2 mm^{-1}$ and $\tau_h = b/(a^2 + b^2) = 0.1 mm^{-1}$, with numerical values for tortuosity metrics presented in Table \ref{tab:phan_tort_vals}.

%Since curvature and torsion for a helix are constant with respect to arc length, and arc length for a helix is $\mathcal{L}_h = 0.8\max\{t\}\sqrt{a^2 + b^2}$, then the expected values for the tortuosity metrics for the helix phantoms are:
%
%\begin{align}
%\mathcal{TC}_h & = \kappa_h\mathcal{L}_h = 7.155 \nonumber \\
%\mathcal{TT}_h& = \tau_h\mathcal{L}_h = 3.578 \nonumber \\
%\mathcal{TCCT}_h & = \sqrt{\kappa_h^2 + \tau_h^2}\mathcal{L} = 8 \nonumber \\
%\mathcal{TC/L}_h & = \kappa_h = 0.2 mm^{-1} \nonumber \\
%\mathcal{TT/L}_h & = \tau_h = 0.1 mm^{-1} \nonumber \\
%\mathcal{TCCT/L}_h & = \sqrt{\kappa_h^2 + \tau_h^2} = 0.224 mm^{-1} \nonumber
%\end{align}

\begin{table*}[!t]
\renewcommand{\arraystretch}{1.3}
% if using array.sty, it might be a good idea to tweak the value of
% \extrarowheight as needed to properly center the text within the cells
\caption{Phantom vessel tortuosity values.}
\label{tab:phan_tort_vals}
\centering
\begin{tabular}{LLLLLLL}
  \hline
 Class & $\mathcal{TC} $ & $\mathcal{TT}$ & $\mathcal{TCCT}$ & $\mathcal{TC/L}$ $(mm^{-1})$ & $\mathcal{TT/L}$ $(mm^{-1})$ & $\mathcal{TCCT/L}$ $(mm^{-1})$ \\ 
  \hline
Semi-Circle & 2.513 & 0 & 2.513 & 0.167 & 0 & 0.167   \\
Helix & 7.155 & 3.578 & 8 & 0.2 & 0.1 & 0.224   \\
 Salkowski & 12.8 & 8.95 & 16.31 & 1 & 0.699 & 1.274   \\ 
   \hline
\end{tabular}
\end{table*}

\subsection{Salkowski Curves}
\label{sec:si_salk}
The Salkowski phantoms are unique to the semi-circles and helices as they have constant curvature but non-constant torsion.  Like circles and helices, the normal vector of a Salkowski curve maintains the same planar orientation as the Frenet-Frame translates along the length of the curve.  For a thorough analysis of Salkowski curves, see Monterde in \cite{s_monterde_compaidgeomdes_2009}.  The parameterized spatial coordinates of a Salkowski curve are given by,

\begin{align}
\hspace{-2mm}
\bv{r}(t)_s & = 
	\frac{n}{m}\begin{cases}
		\frac{(n-1)\sin\left((1+2n)t\right)}{4(1+2n)} - \frac{(1+n)\sin\left((1-2n)t\right)}{4(1-2n)} - \frac{\sin(t)}{2}\buv{x}\\
		\vspace{-2mm} \\
		\frac{(1-n)\cos\left((1+2n)t\right)}{4(1+2n)} + \frac{(1+n)\cos\left((1-2n)t\right)}{4(1-2n)} + \frac{\cos(t)}{2}\buv{y}\\
		\vspace{-2mm}\\
		\frac{\cos(2nt)}{4m}\buv{z}
	\end{cases}
\end{align}

\noindent where $n = m/\sqrt{1+m^2}$, and $m \ne \pm 1/\sqrt{3},$ or $0$.  For each Salkowski phantom, $m = 1/16$, $t \in [0, \pi/2n]$, and $\sigma = 0.001$, resulting in an average effective vessel radius of $\overline{r}_{s, eff} = 2.386 mm\pm 0.008 mm$ and an average sampling rate of 3$points/mm$.  Analytic tortuosity measures for the Salkowski phantoms are $\kappa_s = 1 mm^{-1}$ and most importantly, the non-constant torsion of $\tau_s(t) = \tan(nt)$.  Numerical values for the tortuosity metrics are presented in Table \ref{tab:phan_tort_vals}, however, as their calculation is less obvious we present more detailed expressions below.

Unlike circles and helices, Salkowski curves have nonlinear speed, expressed as, $ds/dt = \cos(nt)/\sqrt{1+m^2}$.  Thus, to account for the 10\% trimming of each end-length of the final interpolated and subsampled phantom vessels, we must identify expressions for the starting and ending arc length positions, $s_0$ and $s_f$ using Eq.~(\ref{eq:arclength}).  These are $s_0 = \frac{1}{n}\sin^{-1}(0.1)$ and $s_f = \frac{1}{n}\sin^{-1}(0.9)$.  Thus, Eqs.~(\ref{eq:tc})-(\ref{eq:tcct}) become

\begin{align}
	\mathcal{TC}_s & = \int_{\frac{1}{n}\sin^{-1}(0.1)}^{\frac{1}{n}\sin^{-1}(0.9)}\frac{\cos(nt)}{\sqrt{1+m^2}}dt \\
	\vspace{-2mm}\nonumber \\
	\mathcal{TT}_s & = \int_{\frac{1}{n}\sin^{-1}(0.1)}^{\frac{1}{n}\sin^{-1}(0.9)}\tan(nt)\frac{\cos(nt)}{\sqrt{1+m^2}}dt \\
	\vspace{-2mm}\nonumber \\
	\mathcal{TCCT}_s & = \int_{\frac{1}{n}\sin^{-1}(0.1)}^{\frac{1}{n}\sin^{-1}(0.9)}\sqrt{1 + \tan(nt)^2}\frac{\cos(nt)}{\sqrt{1+m^2}}dt
\end{align}

\noindent Resulting numerical values for the above expressions are presented in Table \ref{tab:phan_tort_vals}.

%Given that the final interpolated and subsampled vessels have 10\% of each end removed, and that the mapping from arc length to 

\section{Description of Updates to Angicart}
\label{angicart_description}

The segmentation and structural characterization of the pulmonary data utilizes an adapted implementation of the original version of Angicart proposed by \cite{s_newberry_etal_ploscompbio_2015}.  The fundamental goals of vascular network segmentation and analysis, as well as the core skeletonization procedures are retained with little modification from this original method. However, an additional preliminary coarse segmentation allows for the accurate treatment of vascular networks that include loops, while offering the supplementary benefits of more adaptable, localized and parallelizable algorithms.  It should be noted that the pulmonary data analyzed with Angicart had already been processed with the automated integer programming routine of Payer et al.~\cite{s_payer_etal_medimganal_2016} to detect the pulmonary arteries and veins.  Thus, the loop detection update of Angicart was not implemented for this work.

A constant threshold is applied to input 2D or 3D images to differentiate vascular and nonvascular voxels.  The preliminary coarse segmentation proceeds serially to construct topologically simple subsegments of vessels.  The topology is characterized by the adjacency relations between spherically bound regions and their complements within the vascular structure.

Construction of each bounding sphere begins with the selection of a random vascular voxel that has not yet been considered or assigned within another sphere.  The radius of the sphere around the seed point is extended until a preset fraction of the outer shell of the sphere consists of nonvascular voxels.  We find 0.16 to be a reasonably sensitive fraction of nonvascular voxels.

Because the initial seed point can be anywhere along a vessel's length or at any distance from a vessel's centerline, the position of the sphere center is translated to the center of mass of the currently bound vascular voxels after each expansion cycle, and the appropriate radius is determined for the new center.  This adjustment in the location of the center is repeated until the center of mass of the final vascular subsegment coincides with the sphere center.

Spherically bound subsegments completely bisect a segment and are disjoint from one another.  If the bounded subsegment of the trial point at any expansion step intersects another already established subsegment, then the current sphere center and any centers included in the progression since the randomly selected seed are marked as invalid seed points to avoid repeated consideration.  When no new valid spherically bounded subsegments are identified after 200 consecutive random seeds, or if no seed candidates remain, then the coarse segmentation concludes.

This coarse segmentation yields topologically simple subsegments with a small number of adjacent subsections for local skeletonization by erosion, both within the spherically bounded regions, and within the disjoint regions that are their complement in the set of vascular voxels.  Most regions include vascular voxels from a single vessel and have exactly two adjacent subsections.  Sections that have more than two adjacencies include at least one branch point.

Not only is the global topology maintained for networks with loops\textemdash which would have been lost through erosion in the previous approach that used global skeletonization\textemdash but the disjoint nature of the subsets allows for parallel processing and reduces the computational resources that would be required to treat all vascular voxels in the image simultaneously.  An additional benefit is that some variation in vessel shape and size from anatomy or imaging artifacts and noise can be treated automatically before explicit definitions of vessel properties are available.

Skeletonization identifies a centerline connecting all adjacent valid subsegments.  The critical endpoints are defined as the points in the shared boundary between subsegments that are farthest from any nonvascular voxel.  Within the bulk of the subsegment's vascular voxels, shorter distances to the nearest nonvascular voxel are eroded first to yield the minimal centerline.

To encourage a more minimal centerline path, deviations of one voxel from the eroded skeleton are allowed if the deviation shortens the length of a vessel.  The error in path information by this final processing is limited to the voxel-level uncertainty in the boundary of each vessel that is unavoidable from finite image resolution.

\section{Five-Point Stencils}
\label{app:differences}

Here we list the five-point centered difference methods used for approximating the first three derivatives of the position vector, $\bv{r}(s)$.  We chose to use five-point centered differences to ensure accuracy in each derivative through the fourth-order.  This is of particular importance in approximating the first derivative of the position vector as torsion is a function of its first three derivatives.  Here we adopt the following notation: $u_{j,k}$ represents a position vector component $(k = \{x, y, z\})$ evaluated at position $s_j$ along the vessel centerline, $u_{j,k} = u_k(s_j)$; $h$ represents the step size between the uniformly spaced, neighboring points, $h = s_{j+1} - s_j$; and $D'_5, D''_5$, and $ D'''_5$ represent the five-point methods used to approximate the first, second, and third derivatives.  These methods are expressed as,

\begin{align}
D'_5u_{j,k} & = \frac{1}{3h}\left[ \frac{1}{4}u_{j+2,k} + 2u_{j+1,k} - 2u_{j-1,k} - \frac{1}{4}u_{j-2,k} \right]
\label{eq:five_methods_1}
\end{align}
\begin{align}
D''_5u_{j,k} & = \frac{1}{h^2} \left[ -\frac{1}{12}u_{j+2,k} + \frac{4}{3}u_{j+1,k} \right. \nonumber \\
&\quad\quad\quad \left. - \frac{5}{2}u_{j,k} +\frac{4}{3}u_{j-1,k} - \frac{1}{12}u_{j-2,k} \right] \label{eq:five_methods_2}
\end{align}
\begin{align}
D'''_5u_{j,k} & = \frac{1}{h^3}\left[ \frac{1}{2}u_{j+2,k} - u_{j+1,k} + u_{j-1,k} - \frac{1}{2}u_{j-2,k} \right]
\label{eq:five_methods_3}
\end{align}

\section{Numerical Integration Methods}
\label{app:numerical}
Here we present further details on our reconstruction procedure.  For formal sources of information on these methods, the reader is directed toward LeVeque's \textit{Finite difference methods for ordinary and partial differential equations: steady-state and time-dependent problems} \cite{s_leveque_finite_difference_2007}.  We assume one has already measured curvature and torsion using Eqs.~(\ref{eq:curv_pos}), (\ref{eq:tors_pos}), and (\ref{eq:five_methods_1}--\ref{eq:five_methods_3}).  Now, recalling the compact version of the Frenet-Serret Theorem in Eq.~(\ref{eq:fs_system}), $d\bm{\Phi}(s_j)/ds_j = \bm{M}(s_j) \bm{\Phi}(s_j)$, and adopting the same notation as in Supplemental Material Section \ref{app:differences}, we use the nonlinear backward Euler method to express $\bm{\Phi}_{j+1}$ as,

\begin{equation}
\bm{\Phi}_{j+1} = \bm{\Phi}_j + h\bm{M}_{j+1}\bm{\Phi}_{j+1}
\label{eq:back_eul}
\end{equation}

\noindent As Eq.~(\ref{eq:back_eul}) is nonlinear in $\bm{\Phi}_{j+1}$, we employ Newton's method to iteratively solve for $\bm{\Phi}_{j+1}$.  The $m+1$ update step in Newton's method is written as,

\begin{equation}
\bm{\Phi}_{j+1}^{[m+1]} = \bm{\Phi}_{j+1}^{[m]} - \left[ \frac{\partial}{\partial \Phi_{j+1,l}^{[m]}} G(\Phi_{j+1,k}^{[m]})^T \right]^{-1}\hspace{-0.125in}G(\bm{\Phi}_{j+1}^{[m]})
\label{eq:newton_method}
\end{equation}

\noindent where $[m]$ represents the iteration count, and $G(\bm{\Phi}_{j+1})$ is a $9 \times 1$ column vector defined as,

\begin{equation}
G(\bm{\Phi}_{j+1}) = \bm{\Phi}_{j+1} - \bm{\Phi}_j - h\bm{M}_{j+1}\bm{\Phi}_{j+1}
\label{eq:big_g}
\end{equation}

\noindent In order to solve Eqs.~(\ref{eq:newton_method}-\ref{eq:big_g}), the initial condition $\bm{\Phi}_{j=0}^{[m=0]}$ is taken directly from the data.  The term within the square brackets in Eq.~(\ref{eq:newton_method}) is the $k,l$ entry in the Jacobian matrix of $G(\bm{\Phi}_{j+1})$, where the indices $k$ and $l$ respectively represent the rows and columns of this matrix and are written explicitly for clarity.  The Jacobian of $G(\bm{\Phi}_{j+1})$ is given by,

\begin{equation}
\frac{\partial}{\partial \bm{\Phi}_{j+1}^{[m]}} G(\bm{\Phi}_{j+1}^{[m]})^T = h\nu_{j+1}
\begin{pmatrix} \frac{1}{h\nu_{j+1}}\textbf{I} & -\kappa_{j+1}\textbf{I} & \textbf{0} \\  \kappa_{j+1}\textbf{I} & \frac{1}{h\nu_{j+1}}\textbf{I} & -\tau_{j+1}\textbf{I} \\ \textbf{0} & \tau_{j+1}\textbf{I} & \frac{1}{h\nu_{j+1}}\textbf{I} \end{pmatrix} 
\label{eq:jacobian}
\end{equation}

\noindent where $\textbf{0}$ and $\textbf{I}$ are the $3\times3$ zero and identity matrices.  Given the tridiagonal, skew-symmetric form of Eq.~(\ref{eq:jacobian}), it is easily invertible with standard programming packages.  Furthermore, all of the entries in Eq.~(\ref{eq:jacobian}) depend only on $h, \nu_{j+1}, \kappa_{j+1},$ and $\tau_{j+1}$, meaning the Jacobian (and its inverse) need only be calculated once for each Frenet-Serret frame $\bm{\Phi}_{j+1}$.  Moreover, these calculations can also be used for the full iteration over $[m]$ until convergence is reached.  Convergence is determined by setting a threshold on the differences between returned values in successive iterations: $\bm{\Phi}_{j+1}^{[m+1]} - \bm{\Phi}_{j+1}^{[m]} \le \delta$.  Tests for dependence on a suitable threshold were performed by examining the reconstruction error for a range of values spanning $0.1$ to $0.0001$.  We settled on a global value of $\delta = 0.0001$ for all vessels.

In order to solve the nonlinear Eq.~(\ref{eq:nonlinear_position}), we first linearize with the following argument.  We can express the normed derivative in Eq.~(\ref{eq:nonlinear_position}) with a forward Euler difference as $||\bv{r}'(s_j)|| = ||(\bv{r}_{j+1} - \bv{r}_j)/(s_{j+1}-s_j)|| = (1/h)||\bv{r}_{j+1} - \bv{r}_j||$, where we factored out the scalar, point-to-point distance $h = s_{j+1} - s_j$ from the norm.  By construction, the vector $\bv{r}_{j+1} - \bv{r}_j$ begins at the point $s_{j+1}$ and terminates at the point $s_j$.  As the point-to-point distance $h$ along the arc length is calculated as a Euclidean length, then the length of the vector $||\bv{r}_{j+1} - \bv{r}_j||$ is exactly equal to the distance from $s_{j+1}$ to $s_j$.  Therefore $(1/h)||\bv{r}_{j+1} - \bv{r}_j|| = 1$, a result that is importantly independent of step size.

Following the linearization of Eq.~(\ref{eq:nonlinear_position}), integration is straightforward as we are left with a constant equation. Using a forward-Euler routine to solve for the reconstructed position vector $\bv{R}$, we have,

\begin{align}
\bv{R}_{j+1} = \bv{R}_j + h\buv{T}_j
\end{align}

\noindent A point-wise error in reconstruction is defined as the normed difference between the original position coordinate vector $\bv{r}(s_j)$ and the reconstructed coordinate vector $\bv{R}(s_j)$ as,

\begin{align}
\epsilon_j = ||\bv{r}(s_j)-\bv{R}(s_j)||
\end{align}

\noindent By comparing the maximum of the point-wise error, $\epsilon_j$, against the minimum vessel radius, we can identify the appropriate sampling rate, $N$, for reconstruction.

\section{Reproducing Results from Previously Published Data}
\label{app:data_reproduction}
The data used in this study required some standardization to reproduce originally published results as per each authors' published instructions.  Here we describe how we processed the different datasets prior to employing our own methods for analysis.

\subsubsection{Bullitt et al.~(2010)}
\label{app:bullitt_recon}

\begin{figure*}[!t]
    \centering
    \subfigimg[width=1.725in, pos = ll]{\textbf{(a)}}{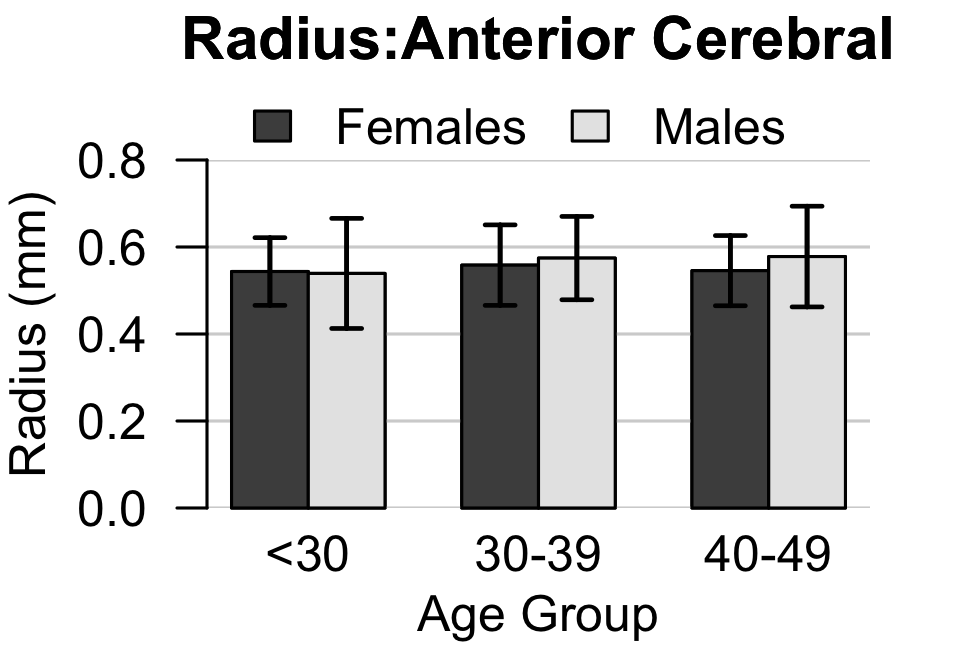}
    \hfill
    \subfigimg[width=1.725in, pos = ll]{\textbf{(b)}}{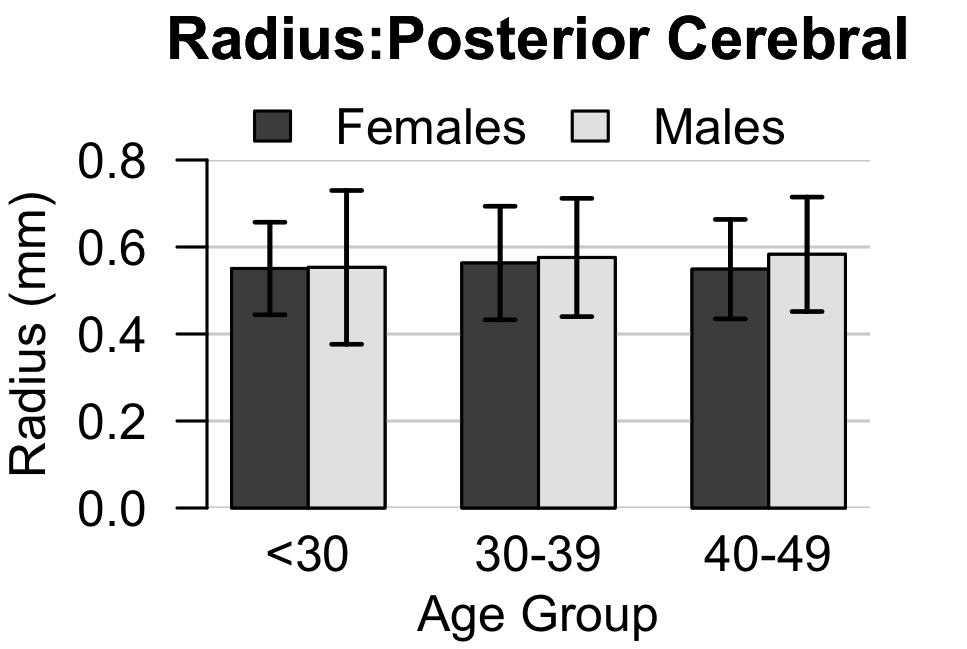}
    \hfill
    \subfigimg[width=1.725in, pos = ll]{\textbf{(c)}}{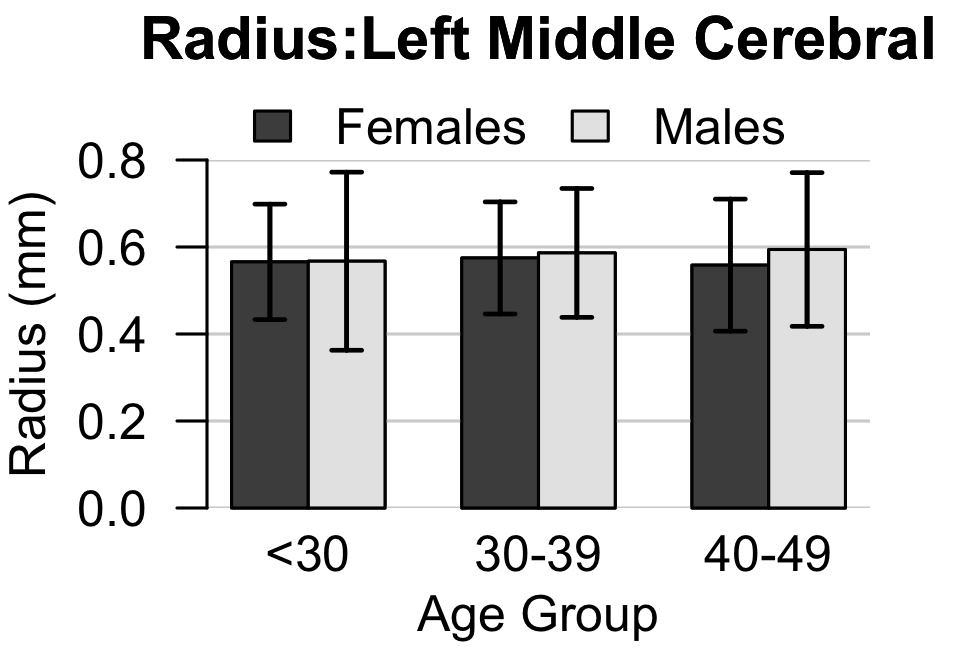}
    \hfill
    \subfigimg[width=1.725in, pos = ll]{\textbf{(d)}}{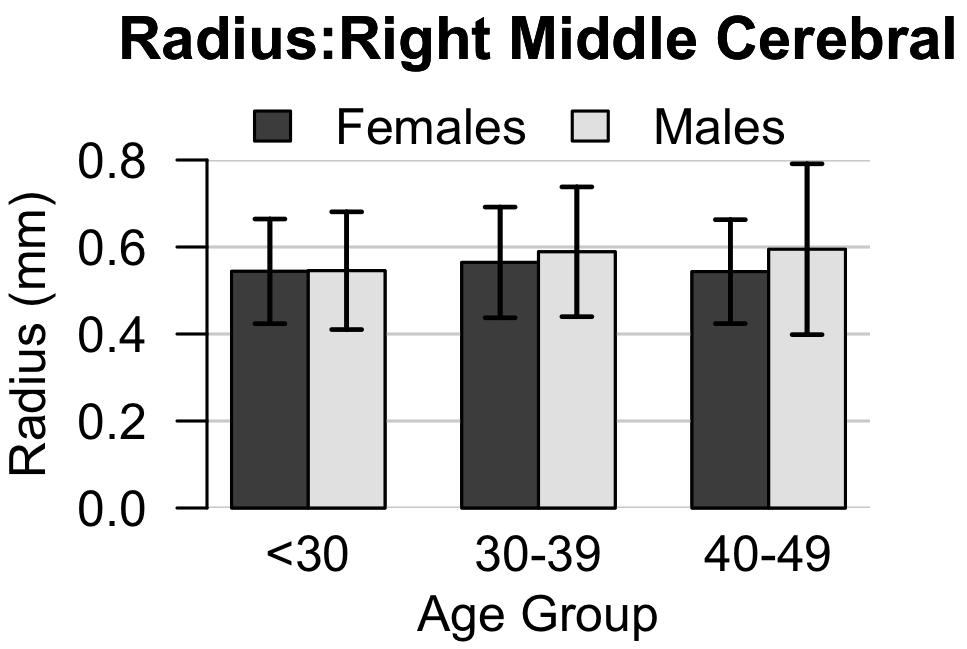}
    \vspace{10pt}
    \\
    \subfigimg[width=1.725in, pos = ll]{\textbf{(e)}}{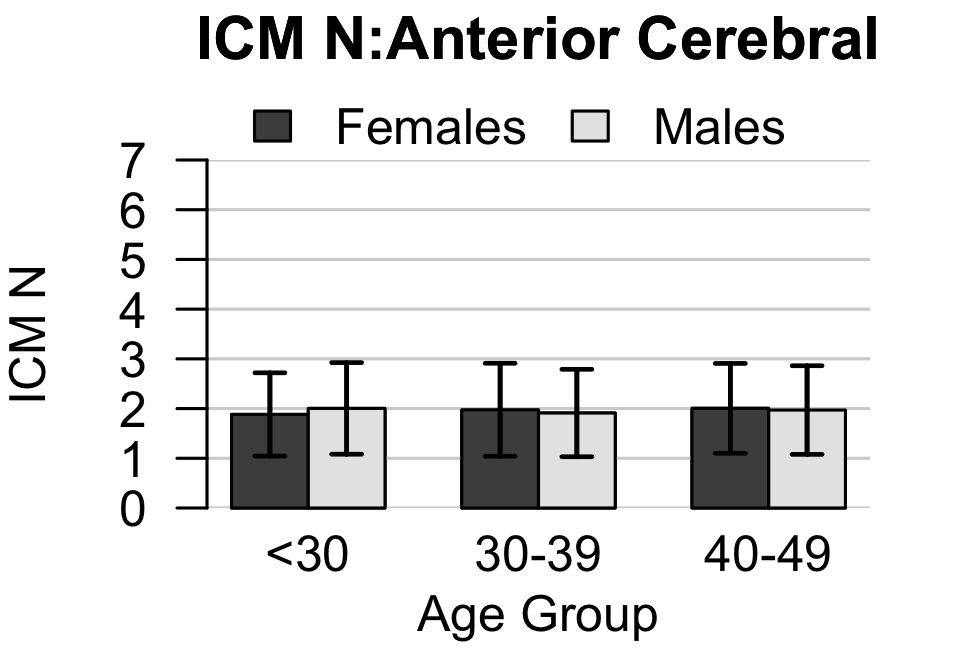}
    \hfill
    \subfigimg[width=1.725in, pos = ll]{\textbf{(f)}}{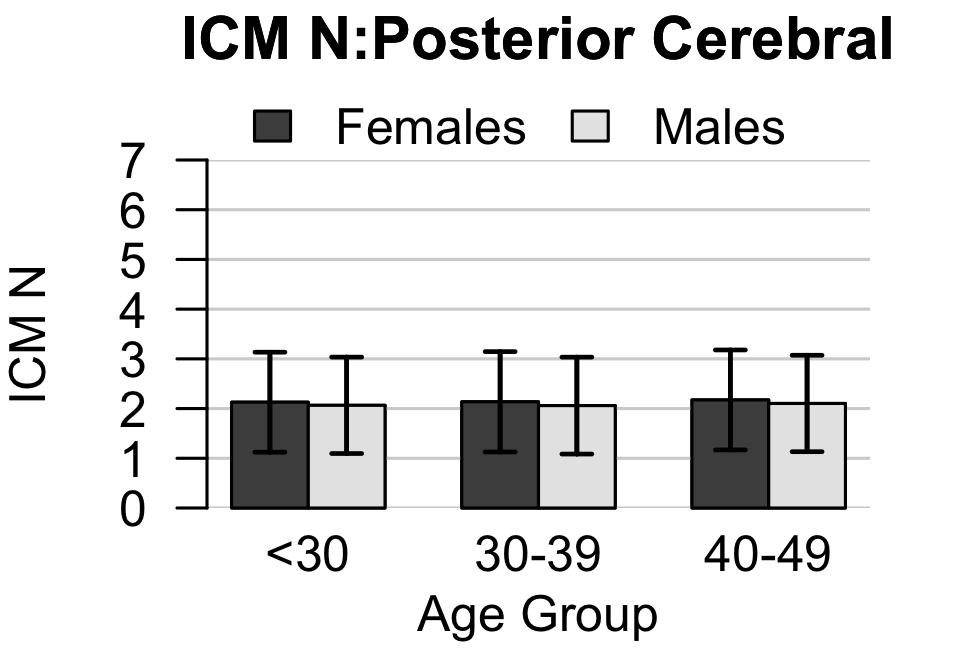}
    \hfill
    \subfigimg[width=1.725in, pos = ll]{\textbf{(g)}}{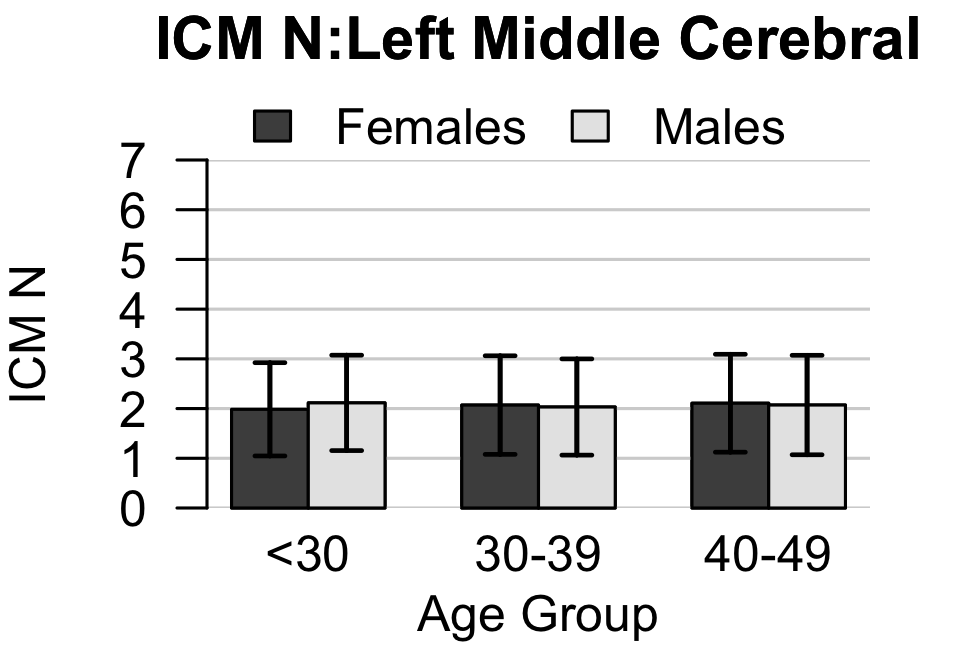}
    \hfill
    \subfigimg[width=1.725in, pos = ll]{\textbf{(h)}}{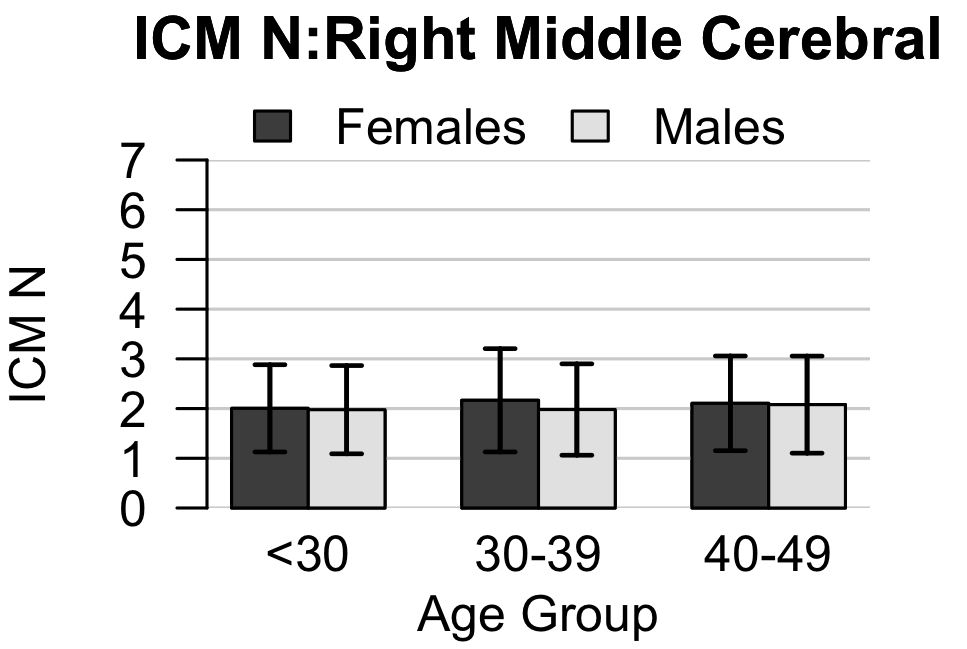}
        \vspace{10pt}
    \\
    \subfigimg[width=1.725in, pos = ll]{\textbf{(i)}}{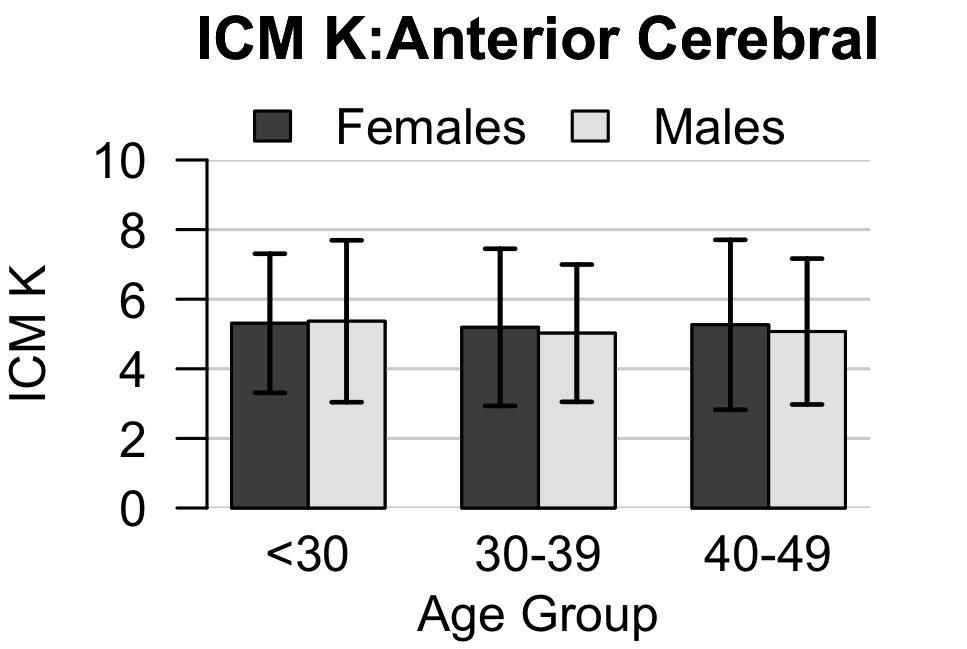}
    \hfill
    \subfigimg[width=1.725in, pos = ll]{\textbf{(j)}}{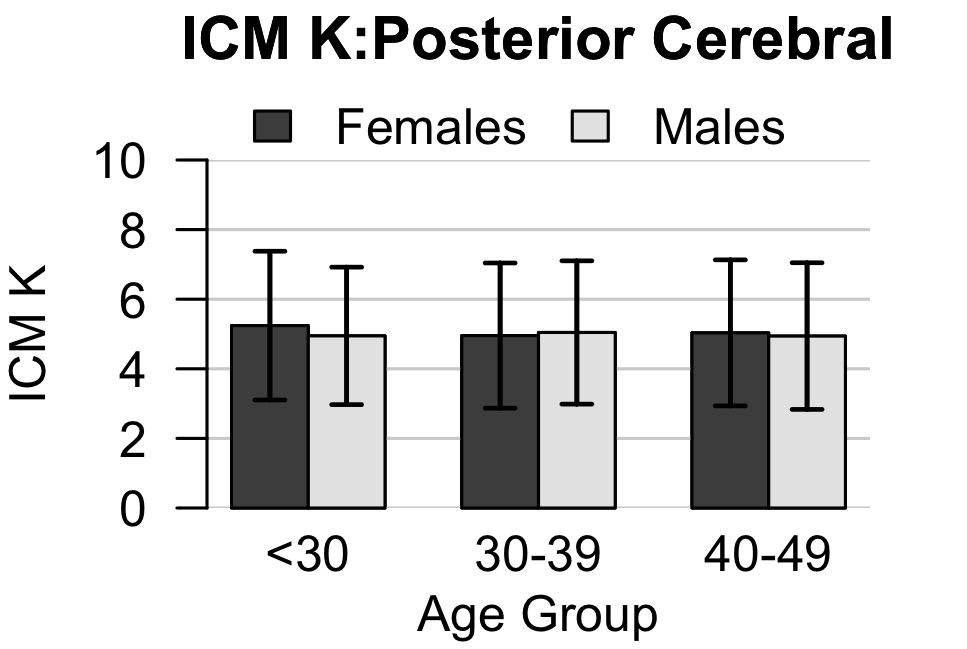}
    \hfill
    \subfigimg[width=1.725in, pos = ll]{\textbf{(k)}}{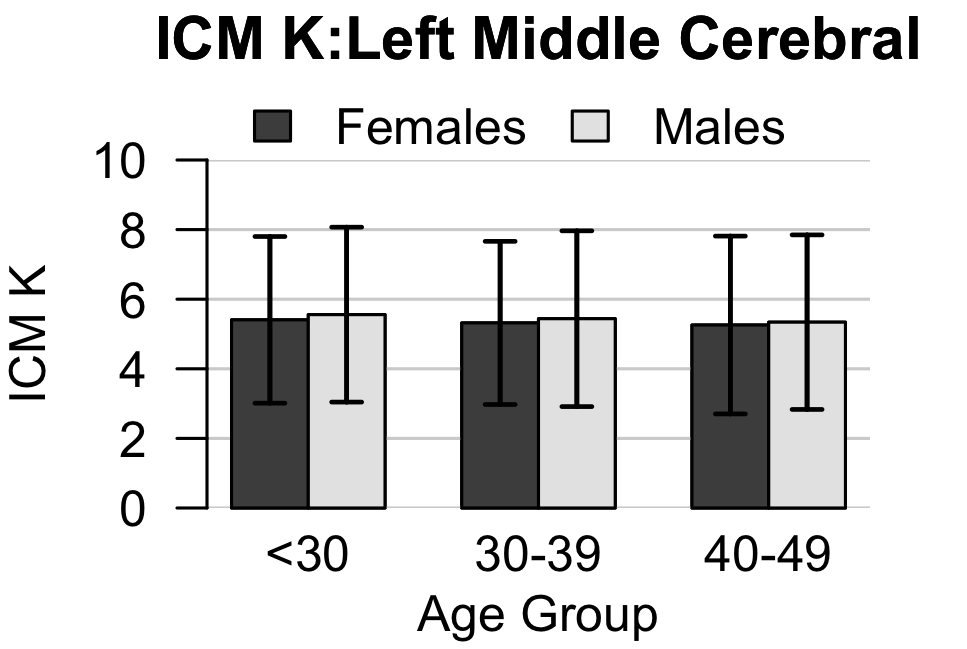}
    \hfill
    \subfigimg[width=1.725in, pos = ll]{\textbf{(l)}}{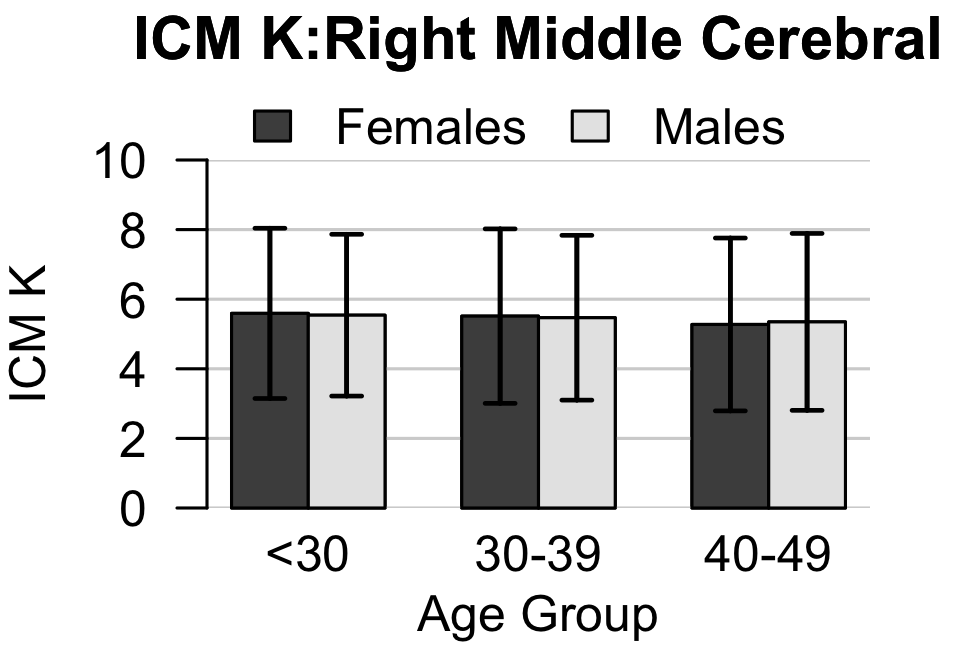}
        \vspace{10pt}
    \\
    \subfigimg[width=1.725in, pos = ll]{\textbf{(m)}}{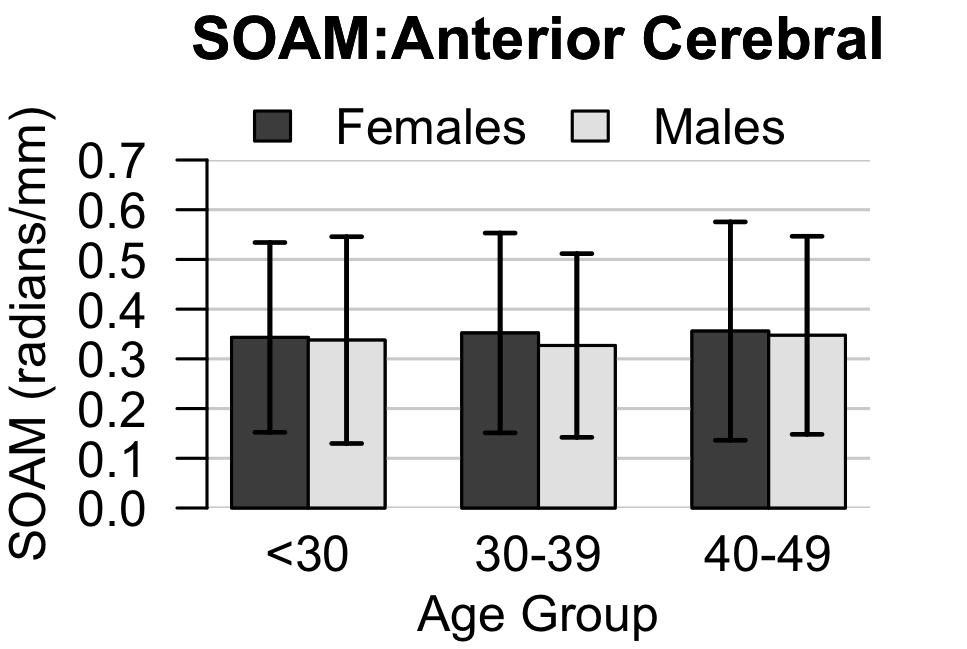}
    \hfill
    \subfigimg[width=1.725in, pos = ll]{\textbf{(n)}}{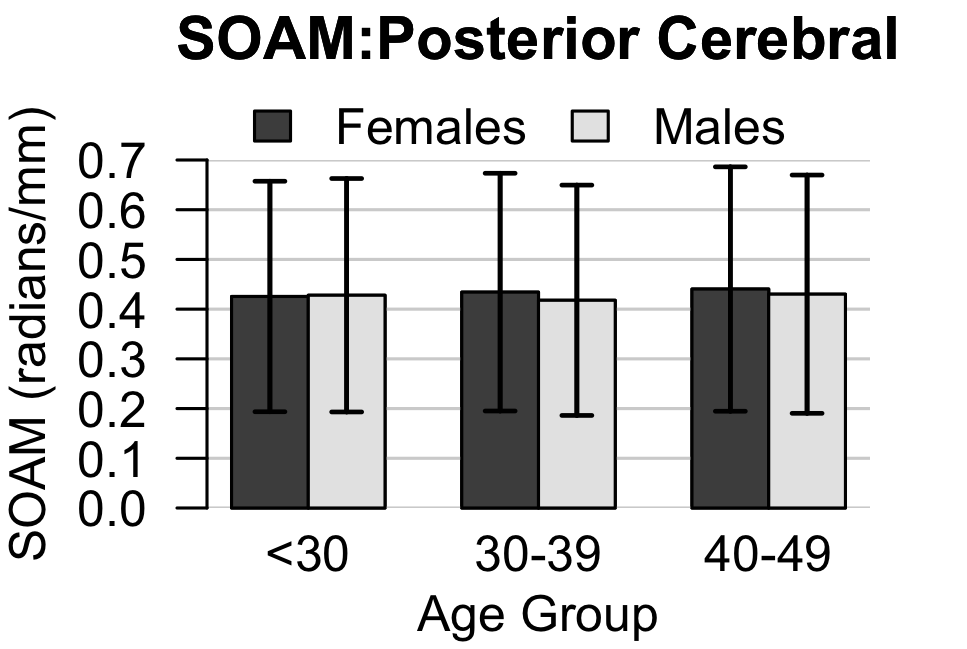}
    \hfill
    \subfigimg[width=1.725in, pos = ll]{\textbf{(o)}}{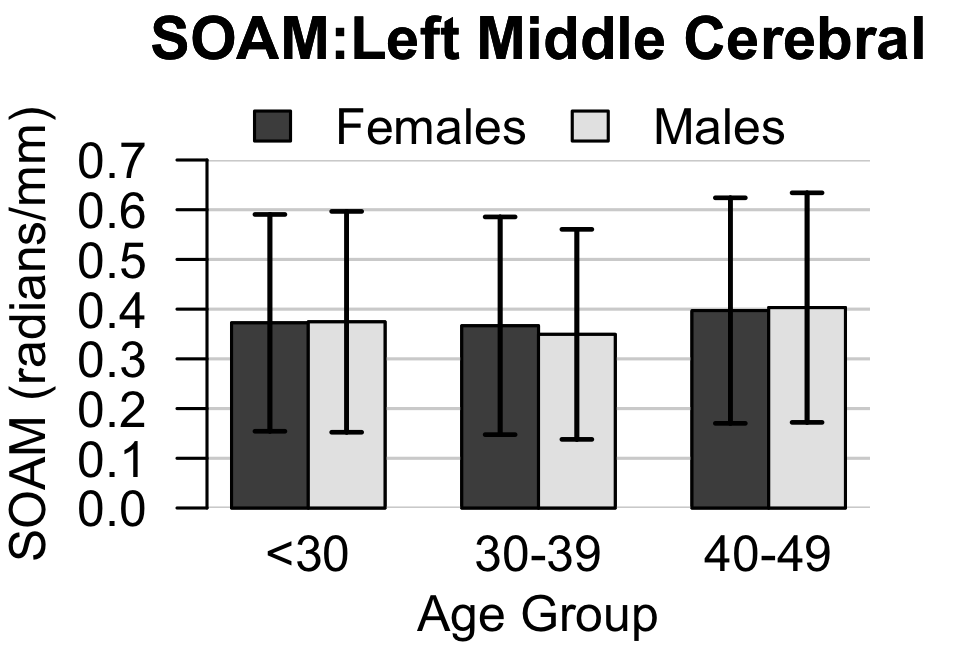}
    \hfill
    \subfigimg[width=1.725in, pos = ll]{\textbf{(p)}}{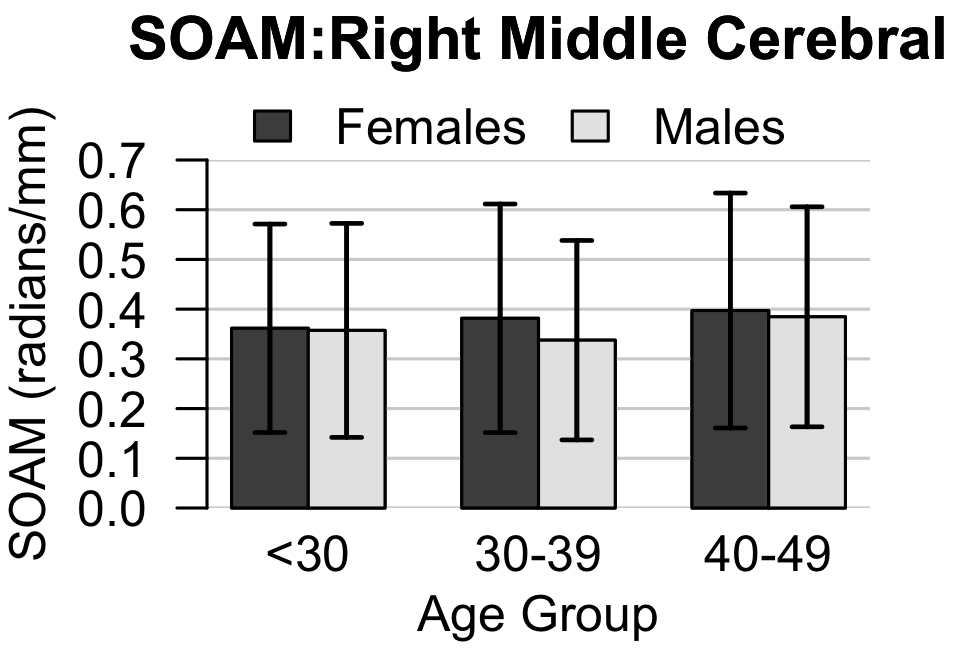}
    \footnotesize
    \caption{Graphs of tortuosity metrics for data from Bullitt et al.~for cerebral arterial vasculature in the anterior, posterior, left middle, and right middle arterial trees.  Metrics measured are average vessel radius, inflection-counts for both methods, $\mathcal{IC}_N$ and $\mathcal{IC}_\kappa$, and the sum-of-angles, $\mathcal{SOA}$.}
    \label{fig:bullitt_data}
    \end{figure*}

Data received from Bullitt et al.~consisted of the original magnetic resonance imaging (MRI) files for all 100 patients' cranial regions, and semi-processed data for 42 patients.  Specifically, for those 42 patients, the MRI files had been run through the ridge-traversal segmentation process described in \cite{s_aylward_etal_ieeetransmedim_2002}, but not through the smoothing and vessel dendritic connection process described in \cite{s_bullitt_etal_medimganalysis_2001}.  Thus, we employed our own interpolation and smoothing routine using penalized splines (P-splines) of degree 2 (cubic splines) with 3 knots.  The degree was chosen as per instructions in \cite{s_aylward_etal_ieeetransmedim_2002}.  P-splines were chosen due to their robustness against over-fitting \cite{s_aguilera_etal_mathcompmod_2013}, and the number of knots was determined iteratively by comparing measured results against those published in \cite{s_bullitt_etal_neurobiologyaging_2010}.  The sampling rate was chosen to match that from \cite{s_bullitt_etal_neurobiologyaging_2010}, specifically 2 points/mm.  Once the measurements for the interpolated vessels were in sufficient agreement with published values (see Fig. \ref{fig:bullitt_data}), the interpolated vessels were then reanalyzed using the methods we describe in the main text.

\subsubsection{Kamenskiy et al.~(2012)}
\label{app:kamen_recon}

\begin{figure}[!t]
    \centering  
    \subfigimg[width=3.45in, pos = ll]{\textbf{(a)}}{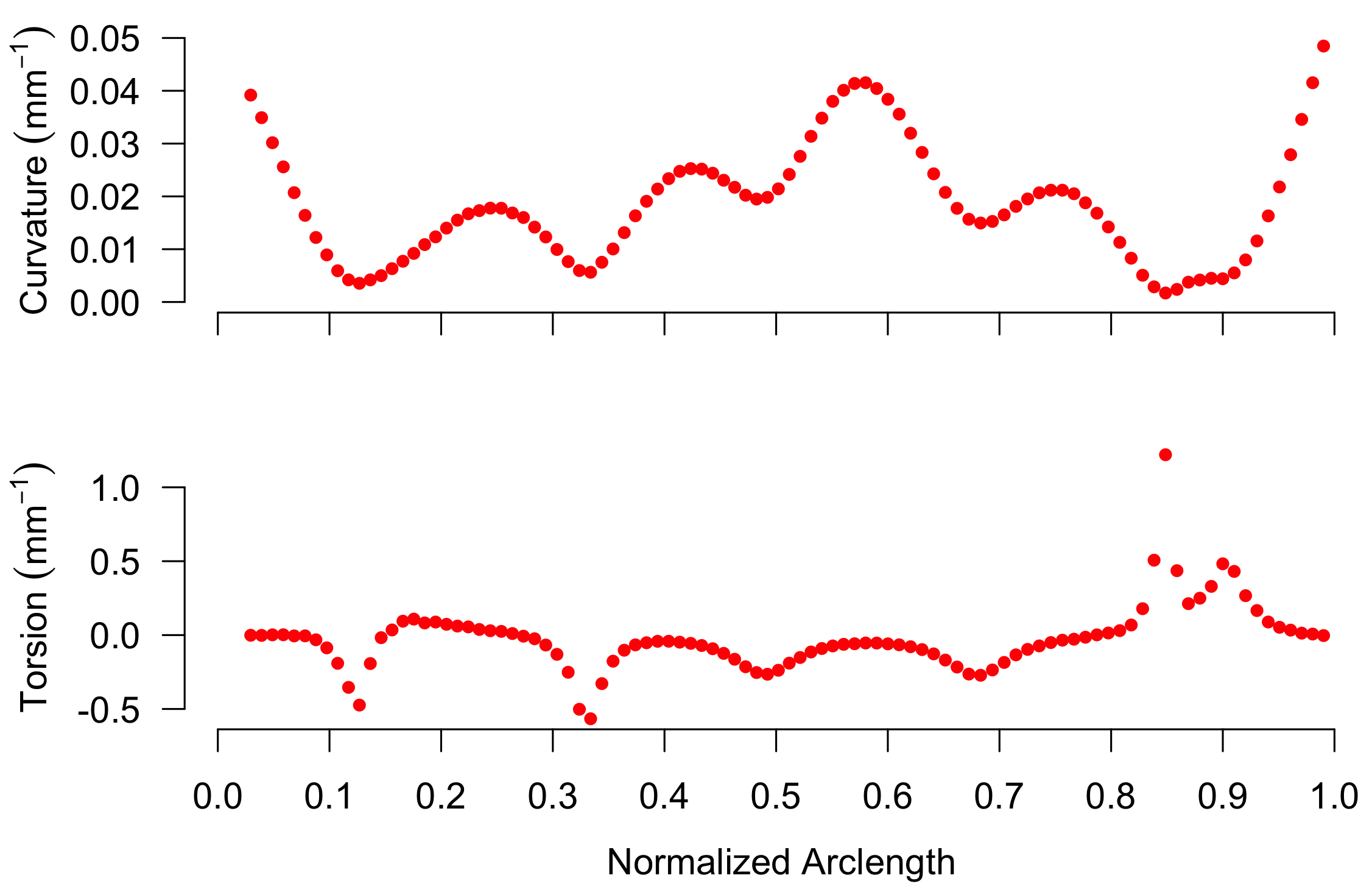}
    \vspace{10pt}
    \\
    \subfigimg[width=3.45in, pos = ll]{\textbf{(b)}}{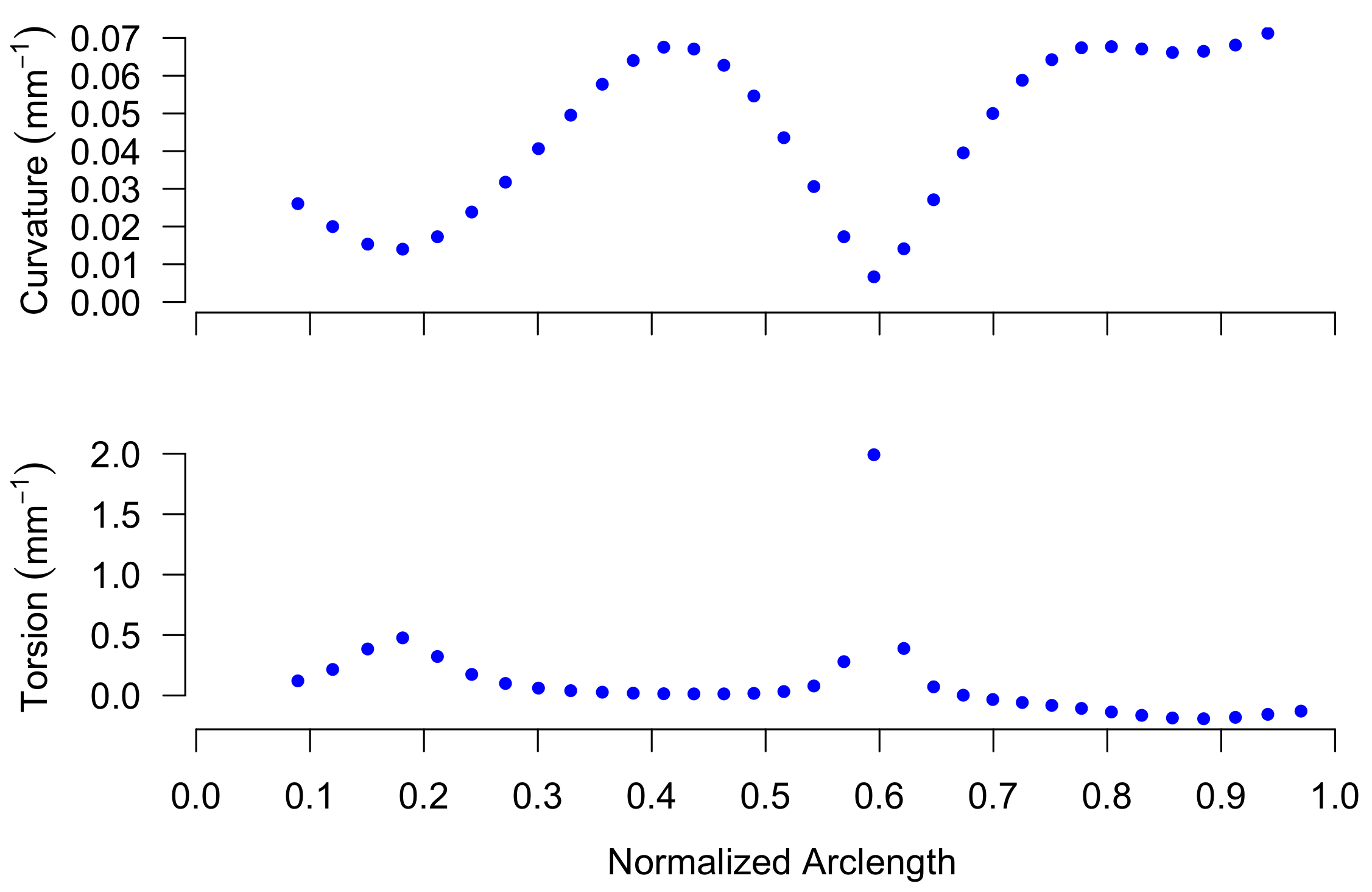}
    \footnotesize
    \caption{Graphs of curvature and torsion versus normalized arc length for data from Kamenskiy et al.~\cite{s_kamenskiy_etal_jbiomecheng_2012}.  \textbf{(a)} is the common carotid artery and internal carotid artery.  \textbf{(b)} is the external carotid artery.  Note that this data is presented using a sampling rate of 1 point/mm as per the original publication.}
    \label{fig:kamen_data}
\end{figure}

The data received from Kamenskiy et al.~consisted of the averaged centerline coordinates for the carotid arteries for all 16 patients from their study \cite{s_kamenskiy_etal_jbiomecheng_2012}.  This data required zero additional effort on our part in terms of interpolation or smoothing to reproduce their published results of curvature and torsion versus arc length (see Fig.~\ref{fig:kamen_data}).  Note that we have graphed these curves in a manner that is reflective of the sampling rate used in the original publication.  Comparison against the original publication will show that our measurements are in agreement to within the previously published standard errors.

\subsubsection{O'Flynn et al.~(2007)}
\label{app:oflynn_recon}

The data received from the authors of O'Flynn et al.~\cite{s_oflynn_etal_annalsbiomedeng_2007} were in sets of three, two-dimensional planar-projections of the $xyz$-coordinates for individual vessels, recovered from scanned images of the author's dissertation.  Thus, for each vessel we had to extract 3 pairs of coordinates from the scanned images and merge into one set of all three coordinates.  After this, we followed the instructions for repeated interpolation as per \cite{s_oflynn_etal_annalsbiomedeng_2007} using 9$\textsuperscript{th}$-degree polynomials with subsampling rates of 5 points/mm.  During this procedure one of the five arteries, the left renal artery (LRA), was not usable due to information loss from the projection.

\begin{figure}[!t]
    \centering  
    \subfigimg[width=3.45in, pos = ll]{\textbf{(a)}}{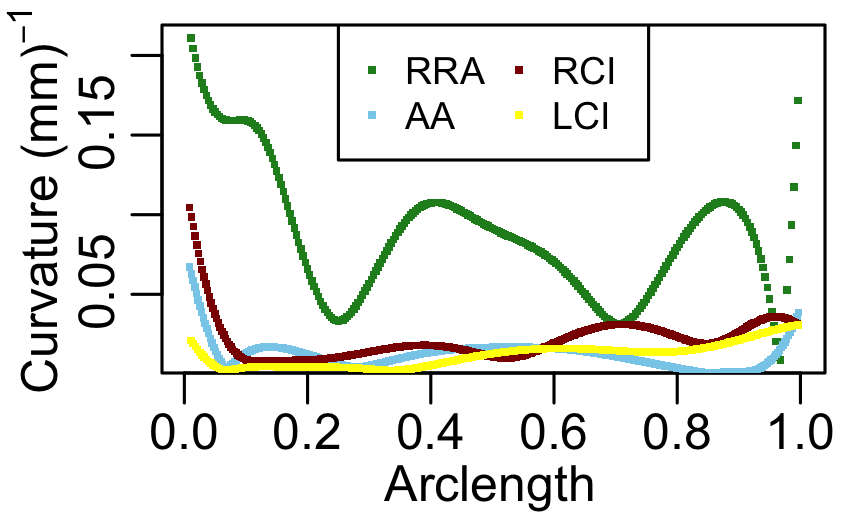}
    \vspace{10pt}
    \\
    \subfigimg[width=3.45in, pos = ll]{\textbf{(b)}}{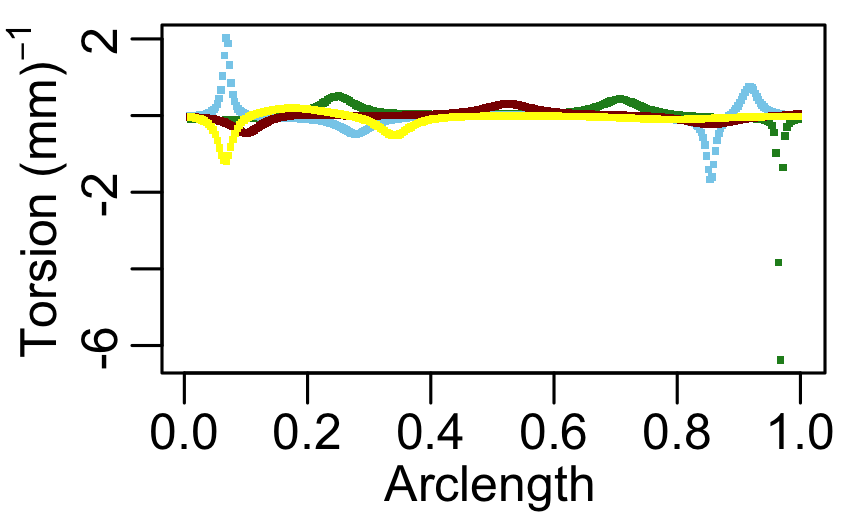}
    \footnotesize
    \caption{Graphs of \textbf{(a)} curvature and \textbf{(b)} torsion versus arc length for data from O'Flynn et al.~\cite{s_oflynn_etal_annalsbiomedeng_2007}.  Measurements are graphed with sampling rates of 5 points per millimeter along the arterial arclengths.  The arteries studied are the right renal artery (RRA), the abdominal artery (AA), the left common iliac artery (LCA), and the right common iliac artery (RCA).}
    \label{fig:oflynn_data}
\end{figure}

In Figure \ref{fig:oflynn_data} we present the curvature and torsion versus arc length graphs for the arteries extracted.  Note that we have graphed these curves in a manner that is reflective of the sampling rate used to calculate associated tortuosity values.  All results match those originally published except for the abdominal aorta.  Specifically, the peaks in torsion located approximately $s_j = 0.05$ and $s_j = 0.85$ should both be inverted.  Despite this, the statistical measurements of curvature torsion presented in Table \ref{tab:oflynn_data} are still in agreement with those originally reported.  An important difference in the tortuosity measures reported by O'Flynn et al.~is that the equations for total curvature, torsion, and combined curvature and torsion are cumulative summations as opposed to integrated along arc length.  Specifically, $\mathcal{TC}^*= \sum_{j=1}^{j=N}|\kappa_j|$, $\mathcal{TT}^*= \sum_{j=1}^{j=N}|\tau_j|$, $\mathcal{TCCT}^*= \sum_{j=1}^{j=N}\sqrt{\kappa_j^2 + \tau_j^2}$.

\subsubsection{Vorobstova et al.~(2016)}
\label{app:vorob_recon}

The data received from Vorobstova et al.~consisted of highly sampled and smoothed centerline coordinates for all coronary arteries involved in their original study \cite{s_vorobstova_etal_annalsbiomedeng_2016}.  This data required zero additional effort on our part in terms of interpolation or smoothing to reproduce their published results (see Table~\ref{tab:vorob_data}).  However, we did have to determine an appropriate sampling rate as it was not explicitly stated in the original paper.  We found the best agreement between the originally published values and our own when using a sampling rate of 10 points/mm.

\section{Demonstrating Equivalence of the $\mathcal{SOA}$ and $\mathcal{ACCT}$ Metrics}
\label{app:metric_equivalence}

Here we present an analytical derivation that demonstrates the equivalence between the sum-of-angles (SOA) and average combined curvature and torsion metrics.  This derivation was motivated by the strength of the correlation observed between the two metrics in Fig.~\ref{fig:soa_acct_samp}. To our knowledge, this result does not exist elsewhere in the literature.

Recall the definition of the $\mathcal{SOA}$ metric from Eq.~(\ref{eq:soam}), 

\begin{align}
\mathcal{SOA}_j = \left \{ \left[ IP_j \right]^2 + \left[TP_j \right]^2 \right\}^{1/2}
\end{align}

\noindent where

\begin{align}
IP_j = \cos^{-1}\left( \frac{\Delta\bv{r}_j}{|\Delta\bv{r}_j|} \cdot \frac{\Delta\bv{r}_{j+1}}{|\Delta\bv{r}_{j+1}|} \right)
\end{align}

\noindent and

\begin{align}
TP_j = \cos^{-1}\left( \frac{\Delta\bv{r}_{j-1} \times \Delta\bv{r}_j}{|\Delta\bv{r}_{j-1} \times \Delta\bv{r}_j|} \cdot \frac{\Delta\bv{r}_{j} \times \Delta\bv{r}_{j+1}}{|\Delta\bv{r}_{j} \times \Delta\bv{r}_{j+1}|} \right)
\end{align}

We focus our attention first on $IP_j$.  Under the assumption of uniform spacing of the vessel coordinates, $|\Delta\bv{r}_j| = \Delta s$ for all $j$, so $IP_j$ can be written as

\begin{align}
IP_j = \cos^{-1}\left( \frac{\Delta\bv{r}_j}{\Delta s} \cdot \frac{\Delta\bv{r}_{j+1}}{\Delta s} \right)
\end{align}

\noindent Recognizing $\Delta \bv{r}_j/\Delta s$ as a first-order discretized difference, in the limit of high sampling rates: $\Delta s \rightarrow 0$ and $\Delta \bv{r}_j/\Delta s \rightarrow d\bv{r}_j/ds$.  By definition, $d\bv{r}_j/ds = \buv{T}_j$ (recalling that in the arc length parameterization $|d\bv{r}_j/ds| = 1$). Therefore

\begin{align}
IP_j = \cos^{-1}\left(\buv{T}_j \cdot \buv{T}_{j+1}\right)
\end{align}

\noindent Using the following dot product identity,

\begin{align}
\bv{a} \cdot \bv{b} = \frac{1}{2}\left( |\bv{a}|^2 + |\bv{b}|^2 - |\bv{a} - \bv{b}|^2 \right)
\end{align}

\noindent and multiplying by a factor of $\Delta s^2/ \Delta s^2$, we have,

\begin{align}
IP_j = \cos^{-1}\left(1 - \frac{\Delta s^2}{2}\left| \frac{\Delta \buv{T}_j}{\Delta s} \right|^2 \right)
\end{align}

\noindent Again, in the limit as $\Delta s \rightarrow 0$, $\Delta \buv{T}_j/\Delta s \rightarrow d\buv{T}_j/ds$.  By definition, $d\buv{T}_j/ds = \kappa_j\buv{N}_j$.  Thus,

\begin{align}
IP_j = \cos^{-1}\left(1 - \frac{\Delta s^2\kappa_j^2}{2}\right)
\end{align} 

Now turning our attention to $TP_j$, we focus first on the numerator of the first term, $\Delta \bv{r}_{j-1} \times \Delta \bv{r}_j$.  The resulting vector can be considered in two parts\textemdash the magnitude and direction.  We will restrict our attention to the magnitude, as the direction will sort itself out during the process.  The magnitude is expressed as $| \Delta \bv{r}_{j-1} \times \Delta \bv{r}_j|$.  Recall the cross-product relation, $|\Delta\bv{r}_{j-1} \times \Delta\bv{r}_j| = |\Delta\bv{r}_{j-1}||\Delta\bv{r}_j|\sin\theta$, where $\theta$ is the angle between the two vectors.  Treating $\Delta \bv{r}_{j-1}$ as the hypotenuse of a right-triangle with legs $\Delta \bv{r}_j$ and $\Delta \bv{r}_j - \Delta \bv{r}_{j-1}$, we can make the substitution $\sin\theta = |\Delta \bv{r}_j - \Delta \bv{r}_{j-1}|/|\Delta\bv{r}_{j-1}|$.  This gives us

\begin{align}
|\Delta \bv{r}_{j-1} \times \Delta \bv{r}_j| = |\Delta \bv{r}_j||\Delta \bv{r}_j - \Delta \bv{r}_{j-1}|
\end{align}

\noindent Noting that $\Delta \bv{r}_j - \Delta \bv{r}_{j-1} = \bv{r}_{j+1} - 2\bv{r}_j + \bv{r}_{j-1}$, and multiplying by a factor of $\Delta s^3/\Delta s^3$ results in

\begin{align}
|\Delta \bv{r}_{j-1} \times \Delta \bv{r}_j| = \Delta s^3 \left( \frac{|\Delta \bv{r}_j|}{\Delta s} \frac{|\bv{r}_{j+1} - 2\bv{r}_j + \bv{r}_{j-1}|}{\Delta s^2} \right)
\end{align}

\noindent Recognizing the first fraction as a first-order difference and the second fraction as a second-order difference, in the limit that $\Delta s \rightarrow 0$, $|\Delta \bv{r}_j|/\Delta s \rightarrow |d\bv{r}_j/ds|$ and $|\bv{r}_{j+1} - 2\bv{r}_j + \bv{r}_{j-1}|/\Delta s^2 \rightarrow |d^2\bv{r}_j/ds^2|$.  By definition, $d\bv{r}_j/ds = \buv{T}_j$, and $d^2\bv{r}_j/ds^2 = \buv{N}_j|d^2\bv{r}_j/ds^2|$, so

\begin{align}
|\Delta \bv{r}_{j-1} \times \Delta \bv{r}_j| = \Delta s^3 \left| \frac{d^2\bv{r}_j}{ds^2} \right| \left(|\buv{T}_j| |\buv{N}_j| \right)
\end{align}

\noindent Despite the equation $|\buv{T}_j| = |\buv{N}_j| = 1$, we will focus on the tangent and normal vectors being perpendicular to each other

\begin{align}
\Delta \bv{r}_{j-1} \times \Delta \bv{r}_j = \Delta s^3 \left| \frac{d^2\bv{r}_j}{ds^2} \right| \buv{T}_j \times \buv{N}_j
\end{align}

\noindent Recalling that: this whole term is divided by its own magnitude in $TP_j$; the same argument can be applied to the second term in $TP_j$; and the definition $\buv{B}_j = \buv{T}_j \times \buv{N}_j$, so that

\begin{align}
TP_j = \cos^{-1} \left( \buv{B}_j \cdot \buv{B}_{j+1} \right)
\end{align}

\noindent Using the same dot-product identity as before gives

\begin{align}
TP_j = \cos^{-1} \left( 1 - \frac{\Delta s^2}{2} \left| \frac{\Delta \buv{B}_j}{\Delta s} \right|^2 \right)
\end{align}

\noindent Again, in the limit as $\Delta s \rightarrow 0$, then $\Delta \buv{B}_j/\Delta s \rightarrow d\buv{B}_j/ds$.  By definition, $d\buv{B}_j/ds = -\tau_j\buv{N}_j$.  Thus,

\begin{align}
TP_j = \cos^{-1}\left( 1 - \frac{\Delta s^2 \tau_j^2}{2} \right)
\end{align}

Returning our attention to the full $\mathcal{SOA}_j$ expression and substituting our approximations based on high sampling rates, we have

\begin{align}
\mathcal{SOA}_j = \sqrt{\cos^{-1}\left( 1 - \frac{\Delta s^2 \kappa_j^2}{2} \right)^2 + \cos^{-1}\left( 1 - \frac{\Delta s^2 \tau_j^2}{2} \right)^2}
\end{align}

\noindent Using the approximation that $\cos^{-1}\left(1- \psi \right)^2 \approx 2 \psi$ for $|\psi| < 1$,

\begin{align}
\mathcal{SOA}_j = \sqrt{ \kappa_j^2 + \tau_j^2 }\Delta s
\end{align}

\noindent Finally, as $\mathcal{SOA} = \sum_{j=1}^N\mathcal{SOA}_j/\sum_{j=1}^N{\Delta s}$, and that $\Delta s$ is a constant for uniform sampling, we have

\begin{align}
\mathcal{SOA} = \frac{1}{N}\sum_{j=1}^{N}\sqrt{\kappa_j^2 + \tau_j^2}
\end{align}

\noindent This is exactly the definition of the average combined curvature-torsion metric\textemdash $\mathcal{ACCT}$.  From this derivation it should be clear how\textemdash in the limit of high sampling rates\textemdash measurements of the sum-of-angles (SOA) and the average combined curvature-torsion metric are not just strongly correlated but exactly equal.  Furthermore, it should be clear that if one is measuring the variation on $\mathcal{SOA}$ that neglects the $TP_j$ term, then the measurements should correlate strongly with the average curvature for low sampling (Lorthois et al.~in  \cite{s_lorthois_etal_microvascres_2014}) and should be exactly equal for high sampling.

\section{Demonstrating Equivalence of Arc Length Normalized Metrics and Average Analogues}
\label{app:all_metric_equivalence}

Let $\mathcal{F}$ represent a measure of vessel tortuosity as a generic function of curvature and torsion, such that $\mathcal{F} = \mathcal{F}(\kappa_j, \tau_j)$.  A total (or integrated) measure of $\mathcal{F}$, denoted as $\mathcal{TF}$, is typically written as $\mathcal{TF} = \int\mathcal{F}(\kappa, \tau)ds$.  However, when evaluating such a function on real data using measured values of curvature $\kappa_j$ and torsion $\tau_j$, one must decide upon a numerical method of integration.  Choosing a left Riemann sum as an example, we have $\mathcal{TF} = \sum_j\mathcal{F}(\kappa_j,\tau_j)(s_{j+1}-s_j)$.  Assuming uniform spacing, then $s_{j+1} - s_j = \Delta s$ is a constant for all $j$, and $\Delta s$ can be factored out of the summation, giving $\mathcal{TF} = \Delta s \sum_j\mathcal{F}(\kappa_j, \tau_j)$.  Expressing total arc length as a summation, $\mathcal{L} = \sum_j\Delta s$, then we again can factor $\Delta s$ from the summation, resulting in $\mathcal{L} = \Delta s N$, for a vessel with $N$ coordinates.  Thus, in the high sampling rate limit, the arc length normalized total metric equates to the average metric, $\mathcal{TF}/\mathcal{L} = 1/N\sum_j\mathcal{F}(\kappa_j, \tau_j).$

\begin{table*}[!t]
\renewcommand{\arraystretch}{1.3}
% if using array.sty, it might be a good idea to tweak the value of
% \extrarowheight as needed to properly center the text within the cells
\caption{Tortuosity measures of data from O'Flynn et al.~(2007) \cite{s_oflynn_etal_annalsbiomedeng_2007}.  Metrics here are calculated as per methods used in cited publication.  Metrics measured are vessel length, $L$, a variation on the distance metric, $\mathcal{D}_1$, a variation on total curvature, $\mathcal{TC}$*, average curvature, $\mathcal{AC}$, a variation on total torsion, $\mathcal{TT}$*, average torsion, $\mathcal{AT}$, a variation on the total combined curvature-torsion, $\mathcal{TCCT}$*, and average combined curvature-torsion, $\mathcal{ACCT}$.  The arteries studied are the right renal artery (RRA), the abdominal artery (AA), the left common iliac artery (LCA), and the right common iliac artery (RCA).}
\label{tab:oflynn_data}
\centering
\begin{tabular}{rrrrrrrrr}
  \hline
 & L (mm) & $\mathcal{D}_1$ & $\mathcal{TC}$* $(mm^{-1})$ & $\mathcal{AC}$ $(mm^{-1})$ & $\mathcal{TT}$* $(mm^{-1})$ & $\mathcal{AT}$ $(mm^{-1})$ & $\mathcal{TCCT}$* $(mm^{-1})$ & $\mathcal{ACCT}$ $(mm^{-1})$ \\ 
  \hline
RRA & 56.64 & 0.25 & 24.18 & 0.09 & 45.56 & 0.16 & 58.69 & 0.21 \\ 
  AA & 71.94 & 0.02 & 4.27 & 0.01 & 56.19 & 0.16 & 57.83 & 0.16 \\ 
  LCIA & 56.63 & 0.01 & 3.21 & 0.01 & 33.48 & 0.12 & 34.23 & 0.12 \\ 
  RCIA & 69.91 & 0.03 & 7.28 & 0.02 & 31.81 & 0.09 & 34.50 & 0.10 \\ 
   \hline
\end{tabular}
\end{table*}

\begin{table}[!t]
%\renewcommand{\arraystretch}{1.3}
% if using array.sty, it might be a good idea to tweak the value of
% \extrarowheight as needed to properly center the text within the cells
\caption{Tortuosity measures of data from Vorobstova et al.~(2016)\cite{s_vorobstova_etal_annalsbiomedeng_2016}.  Metrics measured are the distance metric, $\mathcal{D}$, total torsion, $\mathcal{TT}$, and total curvature, $\mathcal{TC}$.}
\label{tab:vorob_data}
\centering
\begin{tabular}{lrrrr}
  \hline
  & Length (mm) & $\mathcal{D}$ (-) & $\mathcal{TT}$ (-) & $\mathcal{TC}$ (-) \\ 
  \hline
  Patient-A &  &  &  &  \\
  LMCA & 7.881 & 1.009 & 7.672 & 0.715 \\ 
  LAD & 42.764 & 1.051 & 12.959 & 1.708 \\ 
  LAD24 & 41.502 & 1.089 & 16.345 & 4.522 \\ 
  LCX & 25.683 & 1.050 & 7.166 & 1.511 \\ 
  LCX123 & 84.027 & 1.320 & 27.764 & 5.275 \\ 
  MARG1 & 37.648 & 1.056 & 10.395 & 3.259 \\ 
  SPT2 & 33.394 & 1.035 & 8.663 & 2.250 \\
  Patient-B  &  &  &  &  \\
  LAD & 33.465 & 1.031 & 7.967 & 1.719 \\ 
  LAD2\_2 & 16.402 & 1.200 & 6.026 & 2.474 \\ 
  LCX & 13.651 & 1.010 & 6.417 & 0.963 \\ 
  LCX1 & 36.428 & 1.070 & 11.626 & 2.249 \\ 
  ATR & 18.994 & 1.133 & 5.672 & 2.277 \\ 
  MARG2 & 18.742 & 1.173 & 8.751 & 5.261 \\ 
  SPT2 & 18.626 & 1.030 & 7.972 & 1.199 \\ 
  SPT3 & 30.647 & 1.123 & 14.477 & 6.425 \\ 
  Patient-C  &  &  &  &  \\
  LAD\_1 & 17.574 & 1.045 & 5.853 & 1.359 \\ 
  LAD\_2 & 20.874 & 1.015 & 9.546 & 0.879 \\ 
  LAD1 & 26.893 & 1.170 & 7.227 & 3.344 \\ 
  LAD24 & 43.672 & 1.228 & 16.706 & 5.588 \\ 
  LCX & 18.675 & 1.014 & 9.030 & 0.805 \\ 
  LCX12 & 54.162 & 1.109 & 10.947 & 2.183 \\ 
  MARG1 & 63.429 & 1.180 & 15.370 & 3.465 \\ 
  SPT2 & 34.560 & 1.043 & 10.009 & 1.698 \\ 
   \hline
\end{tabular} 
\end{table}

\begin{table*}[!t]
\renewcommand{\arraystretch}{1.3}
\caption{Coefficients of covariance for tortuosity metrics of all 9,000 vessels studied reconstructed at $100X$ resampling rate.  Metrics considered are: average curvature, torsion, and combined curvature and torsion ($\mathcal{AC}$, $\mathcal{AT}$, $\mathcal{ACCT}$); total curvature, torsion, and combined curvature and torsion ($\mathcal{TC}$, $\mathcal{TT}$, $\mathcal{TCCT}$); length-normalized total curvature, torsion, and combined curvature and torsion ($\mathcal{TC}/\mathcal{L}$, $\mathcal{TT}\mathcal{L}$, $\mathcal{TCCT}\mathcal{L}$); curvature-based and normal-vector-based inflection point counts ($\mathcal{IC}_\kappa$, $\mathcal{IC}_\mathcal{N}$); and sum of all angles ($\mathcal{SOA}$).  All correlation coefficients are rounded to two significant digits.  All correlations except those labeled with $\dag$ were significant ($p<0.001$).}
\label{tab:cor_coeff_norm}
\centering
\begin{tabular}{lllllllllllll}
  \hline
 Variable & 1. & 2. & 3. & 4. & 5. & 6. & 7. & 8. & 9. & 10. & 11. & 12. \\ 
  \hline
1. $\mathcal{TT}$ & -- &  &  &  &  &  &  &  &  &  &  &  \\ 
2. $\mathcal{AT}$ & 0.74 & -- &  &  &  &  &  &  &  &  &  &  \\ 
3. $\mathcal{TT}/\mathcal{L}$ & 0.74 & 1 & -- &  &  &  &  &  &  &  &  &  \\ 
4. $\mathcal{AC}$ & 0.18 & 0.4 & 0.4 & -- &  &  &  &  &  &  &  &  \\ 
5. $\mathcal{TC}/\mathcal{L}$ & 0.18 & 0.4 & 0.4 & 1 & -- &  &  &  &  &  &  &  \\ 
6. $\mathcal{TC}$ & 0.51 & 0.23 & 0.23 & 0.59 & 0.59 & -- &  &  &  &  &  &  \\ 
7. $\mathcal{IC_N}$ & 0.38 & 0.19 & 0.19 & 0.53 & 0.53 & 0.87 & -- &  &  &  &  &  \\ 
8. $\mathcal{TCCT}/\mathcal{L}$ & -0.16 & 0.21 & 0.21 & 0.38 & 0.38 & -0.28 & -0.38 & -- &  &  &  &  \\ 
9. $\mathcal{ACCT}$ & -0.16 & 0.21 & 0.21 & 0.38 & 0.38 & -0.28 & -0.38 & 1 & -- &  &  &  \\ 
10. $\mathcal{SOA}$ & -0.16 & 0.21 & 0.21 & 0.38 & 0.38 & -0.28 & -0.38 & 1 & 1 & -- &  &  \\ 
11. $\mathcal{TCCT}$ & 0.33 & 0.069 & 0.068 & 0.19 & 0.18 & 0.35 & 0.082 & 0.41 & 0.41 & 0.41 & -- &  \\ 
12.  $\mathcal{IC_\kappa}$ & 0.18 & -0.0014\dag & -0.0018\dag & 0.29 & 0.29 & 0.55 & 0.49 & 0.07 & 0.072 & 0.07 & 0.55 & -- \\ 
   \hline
\end{tabular}
\end{table*}

%\section*{Acknowledgment}


\ifCLASSOPTIONcaptionsoff
  \newpage
\fi

% \IEEEtriggeratref{8}


\bibliographystyle{IEEEtran}

\end{document}